\documentclass[usenatbib]{mn2e}

\usepackage{amsmath,amsfonts,amssymb,
  graphicx,times,textcomp,verbatim,url}

\newcommand{\standardmu}{$10.70\pm0.03$}

\newcommand{\standardcovar}{\begin{matrix}
1.00&-0.20&-0.54&-0.23\\ 
-0.20&1.00&-0.23&-0.31\\ 
-0.54&-0.23&1.00&0.57\\ 
-0.23&-0.31&0.57&1.00\\ 
\end{matrix}}
\newcommand{\standardrms}{\begin{matrix}B&V&I&H\\0.12&0.03&0.11&0.29\end{matrix}}
\newcommand{\extlawmu}{$10.60\pm0.03$}

\newcommand{\extlawcovar}{\begin{matrix}
1.00&-0.14&-0.49&-0.15\\ 
-0.14&1.00&-0.38&-0.38\\ 
-0.49&-0.38&1.00&0.46\\ 
-0.15&-0.38&0.46&1.00\\ 
\end{matrix}}
\newcommand{\extlawrms}{\begin{matrix}B&V&I&H\\0.13&0.03&0.10&0.29\end{matrix}}
\newcommand{\Zstandardonemu}{$10.75\pm0.05$}
\newcommand{\Zstandardonegammaa}{$-0.18\pm0.20$}

\newcommand{\Zstandardtwomu}{$10.80\pm0.11$}
\newcommand{\Zstandardtwogammaa}{$-0.32\pm0.35$}

\newcommand{\Bstandardmu}{$10.74\pm0.05$}
\newcommand{\Bstandardgammaa}{$-0.50\pm0.54$}

\newcommand{\allstandardmu}{$10.73\pm0.01$}

\newcommand{\allstandardrchi}{$ 1.00$}
\newcommand{\allstandardcovar}{\begin{matrix}
1.00&-0.02&-0.66&-0.25\\ 
-0.02&1.00&-0.38&-0.23\\ 
-0.66&-0.38&1.00&0.75\\ 
-0.25&-0.23&0.75&1.00\\ 
\end{matrix}}
\newcommand{\allstandardrms}{\begin{matrix}B&V&I&H\\0.11&0.04&0.09&0.30\end{matrix}}

\newcommand{\allZstandardonemu}{$10.81\pm0.03$}
\newcommand{\allZstandardonegammaa}{$-0.24\pm0.08$}

\newcommand{\allZstandardonerchi}{$ 0.97$}

\newcommand{\allZstandardtwomu}{$10.89\pm0.06$}
\newcommand{\allZstandardtwogammaa}{$-0.42\pm0.14$}

\newcommand{\allBstandardmu}{$10.81\pm0.03$}
\newcommand{\allBstandardgammaa}{$-0.65\pm0.22$}

\newcommand{\allZstandardonegtwogammaa}{$-0.00\pm0.15$}
\newcommand{\allZstandardonegtwogammab}{$-0.44\pm0.24$}

\newcommand{\allZstandardtwogtwogammaa}{$-0.20\pm0.18$}
\newcommand{\allZstandardtwogtwogammab}{$-0.44\pm0.22$}

\newcommand{\allBstandardgtwogammaa}{$0.02\pm0.43$}
\newcommand{\allBstandardgtwogammab}{$-1.23\pm0.68$}

\newcommand{\allZstandardonegtwoBVImu}{$10.83\pm0.04$}
\newcommand{\allZstandardonegtwoBVIgammaa}{$-0.61\pm0.33$}
\newcommand{\allZstandardonegtwoBVIgammab}{$-0.20\pm0.18$}

\newcommand{\allZstandardtwogtwoBVImu}{$10.91\pm0.06$}
\newcommand{\allZstandardtwogtwoBVIgammaa}{$-0.84\pm0.35$}
\newcommand{\allZstandardtwogtwoBVIgammab}{$-0.22\pm0.19$}

\newcommand{\allBstandardgtwoBVImu}{$10.83\pm0.03$}
\newcommand{\allBstandardgtwoBVIgammaa}{$-1.69\pm0.91$}

\newcommand{\bootstandardmu}{$10.71\pm0.06$}
\newcommand{\bootallstandardmu}{$10.73\pm0.02$}
\newcommand{\bootextlawmu}{$10.60\pm0.07$}

\newcommand{\bootZstandardtwomu}{$10.79\pm0.12$}
\newcommand{\bootZstandardtwogammaa}{$-0.29\pm0.39$}
\newcommand{\bootZextlawtwomu}{$10.69\pm0.14$}
\newcommand{\bootZextlawtwogammaa}{$-0.34\pm0.42$}
\newcommand{\bootBstandardmu}{$10.74\pm0.09$}
\newcommand{\bootBstandardgammaa}{$-0.45\pm0.62$}
\newcommand{\bootBextlawmu}{$10.64\pm0.11$}
\newcommand{\bootBextlawgammaa}{$-0.54\pm0.69$}

\newcommand{\bootallZstandardtwomu}{$10.98\pm0.10$}
\newcommand{\bootallZstandardtwogammaa}{$-0.61\pm0.21$}
\newcommand{\bootallZextlawtwomu}{$10.94\pm0.13$}
\newcommand{\bootallZextlawtwogammaa}{$-0.89\pm0.27$}
\newcommand{\bootallBstandardmu}{$10.88\pm0.09$}
\newcommand{\bootallBstandardgammaa}{$-0.95\pm0.35$}
\newcommand{\bootallBextlawmu}{$10.79\pm0.13$}
\newcommand{\bootallBextlawgammaa}{$-1.42\pm0.45$}

\newcommand{\bootZstandardtwogtwomu}{$10.79\pm0.12$}
\newcommand{\bootZstandardtwogtwogammaa}{$-0.08\pm0.59$}
\newcommand{\bootZstandardtwogtwogammab}{$-0.32\pm0.60$}
\newcommand{\bootZextlawtwogtwomu}{$10.69\pm0.14$}
\newcommand{\bootZextlawtwogtwogammaa}{$-0.62\pm0.89$}
\newcommand{\bootZextlawtwogtwogammab}{$ 0.31\pm0.79$}
\newcommand{\bootBstandardgtwomu}{$10.74\pm0.09$}
\newcommand{\bootBstandardgtwogammaa}{$-0.20\pm1.31$}
\newcommand{\bootBstandardgtwogammab}{$-0.44\pm1.49$}
\newcommand{\bootBextlawgtwomu}{$10.63\pm0.10$}
\newcommand{\bootBextlawgtwogammaa}{$-1.10\pm2.01$}
\newcommand{\bootBextlawgtwogammab}{$ 0.62\pm1.96$}
\newcommand{\bootallZstandardtwogtwomu}{$10.98\pm0.10$}
\newcommand{\bootallZstandardtwogtwogammaa}{$-0.35\pm0.26$}
\newcommand{\bootallZstandardtwogtwogammab}{$-0.47\pm0.24$}
\newcommand{\bootallZextlawtwogtwomu}{$10.94\pm0.13$}
\newcommand{\bootallZextlawtwogtwogammaa}{$-1.22\pm0.40$}
\newcommand{\bootallZextlawtwogtwogammab}{$ 0.44\pm0.35$}
\newcommand{\bootallBstandardgtwomu}{$10.88\pm0.10$}
\newcommand{\bootallBstandardgtwogammaa}{$-0.43\pm0.56$}
\newcommand{\bootallBstandardgtwogammab}{$-0.95\pm0.63$}
\newcommand{\bootallBextlawgtwomu}{$10.79\pm0.13$}
\newcommand{\bootallBextlawgtwogammaa}{$-2.31\pm1.00$}
\newcommand{\bootallBextlawgtwogammab}{$ 1.17\pm1.09$}
\newcommand{\bootZstandardtwogtwoBVImu}{$10.80\pm0.11$}
\newcommand{\bootZstandardtwogtwoBVIgammaa}{$-0.82\pm0.61$}
\newcommand{\bootZstandardtwogtwoBVIgammab}{$-0.29\pm0.18$}
\newcommand{\bootZextlawtwogtwoBVImu}{$10.69\pm0.13$}
\newcommand{\bootZextlawtwogtwoBVIgammaa}{$-0.12\pm0.55$}
\newcommand{\bootZextlawtwogtwoBVIgammab}{$ 0.11\pm0.11$}
\newcommand{\bootBstandardgtwoBVImu}{$10.75\pm0.07$}
\newcommand{\bootBstandardgtwoBVIgammaa}{$-1.14\pm1.49$}
\newcommand{\bootBstandardgtwoBVIgammab}{$-0.42\pm0.60$}
\newcommand{\bootBextlawgtwoBVImu}{$10.62\pm0.09$}
\newcommand{\bootBextlawgtwoBVIgammaa}{$-0.08\pm1.10$}
\newcommand{\bootBextlawgtwoBVIgammab}{$ 0.24\pm0.29$}
\newcommand{\bootallZstandardtwogtwoBVImu}{$10.98\pm0.09$}
\newcommand{\bootallZstandardtwogtwoBVIgammaa}{$-1.03\pm0.37$}
\newcommand{\bootallZstandardtwogtwoBVIgammab}{$-0.23\pm0.13$}
\newcommand{\bootallZextlawtwogtwoBVImu}{$10.95\pm0.13$}
\newcommand{\bootallZextlawtwogtwoBVIgammaa}{$-0.64\pm0.35$}
\newcommand{\bootallZextlawtwogtwoBVIgammab}{$ 0.13\pm0.08$}
\newcommand{\bootallBstandardgtwoBVImu}{$10.88\pm0.10$}
\newcommand{\bootallBstandardgtwoBVIgammaa}{$-1.87\pm0.92$}
\newcommand{\bootallBstandardgtwoBVIgammab}{$-0.50\pm0.35$}
\newcommand{\bootallBextlawgtwoBVImu}{$10.80\pm0.14$}
\newcommand{\bootallBextlawgtwoBVIgammaa}{$-0.81\pm0.77$}
\newcommand{\bootallBextlawgtwoBVIgammab}{$ 0.30\pm0.22$}

\newcommand{\bootZstandardtwogtwopriormu}{$10.84\pm0.07$}
\newcommand{\bootZstandardtwogtwopriorgammaa}{$-0.24\pm0.47$}
\newcommand{\bootZstandardtwogtwopriorgammab}{$-0.33\pm0.61$}
\newcommand{\bootBstandardgtwopriormu}{$10.77\pm0.08$}
\newcommand{\bootBstandardgtwopriorgammaa}{$-0.34\pm1.18$}
\newcommand{\bootBstandardgtwopriorgammab}{$-0.47\pm1.52$}
             
\newcommand{\bootallZstandardtwogtwopriormu}{$10.98\pm0.09$}
\newcommand{\bootallZstandardtwogtwopriorgammaa}{$-0.34\pm0.25$}
\newcommand{\bootallZstandardtwogtwopriorgammab}{$-0.48\pm0.24$}
\newcommand{\bootallBstandardgtwopriormu}{$10.88\pm0.09$}
\newcommand{\bootallBstandardgtwopriorgammaa}{$-0.43\pm0.54$}
\newcommand{\bootallBstandardgtwopriorgammab}{$-0.94\pm0.63$}

\newcommand{\bootZstandardtwogtwoBVIpriormu}{$10.85\pm0.06$}
\newcommand{\bootZstandardtwogtwoBVIpriorgammaa}{$-0.96\pm0.50$}
\newcommand{\bootZstandardtwogtwoBVIpriorgammab}{$-0.30\pm0.18$}
\newcommand{\bootBstandardgtwoBVIpriormu}{$10.78\pm0.07$}
\newcommand{\bootBstandardgtwoBVIpriorgammaa}{$-1.37\pm1.58$}
\newcommand{\bootBstandardgtwoBVIpriorgammab}{$-0.42\pm0.63$}

\newcommand{\bootallZstandardtwogtwoBVIpriormu}{$10.98\pm0.09$}
\newcommand{\bootallZstandardtwogtwoBVIpriorgammaa}{$-1.02\pm0.37$}
\newcommand{\bootallZstandardtwogtwoBVIpriorgammab}{$-0.24\pm0.13$}
\newcommand{\bootallBstandardgtwoBVIpriormu}{$10.88\pm0.09$}
\newcommand{\bootallBstandardgtwoBVIpriorgammaa}{$-1.86\pm0.92$}
\newcommand{\bootallBstandardgtwoBVIpriorgammab}{$-0.50\pm0.35$}

\newcommand{\radinmu}{$10.72\pm0.06$}

\newcommand{\radmidmu}{$10.62\pm0.05$}

\newcommand{\radoutmu}{$10.78\pm0.04$}

\newcommand{\macrimetal}{ $-0.29 \pm0.09_{stat}  \pm 0.05_{sys}$}

\newcommand{\Zzero}{$9.06\pm0.03$}
\newcommand{\Bzero}{$8.51\pm0.02$}
\newcommand{\Zslope}{$-0.30\pm0.05$}
\newcommand{\Bslope}{$-0.19\pm0.04$}

\newcommand{\RVbest}{$4.9_{-0.7}^{+0.9}$}
\newcommand{\RVbestshort}{$4.9$}
\newcommand{\bootRV}{$4.8 \pm 1.7$}
\newcommand{\RVbestM}{$4.7_{-0.4}^{+0.5}$}
\newcommand{\allextlawmuM}{$10.60 \pm 0.02$}

\newcommand{\gammamin}{$-0.24\pm0.08$}
\newcommand{\gammamax}{$-1.4\pm0.45$}

\newcommand{\muLBT}{$10.70 \pm 0.08_{stat} \pm 0.06_{sys}$}
\newcommand{\muM}{$10.83   \pm 0.08_{stat} \pm 0.09_{sys}$}
\newcommand{\muLMCLBT}{$18.70 \pm 0.12$}
\newcommand{\muLMCM}{$18.57   \pm 0.14$}

\newcommand{\distLMCM}{$51.82  \pm 3.23$}

\title[The Cepheid distance to NGC 4258]{The Cepheid distance to the maser-host galaxy NGC 4258:  Studying systematics with the Large Binocular Telescope}
\author[M. M. Fausnaugh et al.]{M. M. Fausnaugh,$^1$  C. S. Kochanek,$^{1,2}$  J. R. Gerke,$^1$
  \newauthor
  L. M. Macri,$^3$  A. G. Riess,$^{4,5}$ and  K. Z. Stanek$^1$\\
  $^1$Department of Astronomy, The Ohio State University, 140 West 18th Avenue, Columbus, OH 43210, USA\\
  $^{2}$Center for Cosmology and AstroParticle Physics, The Ohio State University, 191 West Woodruff Avenue, Columbus, OH 43210,USA\\
  $^{3}$George P. and Cynthia Woods Mitchell Institute for Fundamental Physics and Astronomy, Department of Physics \& Astronomy, Texas A\&M\\
  University, 4242 TAMU, College Station, TX 77843-4242, USA\\
  $^{4}$Department of Physics and Astronomy, Johns Hopkins University,  Baltimore, MD 21218, USA\\
  $^{5}$Space Telescope Science Institute, 3700 San Martin Drive,
  Baltimore, MD 21218, USA\\
}
\begin{document}

\maketitle

\begin{abstract}
  We identify and phase a sample of 81 Cepheids in the maser-host
  galaxy NGC 4258 using the Large Binocular Telescope (LBT), and
  obtain calibrated mean magnitudes in up to 4 filters for a subset of
  43 Cepheids using archival \emph{HST} data.  We employ 3 models to
  study the systematic effects of extinction, the assumed extinction
  law, and metallicity on the Cepheid distance to NGC 4258.  We find a
  correction to the Cepheid colors consistent with a grayer extinction
  law in NGC 4258 compared to the Milky Way ($R_V =$ \RVbest),
  although we believe this is indicative of other systematic effects.
  If we combine our Cepheid sample with previously known Cepheids, we
  find a significant metallicity adjustment to the distance modulus of
  $\gamma_1 =$ \bootallZstandardtwogammaa\ mag/dex for the
  \citet{Zaritsky1994} metallicity scale, as well as a weak trend of
  Cepheid colors with metallicity.  Conclusions about the absolute
  effect of metallicity on Cepheid mean magnitudes are limited by the
  available data on the metallicity gradient in NGC 4258, but our
  Cepheid data require at least some metallicity adjustment to make
  the Cepheid distance consistent with independent distances to the
  LMC and NGC 4258.  From our ensemble of models and the geometric
  maser distance of NGC 4258 ($\mu_{N4258} = 29.40 \pm 0.06$ mag), we
  estimate $\mu_{LMC} =$\muLMCM\ mag (\distLMCM\ kpc), including the
  uncertainties due to metallicity.
\end{abstract}
\section{INTRODUCTION}
Cepheid variables remain important for cosmological studies because
they anchor the local cosmological distance scale (see the review by
\citealt{Freedman2010}).  Recent measurements of the Hubble constant
$H_0$ from Cepheids (\citealt{Riess2011}, \citealt{Freedman2012},
\citealt{Efstathiou2014}) are in moderate tension with determinations
from from the cosmic microwave background (\citealt{Planck2013}) and
baryon acoustic oscillations (\citealt{BAO2014}).  If these
discrepancies are confirmed at higher significance, they could be
evidence of `new Physics,' for example, an additional relativistic
species in the early Universe. However, before such claims can be
made, it is critical to have a better understanding of systematic
uncertainties in the local distance scale.  These uncertainties
include calibration of the Cepheid period-luminosity (PL) relation,
and any dependence of the mean magnitudes and colors on extinction,
metallicity, and blending.

Determining the absolute zero point of the PL relation requires either
a sample of Galactic Cepheids at known distances, or (at least) one
independently determined distance to an external galaxy.  The Large
Magellanic Cloud (LMC), as the closest galaxy to the Milky Way, has
traditionally served as the calibrating galaxy
(\citealt{Freedman2012}).  Several independent distances to the LMC
exist, such as those derived from eclipsing binaries
(\citealt{Bonanos2011}, \citealt{Pietrzy2013}) or red-clump stars
(\citealt{Subramanian2013}).  However, uncertainties in the distance
to the LMC continue to be a significant source of systematic error for
the Cepheid distance scale.  Recently, an alternative calibrating
galaxy has been provided by NGC 4258.  A precise geometric distance to
this galaxy (3\%) has been determined by \citet{Humphreys2013} based
on the kinematics of water masers near the galaxy's nucleus.  Such a
high precision measurement makes NGC 4258 a good candidate for
calibrating the Cepheid PL relation.  If NGC 4258 is to serve as the
calibrating galaxy, it is extremely important to understand the
systematic effects influencing the PL relation and Cepheid mean
magnitudes in this galaxy.  Moreover, if the independent distances to
the LMC and NGC 4258 are correct, they provide a powerful check on
systematic effects in the Cepheid distance scale.

For Cepheids, the standard approach for treating extinction is to
obtain two-band photometry, from which it is trivial to estimate an
extinction correction given a known extinction law (the so-called
``Wesenheit'' indices).  Recent work has focused on expanding
observations of Cepheids to the near and mid-infrared (IR), where the
effects of extinction are significantly smaller than in the optical.
However, it is usually assumed that the form of the extinction law
follows the \citet{Cardelli1989} parameterization, with the ratio of
total to selective extinction $R_V = A_V/E(B-V)$ chosen to be either
$3.1$ or 3.3 (e.g., \citealt{Macri2006}, \citealt{Shappee2011},
\citealt{Gerke2011}, \citealt{Riess2011}, and \citealt{Freedman2012}).
While $R_V=3.1$ is a reasonable average for sight lines within our own
galaxy, it is also known that the extinction law varies between sight
lines and galaxies, presumably due to variations in the physical
properties of the dust grains (\citealt{Cardelli1989}).  Most Cepheid
studies approach this problem by simply adding a small contribution
($\le 1\%$) to the systematic error budget for the uncertainty in
$R_V$ (e.g., \citealt{Riess2009} and \citealt{Shappee2011}), although
a few studies measure $R_V$ directly or explore its effects on the
distance modulus. For example, \citet{Pejcha2012} were able to measure
the mean extinction law for a large sample of Cepheids drawn from the
Galaxy, LMC, and Small Magellanic Cloud, and they found $R_V = 3.127$,
in good agreement with the canonical value.  Nevertheless, even in the
near and mid-IR, the extinction law exhibits variations of shape along
different sight lines (\citealt{Flaherty2007},
\citealt{Nishiyama2009}), and it is an open question whether
$R_V=3.1$\ is a reasonable estimate of this parameter for all
galaxies.

Metallicity is also expected to have an important effect on Cepheid
mean magnitudes and colors (e.g., \citealt{Romaniello2008},
\citealt{Bono2010}, \citealt{Freedman2011a}).  Studies to date depend
on galaxies with significant metallicity gradients, for example M101
(\citealt{Kennicutt1998} and \citealt{Shappee2011}) or M81
(\citealt{Gerke2011}), but the gradients (and hence the impact of
metallicity on Cepheid distances) depend sensitively on the method
used to measure the metallicity of H II regions in the host galaxy
(e.g., \citeauthor {Bresolin2011} 2011ab).  Previous empirical
measurements of the metallicity effect on distances have ranged from
non-detections to $-0.89$ mag/dex, with typical values of about
$-0.27$ mag/dex.  The general consensus is that metal-rich Cepheids
are brighter and redder than their metal-poor counterparts
(\citealt{Gould1994}, \citealt{Kochanek1997}, \citealt{Kennicutt1998},
\citealt{Macri2006}, \citealt{Shappee2011}, \citealt{Gerke2011},
\citealt{Mager2013}).  Furthermore, stellar pulsation models indicate
that the metallicity dependence varies across pass bands, and may not
be a monotonic function of wavelength (\citealt{Bono2008},
\citealt{Bono2010}).  Improvements in our understanding of the
metallicity effect require data to be gathered in a wide range of
photometric bands, as well as obtaining better estimates of Cepheid
metallicities or their proxies.

Finally, Cepheid mean magnitudes may be biased due to blending
(\citealt{Stanek1999}, \citealt{Mochejska2000}, \citealt{Macri2001},
\citealt{Chavez2012}).  As massive stars, a sizable fraction of
Cepheids are expected to have nearby unresolved companions, which will
bias the Cepheid mean magnitudes, reduce their apparent amplitudes,
and (typically) make them appear bluer.  The effects of blending have
been estimated by injecting artificial stars into the PSF of known
Cepheids, and looking for changes in the recovered photometry
(e.g., \citealt{Riess2009}, \citealt{Riess2011}).  However, no study
has systematically determined the magnitude of this effect,
particularly as a function of distance, and existing corrections do
not take into account the strong clustering of massive stars
(\citealt{Harris1999}).

In this study, we redetermine the Cepheid distance to NGC 4258 and
examine the effects of extinction, the assumed extinction law, and
metallicity on the measured distance.  The last independent selection
of a Cepheid sample in NGC 4258 was by \citeauthor{Macri2006} (2006,
hereafter M06).  Their large Cepheid sample (89 Cepheids were used in
the final fit) was identified with the \emph{Hubble Space Telescope}
(\emph{HST}) in two fields at different galactocentric radii -- an
``inner'' field at 6.3 kpc and ``outer'' field at 17.1 kpc. They found
a distance modulus relative to the LMC of $10.71\pm 0.04_{stat}\pm
0.05_{sys} $ mag for the inner field and
$10.87\pm0.05_{stat}\pm0.05_{sys}$ mag for the outer field.  Based on
the metallicity gradient determined by \citeauthor{Zaritsky1994}
(1994, hereafter Z94), they interpreted this difference as a
metallicity effect, with a dependence of \macrimetal\ mag/dex.

Here, we identify a new Cepheid sample in NGC 4258, drawn from a wide
range of galactocentric radii and azimuthal angles using the
Large Binocular Telescope (LBT), and calibrate the
Cepheid mean magnitudes in four bands using \emph{HST}.  In \S 2 we
describe our observations, period search, and criteria for identifying
Cepheids.  In \S 3 we explain our procedure for calibrating the
Cepheid mean magnitudes from \emph{HST} observations.  In \S 4 we
describe three models for the distance to NGC 4258, in which we
sequentially explore the effects of extinction, the assumed extinction
law, and metallicity on the estimated distance modulus.  In \S 5, we
present our results and compare them to previous studies.  Finally, in
\S 6, we review the systematic effects associated with this study, and
provide a calibration of the absolute PL relations.

\section{OBSERVATIONS AND DATA REDUCTION}
NGC 4258 was observed on 32 nights between March of 2008 and June of
2013 with the Large Binocular Cameras (LBC, \citealt{Giallongo2008})
on the LBT (\citealt{Hill2010}), as part of an observational
search for failed supernovae \citep{Gerke2014}.  Each camera has an
approximate field of view of $23'$x$23'$, easily framing the entire
disk of NGC 4258.  The LBC consists of 2 cameras, one for each primary
mirror, with LBC/Blue optimized for wavelengths of 320-500 nm, and
LBC/Red for wavelengths of 500-1000 nm.  Four to nine exposures were
obtained each night, each of 200 seconds.  The blue-side observations
cycled through the $UBV$ filters, while the red side only used the R
band.  Images were over-scan corrected, bias subtracted, and flat
fielded using the IRAF MSCRED package.  The nightly exposures were
then averaged into a single image, or averaged into 2 images if more
than 8 exposures were available.  These procedures yielded 32--35
images in each of the Johnson/Cousins $UBVR$ bands.  After excluding
images where the full width at half the maximum (FWHM) of the point
spread function (PSF) exceeded $1\farcs6$, we were left with 20--26
images per filter.

Following \citeauthor{Gerke2011} (2011, hereafter G11), we searched
for variable sources using the ISIS image subtraction package
(\citealt{Alard1998}).  We first built a reference image from the 4--5
images with the best seeing and lowest sky levels in all filters.  All
images were then registered and aligned to the frame of the $R$ band
reference image.  The reference image was scaled and convolved with a
spatially variable kernel to match the PSF for each epoch, and then
subtracted to leave only sources with variable flux.  We next
constructed a ``variability'' image, equal to the root-mean-square
(rms) of the subtracted images, and identified variable sources using
Sextractor (\citealt{Bertin1996}).  This procedure yielded
approximately 2000 variable sources in each band.  We extracted
lightcurves for these sources using ISIS's photometry package.

\subsection{Period search}

In order to identify Cepheid variables, we adopted the Cepheid
lightcurve templates constructed by \citet{Pejcha2012}, and employed a
brute force fitting routine.  These templates have the virtues of
being physically motivated and derived from a large data set--177,314
data points from 287 Cepheids in 29 different bands.  The templates
parameterize a Cepheid lightcurve as variation in the star's
temperature and radius.  The time dependence is modeled by a 20-term
Fourier series, and the flux in a given filter is calculated directly
from the physical parameters.  We used the resulting template
lightcurves\footnote{ The templates can be downloaded at
  \url{http://www.astronomy.ohio-state.edu/~pejcha/cepheids/} and are
  available as tables in \citet{Pejcha2012}. } $T_F(\phi)$ for filter
$F$ at phase $\phi$, each with self-consistently scaled amplitudes for
fundamental mode Cepheids with periods $P$ between 10 and 100 days.
We restricted our Cepheid sample to the same range of periods, fitting
the lightcurves to the templates by phasing the data to 415 different
periods between 10 and 100 days.  This has no practical consequences,
since $P\leq 10$ day Cepheids are too faint for the LBT survey and
$P>100$ day Cepheids are both rare and likely to lie on a different PL
relation \citep{Bird2009}.  The phase at an epoch $t_i$
\begin{align}
\phi_i = \frac{t_i-t_0}{P}
\end{align}
is determined by the period $P$ and a reference time $t_0$.  The
periods were chosen so that the phase shift between sequential periods
over the span of the data ($\Delta t = 1919.8$ days, about 5 years) was
\begin{align}
\delta \phi = - \frac{\Delta t}{P^2}\delta P = 0.4\text{ radians}.
\end{align}
The value of $0.4$ radians (6\% of a full cycle) was empirically
tested by applying the method to the known Cepheids in M81 from
G11.

For each LBT $B$, $V$, and $R$ band lightcurve, we converted the
differential counts and their uncertainties to fiducial magnitudes
arbitrarily centered at 13.5 mag (the calibration of the mean
magnitudes is discussed in \S3), and fit the observed lightcurves to
the templates by minimizing
\begin{align}
  \chi^2 &= \sum_i\left(\frac{m_{Fi} - ( \hat{m} + AT_F(\phi_i)
      )}{\sigma_i} \right)^2
\end{align}
where $\hat m$ and $A$ are the mean magnitude and the
amplitude of the Cepheid model, and $\sigma_i$ is is the uncertainty
in the measurement.  We made no attempt to match lightcurves in the
three different bands until after the period search, so as to impose
an additional check on our procedure.  Using an implementation of the
Levenberg-Marquardt $\chi^2$--minimization algorithm (MPFIT,
\citealt{Markwardt2009}), we allowed $\hat m$, $A$, and
$t_0$ to freely vary and calculated the minimum $\chi^2$ for our grid
of periods.  We then sampled an additional 100 periods spanning the
interval around the best-fit period.  The period with the over-all
minimum value of $\chi^2$ was taken as an initial estimate.  This
approach has the advantage over periodograms of using the
period-dependent shape of the lightcurve to help break period
degeneracies (aliases).

To eliminate variables that are not Cepheids, we first compared each
source's goodness of fit as a Cepheid, $\chi_C^2$, to that for a
linear trend, $\chi_{lin}^2$, using the F-test.  The F statistic is
defined as
\begin{align}
F = \frac{\chi^2_{\text{lin}}/dof}{\chi^2_{\text{C}}/dof},
\end{align}
where $dof$ is the number of degrees of freedom in the fit.  We
eliminated all light curves with $F < 2.5$ from our sample---for
Gaussian statistics, with $F>2.5$ and $dof\sim 25$ (depending on the
fit, the lightcurve, and the filter), the hypothesis that both models
fit the data equally well can be ruled out at $>99\%$ confidence. In
practice, we only use this cut to reduce the number of candidates,
since contaminants such as single epoch novae also pass the F test.
This left us with 156 lightcurves to examine by eye, both for the
quality of the fits and for any obvious problems in the subtracted
images.  After verifying that the lightcurves followed the typical
saw-tooth pattern characteristic of longer-period Cepheid variables,
and that all objects had clean subtractions, we matched lightcurves
extracted from different filters by spatial coordinates.  This allows
us to check the recovered periods of unique Cepheids in different
filters.  Our procedure yielded 81 unique Cepheids, 40 of which were
matched in two or more filters.  We found the periods from different
filters to be in excellent agreement -- the average difference in
period was 0.005 days.  The coordinates and periods of the Cepheids
are reported in Tables \ref{tab:cephdata} and
\ref{tab:cephdata_nocal}.

An additional complication arose from the systematic underestimation
of lightcurve uncertainties by ISIS.  This underestimation does not
affect the F-test because the F-test only compares the relative
ability of two models to describe a given data set.  However, for our
determination of the mean magnitudes (\S 3), it is useful to adjust
the formal errorbars so that they are consistent with the observed
scatter.  The retained lightcurves typically had formal
$\chi^2_{C}/dof$ values of $1.2 - 9.7$, with a median value of $4.0$.
Three of the brightest Cepheids had $\chi^2_{C}/dof$ greater than
10.0, with the maximum being $22.1$, due to the small fractional
uncertainty estimates of ISIS.  In all cases, we broadened the
photometric uncertainties so that $\chi^2_{C}/dof =1$ for each
individual lightcurve.

\section{HST CALIBRATION}
The next step is to measure the Cepheid's apparent magnitudes and
determine the true value of $\hat m$.  However, photometric
measurements in the LBT reference images are subject to considerable
uncertainty due to crowding and confusion with other sources.
Instead, we calibrate the Cepheid lightcurves and determine $\hat m $
from higher-resolution \emph{HST} data. NGC 4258 was observed as part
of the Supernovae and $H_0$ for the Equation of State (SH0ES) project
(\citealt{Riess2009}).  There were 17 observations of NGC 4258 between
December of 2009 and May of 2010 using the Advanced Camera for Surveys
(ACS) and Wide Field Camera 3 (WFC3) in the F435W, F555W, F814W, and
F160W filters.  These filters roughly correspond to the
Johnson/Cousins $BVIH$ bands, and the observations spanned most of the
galaxy's disk.  PSF photometry was performed on the images, with
fluxes calculated in the \emph{HST} VEGAMAG system.  A full
description of the data reduction and PSF photometry procedures can be
found in \citet{Riess2009} and \citet{Riess2011}.

The LBT Cepheid candidates were matched with \emph{HST} sources by
calculating the mean offsets of the brightest stars in the \emph{HST} F555W
images from the LBT $V$ band images, and shifting all LBT sources by
this amount.  Any \emph{HST} source within $0\farcs 23$ ($\sim 1$ LBC pixel)
was selected as a potential match.  All matches were verified by eye,
and sources that lacked clean, isolated matches were cut from our
sample.  In practice, the mean shifts were less than $0\farcs05$.  A
few sources were cut due to crowded/confused matches, but roughly a
quarter of our sample did not match any bright \emph{HST} sources.  It is
likely that these sources lacked sufficient contrast with the galaxy's
surface brightness to be identified in the \emph{HST} images, even though we
detected them in the subtracted LBT images.  While \emph{HST} provides less
crowded direct images, the LBT variability image is even less crowded,
allowing the robust identification of variables even in very dense
stellar fields.  An additional check was made on the photometric
sharpness of each source.  Anomalously high or low sharpness
measurements indicate blends, extended sources, or image processing
artifacts (cosmic rays, etc.).  All sources met our sharpness
criterion of $-1\le S \le 1$.

Of our 81 Cepheid candidates, 16 were outside of the \emph{HST} footprint.
Of the remaining 65, we found 49 unambiguous matches in the \emph{HST}
fields.  Through this point, each Cepheid had some mixture of $B$, $V$, and
$R$ band LBT lightcurves.  However, we required $B$ and $V$ band LBT
lightcurves for our calibration procedure. If either lightcurve was
missing from the LBT data, we forced ISIS to extract photometry at the
position of the source in the subtracted LBT images for the missing
filter, and visually inspected the phased lightcurves.  This step was
necessary for 20 sources, which were primarily missing $B$ band
lightcurves.  Six sources were rejected because the newly extracted
lightcurves were poorly phased, bringing our sample to 43 Cepheids.
We then re-scaled the lightcurve uncertainties as described above.

We next attempted to identify these sources in the \emph{HST} F160W
filter images.  Sources in the $BVI$ images do not always have obvious
near IR counterparts (\citealt{Riess2011}), and the footprint of the
F160W band is not identical to that of the optical data.  We extracted
F160W band photometry for the expected position of the Cepheids based
on their F555W band positions.  Of our \emph{HST}-matched sample, 11
Cepheids were outside of the \emph{HST} F160W band coverage of NGC
4258.  For 8 other Cepheids, the F160W band measurements were
unreliable, with uncertainties $>1.0$ magnitudes.  Our final
\emph{HST}-calibrated sample consists of 43 Cepheids, 24 of which have
usable $BVIH$ photometry, 17 have $BVI$ photometry, and 2 Cepheids
only have $VI$ photometry.

\subsection{Calibration}
We do not want to simply use the \emph{HST} magnitudes as a random
phase estimate of $\hat m $.  Rather, we use the LBT data and the
\cite{Pejcha2012} templates described in \S2.1 to determine the phases
and amplitudes of the Cepheids at each epoch of the \emph{HST}
observations.  At the time of the \emph{HST} observation, the observed
magnitude in a filter $F$ is given by
\begin{align}
m_F = \hat m_F + A_FT_F(\phi_{HST}-\phi_0).
\end{align}
where $T_F$ is the same template defined in \S2.1.  With knowledge of
the amplitude and the phase, we can determine the mean magnitude $\hat
m_F$ by fitting $m_F$ to the observed \emph{HST} data.  The phase
difference $\phi_{HST} - \phi_0 = (t_{HST} - t_0)/P$ is defined by the
phase $\phi_0$ at the time of the first LBT observation $t_0$, as
compared to the time of the \emph{HST} observation $t_{HST}$.  Since
the epochs of the LBT and \emph{HST} data overlap, there is little
ambiguity about the relative phasing.

To accurately determine $A_F$ and $\phi_0$, we model the LBT
differential lightcurves by recasting the flux of the template
lightcurves in terms of differential counts
\begin{align}
\Delta C_{Fi} = 10^{-0.4(\hat {m}_{F} + A_FT_F(\phi_i -\phi_0) - Z_F)} - C_{0F}
\end{align}
where $\Delta C_{Fi}$ is the differential counts measured by the LBT
in filter $F$ at phase $\phi_i - \phi_0$, $C_{0F}$ is the (unknown)
counts of the Cepheid in the LBT reference image of filter $F$, and
$Z_F$ is the photometric zeropoint of the LBT reference image.  As a
reminder, $C_{0F}$ cannot be reliably determined from the LBT data
alone, due to crowding, which is why make use of higher resolution
HST data.  We determined all of the parameters by optimizing
\begin{align}
  \chi^2 &=\sum_F \sum_i \left(\frac{\Delta C_{Fi}^{LBT} - \Delta
      C_{Fi}}{\sigma_{Fi}^{LBT}} \right)^2 \nonumber\\
  &+ \sum_F\sum_j\left(\frac{m_{Fj}^{HST}-m_{Fj}}{\sigma_{Fj}^{HST}}
  \right)^2
\end{align}
where $i$ runs over the LBT observations and $j$ runs over the
\emph{HST} observations.  The problem is to simultaneously fit for
$\hat m_F$, $A_F$, $C_{0F}$, and $\phi_0$.  Given the non-linear
nature of this task, we performed the calculation using Markov Chain
Monte Carlo methods, and estimated the parameter uncertainties from
the marginal distributions.  We simultaneously fit the $B$ and $V$
band LBT lightcurves and all available \emph{HST} F435W and
F555W observations.  Given the amplitudes estimated from the data for
the $V$ and $B$ bands, the \citet{Pejcha2012} templates determine the
amplitudes and phases for all other wavelengths, allowing us to
determine $\hat m$ for the F814W and F160W bands as well.

The photometric zeropoints $Z_F$ of the LBT reference images were
calculated in four steps.  First, we found instrumental magnitudes in
the reference image using DAOPHOT.  Next, we matched the brightest
stars in the reference image to SDSS DR7 (\citealt{SDSSDR72009}).  We
then solved for the zeropoints $Z_F$ that placed our instrumental
magnitudes on the Johnson $UBV$ system, following the transformations
given by \citet{Fukugita1996}.  In all, about 60 stars were used to
calculate the zeropoints.  Finally, the zeropoints were shifted to the
\emph{HST} VEGAMAG system, following the zeropoint offsets provided by
\citet{Sirianni2005}.  After accounting for the uncertainties in each
transform, we estimate a 0.07 mag error on $Z_F$ in each reference
image.  However, because we must solve for $C_0$, the LBT zeropoints
only weakly affect the magnitude calibration.  G11 found that shifting
the zeropoint by as much as 0.30 mag had little effect on the mean
magnitudes, changing the final determination of the distance modulus
by no more than 0.01 mag.  Instead, $Z_F$ primarily influences the
determination of the amplitude $A$, since these parameters are
correlated (see G11).  However, we usually have 2-3 \emph{HST}
calibrating points at differing phases, which helps to constrain the
amplitudes and minimizes this problem.

\begin{figure*}
  \begin{tabular}{cc}
    \includegraphics[width=0.5\textwidth]{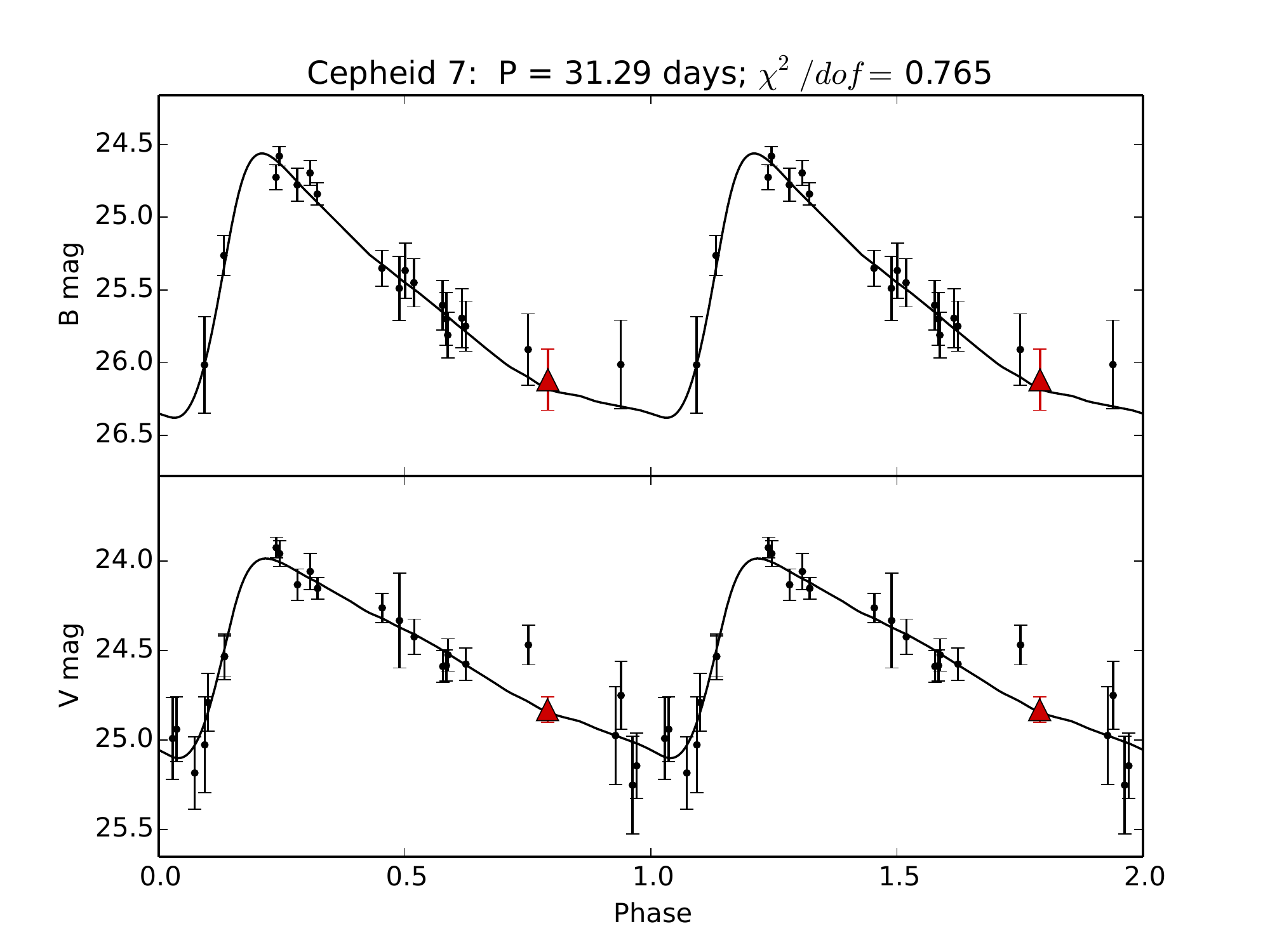}   &\includegraphics[width=0.5\textwidth]{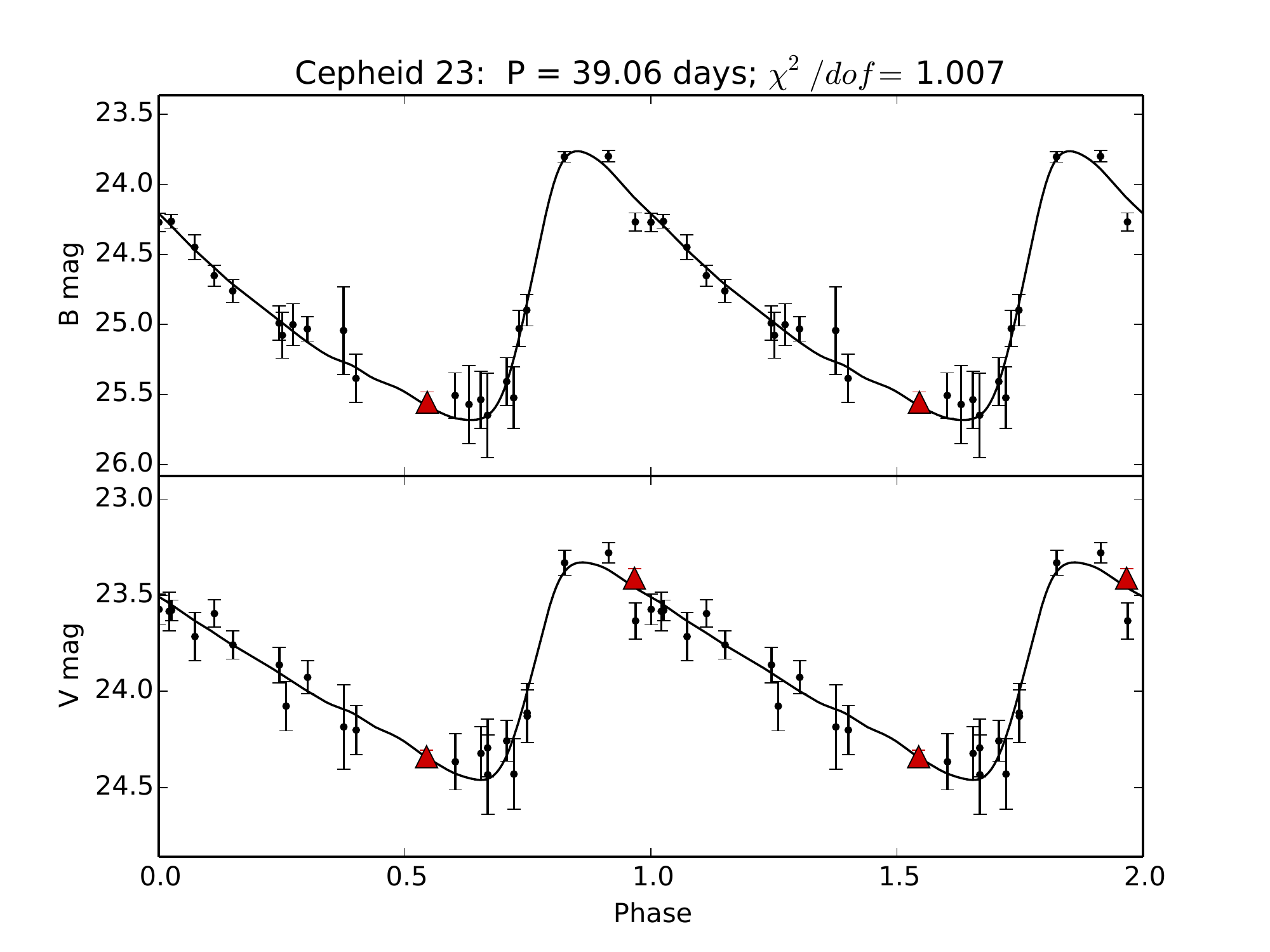}\\
    \includegraphics[width=0.5\textwidth]{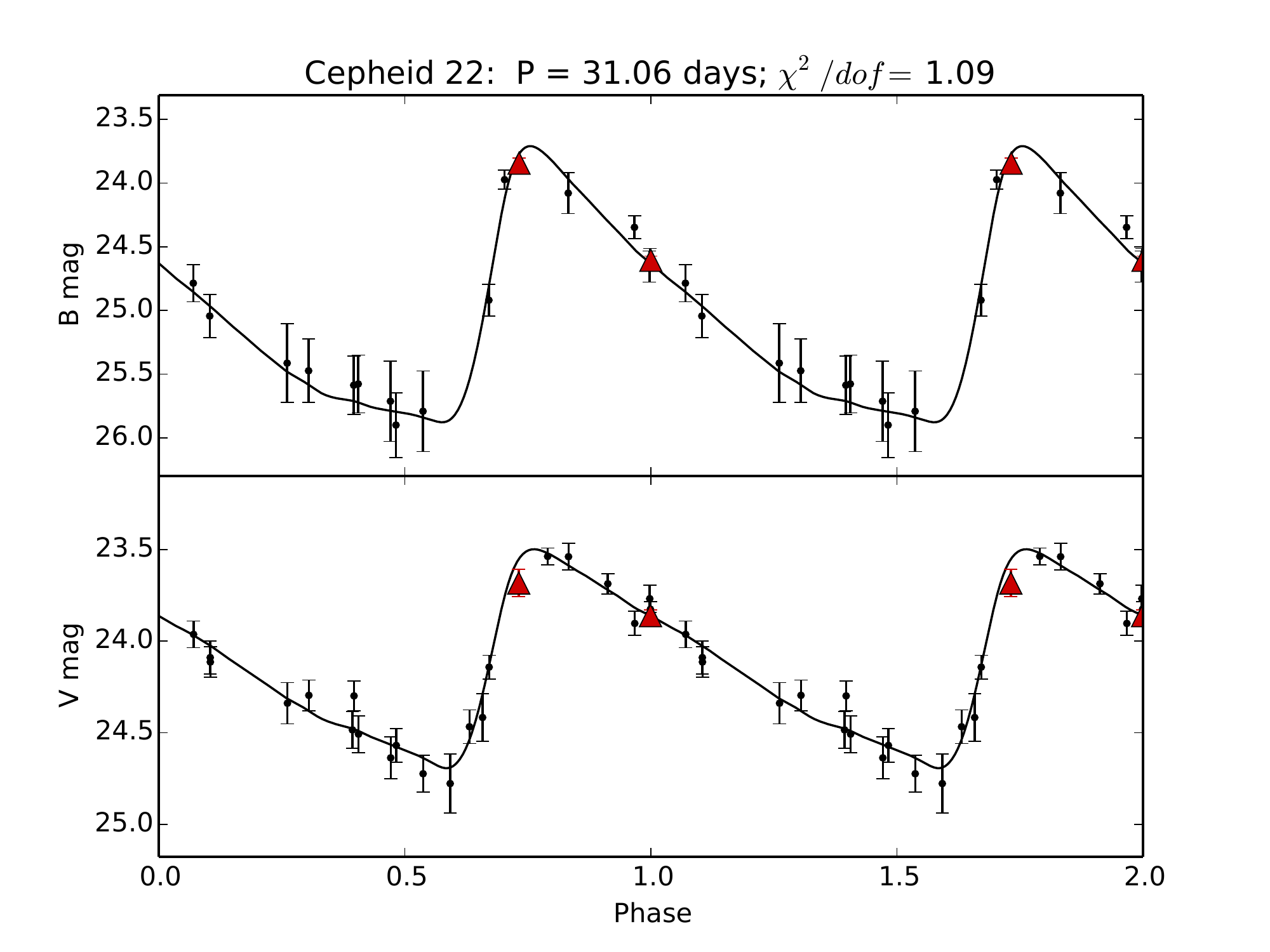}   &\includegraphics[width=0.5\textwidth]{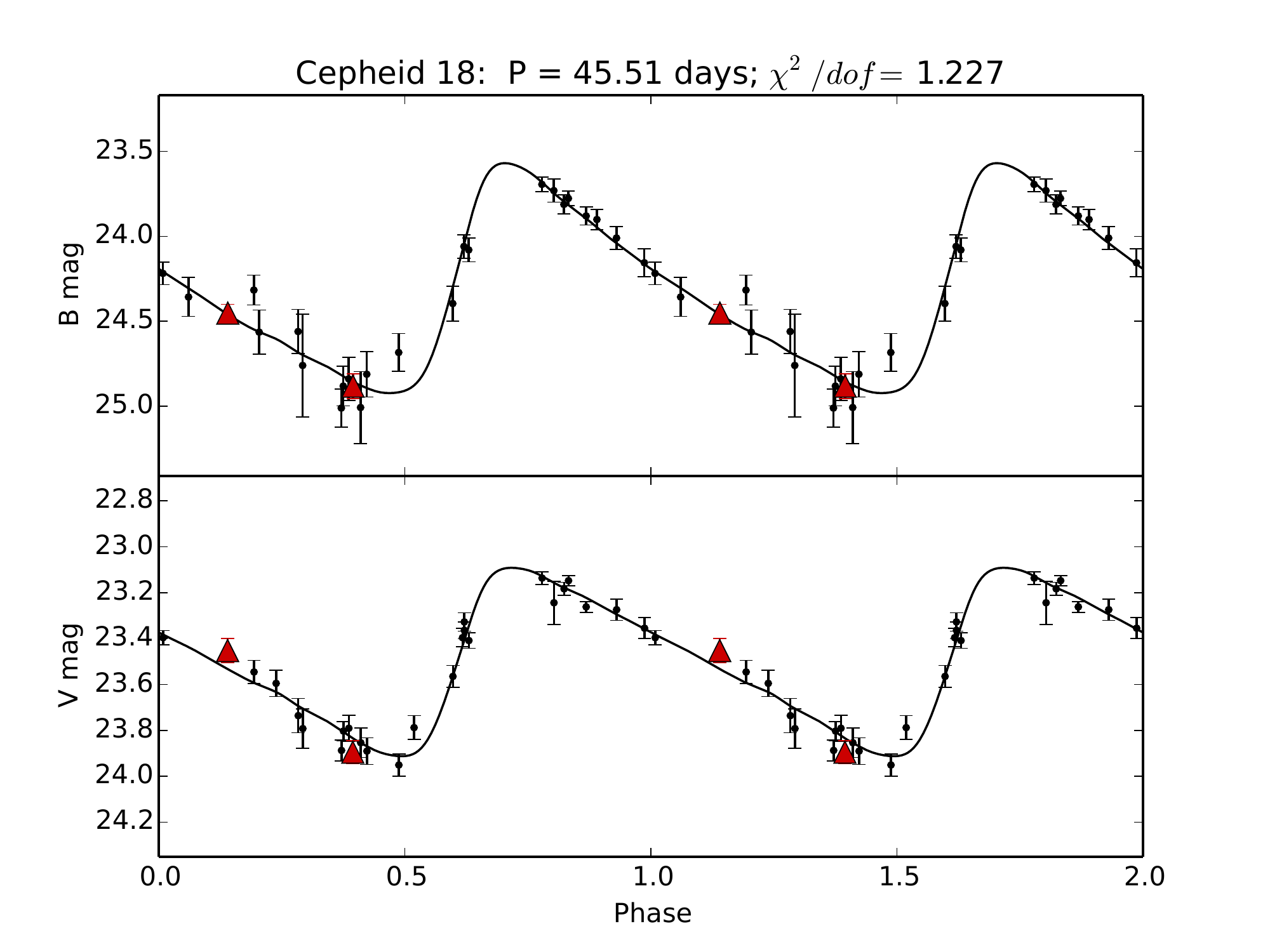}\\
    \includegraphics[width=0.5\textwidth]{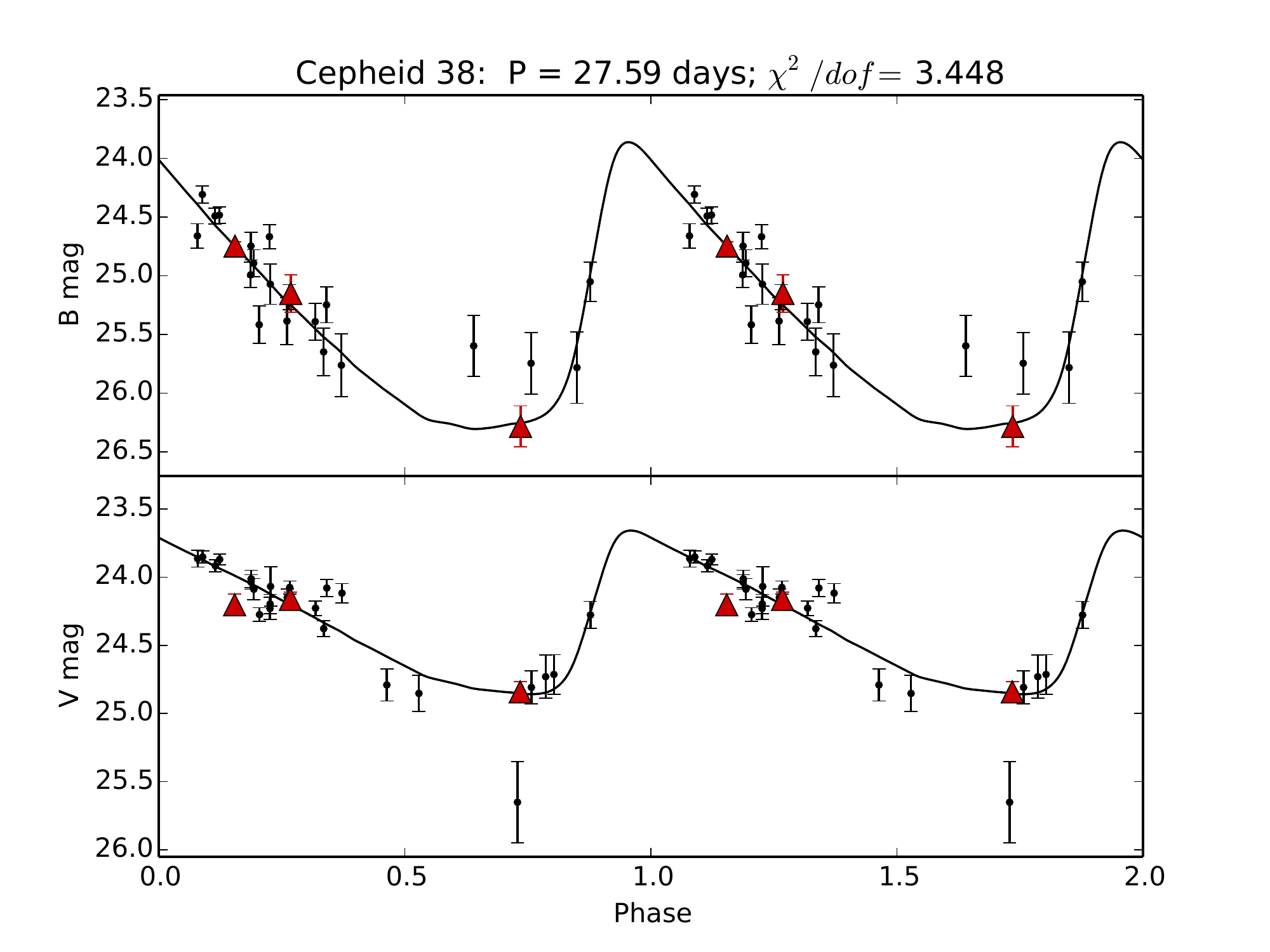}   &\includegraphics[width=0.5\textwidth]{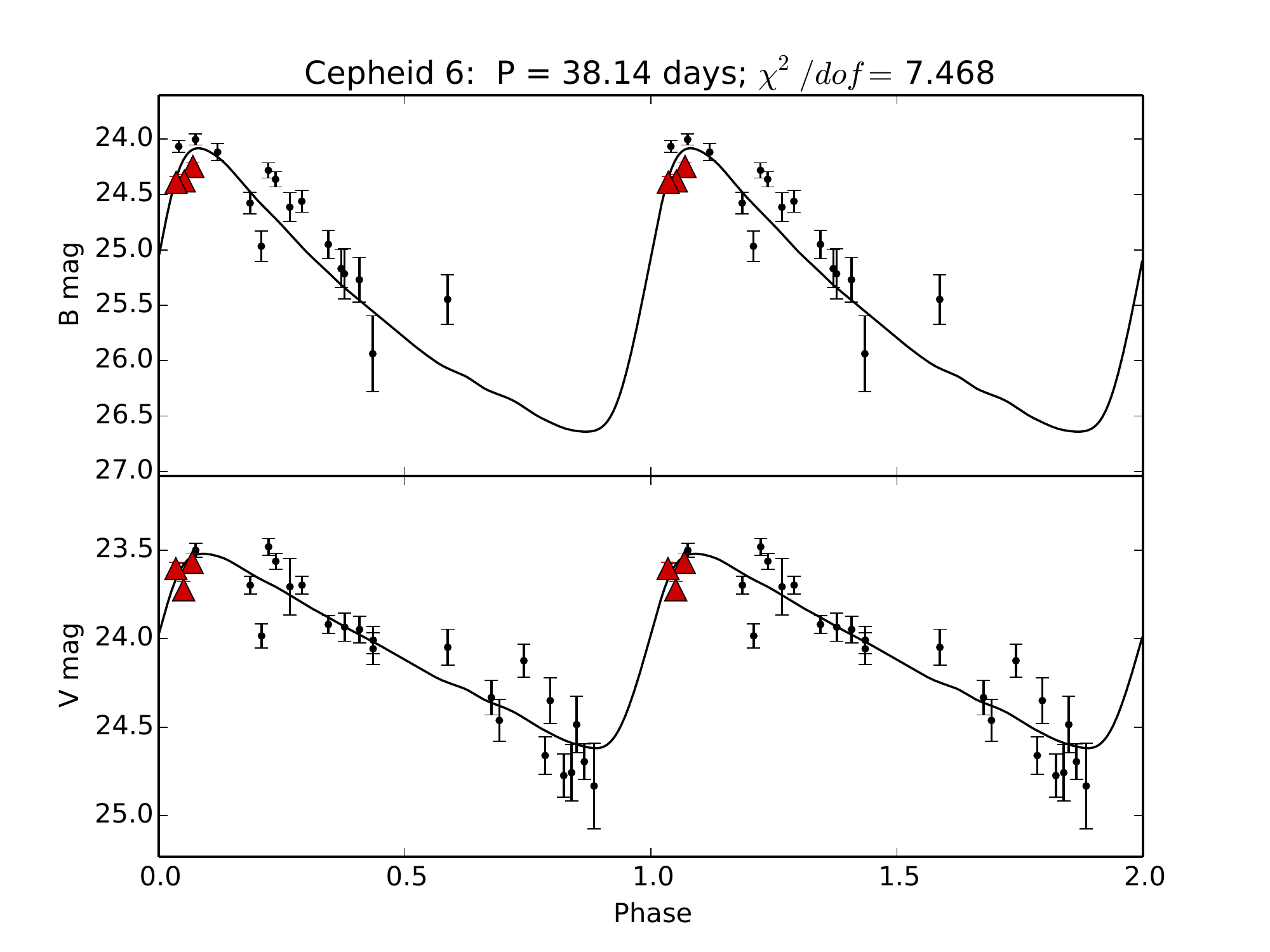}\\
  \end{tabular}
  \caption{Examples of calibrated lightcurves and their fitted
    templates.  The large red triangles are the \emph{HST} calibration points
    and the smooth black lines are the empirical templates.  The fits
    degrade from left to right and top to bottom based on the value of
    $\chi^2/dof$.  Cepheid 7 has the smallest value of $\chi^2/dof$
    among all Cepheids, while Cepheids 38 and 6 have the largest two
    values.}
    \label{lightcurves}
\end{figure*}

After fitting for the mean magnitudes, we visually checked the
calibrated lightcurves and the posterior distributions of their
parameters to ensure that the fits had converged and that the
parameter space was well-sampled.  Because the Cepheids have been
calibrated to the \emph{HST} VEGAMAG system, we then converted their
mean magnitudes to the Johnson/Cousins $BVI$ system, again following
the prescription of \citet{Sirianni2005}.  These authors determined
empirical conversions for \emph{HST} VEGAMAG/Johnson $UBV$ using a
zeropoint correction and a single color term.  Coefficients for the
zeropoint and color term were taken from their Table 18.  The typical
uncertainty of these transforms is $\leq 0.03$ mag, which makes a
small contribution to the final calibration uncertainty, typically
0.06--0.10 mag. As already noted, the F160W filter is similar to $H$
band, and we leave these measurements in the native \emph{HST}
filter/detector photometric system.  Table \ref{tab:cephdata} gives
the calibrated mean magnitudes of the Cepheids (with uncertainties
including those of the final photometric transform), and Figure
\ref{lightcurves} provides examples of calibrated $B$ and $V$ band LBT
lightcurves.

\subsection{Comparison to previous studies}
\begin{figure*}
      \includegraphics[width=\textwidth]{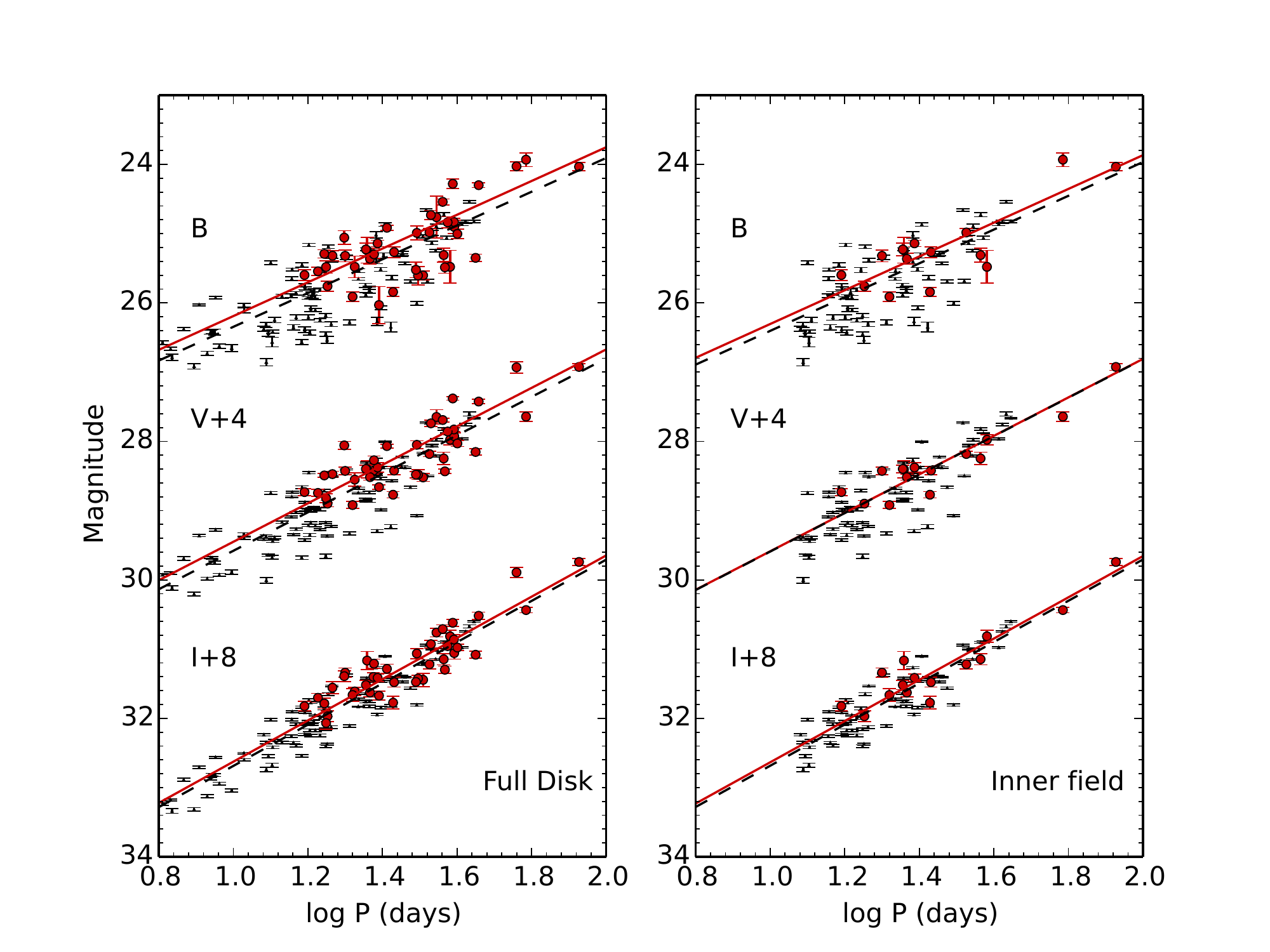}
      \caption{Mean magnitude BVI PL relations for NGC 4258 with no
        extinction corrections.  The large red circles are the
        LBT Cepheids, and the small points are the Cepheids
        from M06.  The solid lines are the PL relations from M06
        shifted to best fit the LBT data, and the dashed lines
        are the same for the M06 sample. The left panel shows all
        Cepheids, while the right panel only shows Cepheids interior
        to the radius of the outer edge of the M06 inner field.  The
        $V$ band and $I$ band results are shifted by 4 and 8 mag,
        respectively, to avoid overlapping the data.}
    \label{fig:macri}
\end{figure*}

M06 identified a large sample of generally shorter period Cepheids in
NGC 4258.  The left panel of Figure \ref{fig:macri} compares the raw
PL relations for the two samples, with no corrections for extinction.
Our Cepheids are systematically brighter at fixed period, with average
shifts of $-0.155\pm 0.011$ mag, $-0.130 \pm 0.007$ mag, and
$-0.061\pm 0.010$ mag in the $B$, $V$, and $I$ bands.  We also
compared only those Cepheids in the M06 sample with periods $P>10$
days, so as to match the range of periods between the two samples.
This criterion removed 12 short-period Cepheids from the M06 sample, and
increased the mean offset between samples to $-0.209$ mag, $-0.139$
mag, and $-0.063$ mag in $B$, $V$, and $I$, respectively.  However,
our sample is drawn from a much larger extent of the galaxy's disk,
and we might reasonably expect a smaller mean value of extinction.  To
test this hypothesis, we selected the subset of our Cepheid sample
interior to the outer edge of M06 inner field, and we found smaller
offsets of $-0.099\pm 0.02$ mag, $-0.003\pm 0.014$ mag, and $-0.050
\pm0.017$ mag in the $B$, $V$, and $I$ bands, respectively, shown in
the right panel of Figure \ref{fig:macri}.  The dispersions of both
samples (the LBT Cepheids and all M06 Cepheids) around their PL
relations are almost identical at $0.36$ mag, $0.30$ mag, and $0.23$
mag for the $B$, $V$, and $I$ bands.

We can also compare individual Cepheids between samples.  We matched
seven Cepheids from the M06 sample in the LBT sample, and
compared the periods and mean magnitudes.  The limited overlap comes
from the difference in period ranges (see Figure \ref{fig:macri}).  As
an experiment, we examined the positions of the M06 Cepheids in our
rms image by eye.  We found two additional Cepheids that matched
sources for which we had extracted lightcurves.  One Cepheid was not
identified because it had a period of 8.96 days, and was therefore
excluded from our period search.  The other M06 Cepheid had a period
of 32.25 days and was identified in the LBT $R$ band with $F =
3.59$ and a period of $32.85$ days.  However, it was cut during our
visual inspection step because of a particularly small amplitude (0.02
mag) compared to its scatter.  The original \emph{HST} light curve
from M06 has an amplitude of 1.2 mag, suggesting that the image
subtractions and extracted photometry for this source are particularly
noisy.

\begin{figure*}
    \begin{tabular}{cc}
      \includegraphics[width = 0.5\textwidth]{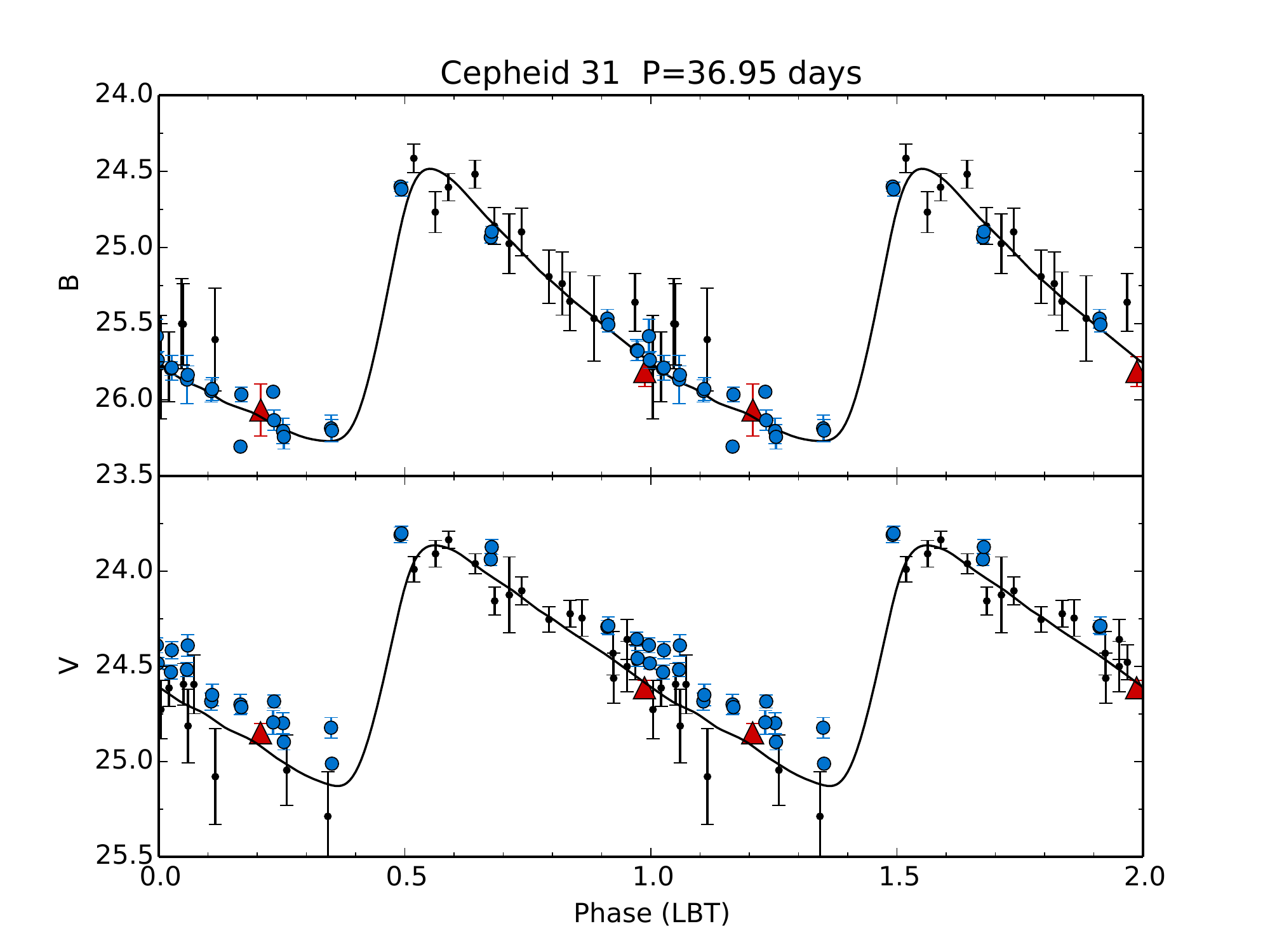}\\
      \includegraphics[width = 0.5\textwidth]{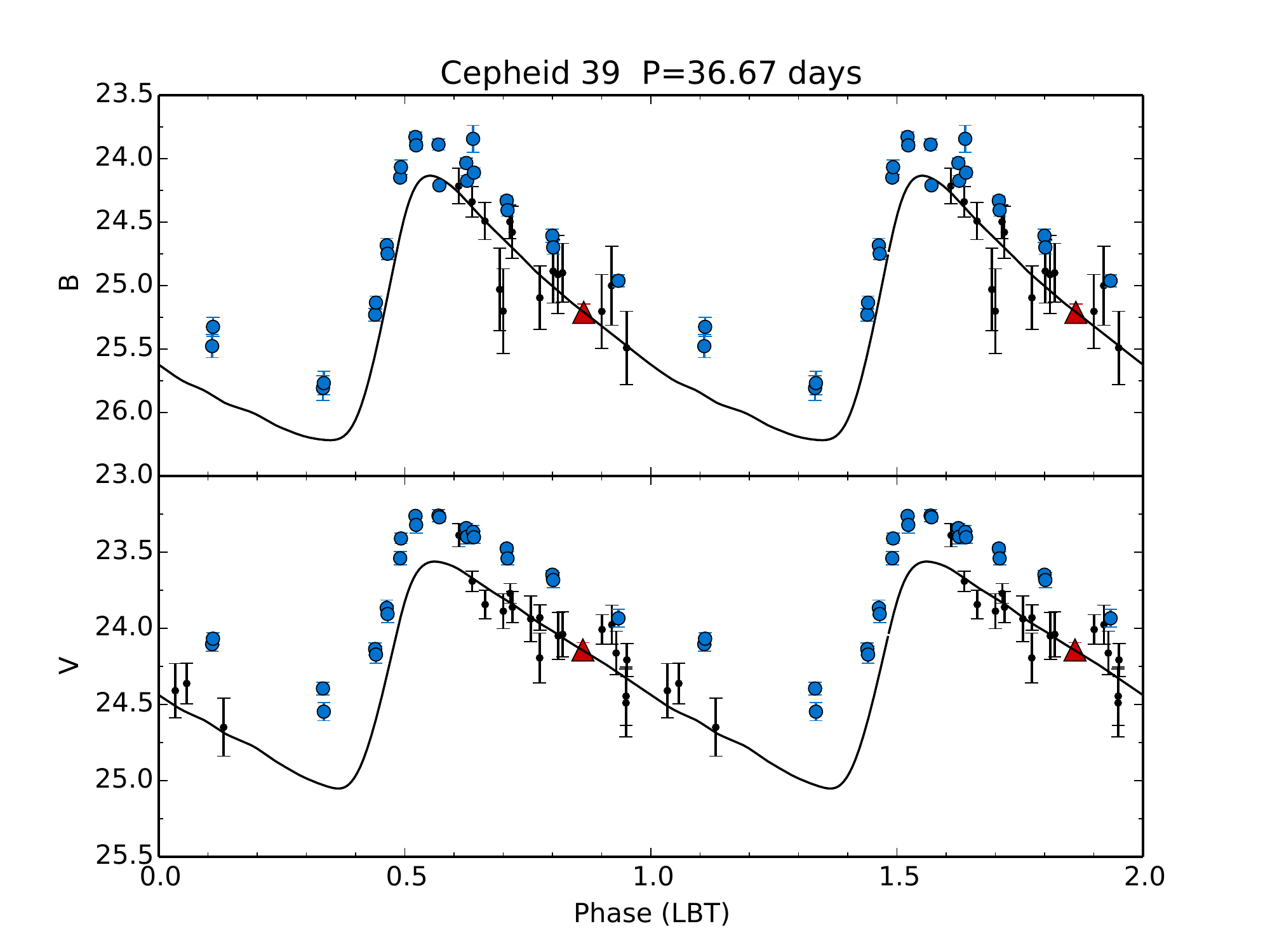}\\
      \includegraphics[width = 0.5\textwidth]{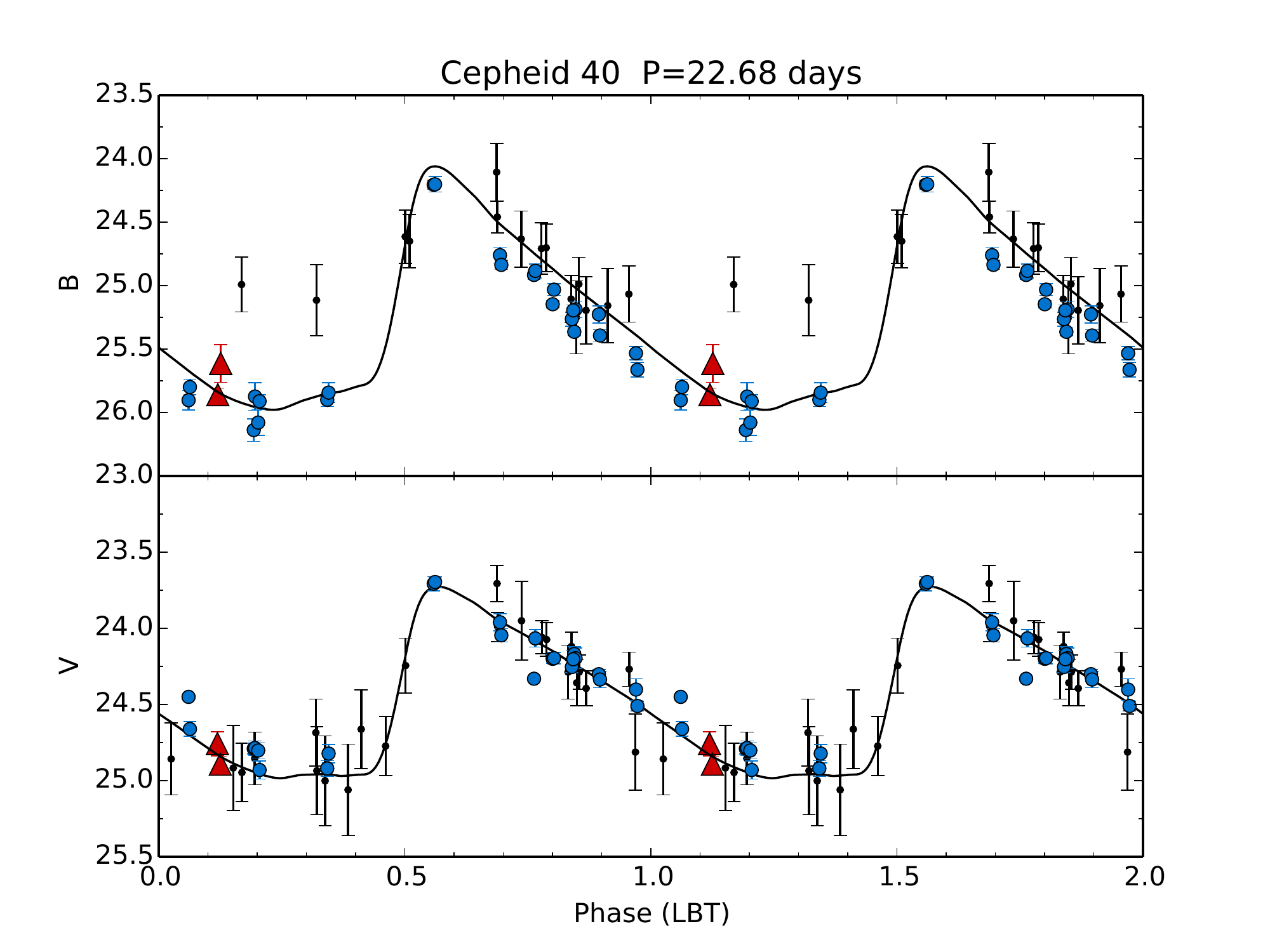}
    \end{tabular}
    \caption{$B$ and $V$ band LBT and M06 lightcurves for the
      two Cepheids with the largest mean magnitude differences
      (Cepheids 31, top, and 39, middle) and the smallest difference
      (Cepheid 40, bottom).  The LBT data, with errors, and
      fitted \citet{Pejcha2012} templates are in black, while the
      \emph{HST} calibration points are the red triangles.  The M06
      data are the larger blue circles.}
    \label{fig:lc_comp}
\end{figure*}

Table \ref{tab:matches} summarizes the differences in period and mean
magnitudes for the seven matches.  The agreement of the periods is
good, with typical differences of a few tenths of a day.  The M06
observations spanned $\Delta t=45$ days, so we would not expect
periods more accurate than $\delta P = 0.4P^2/\Delta t \sim
0.9(P/10\text{ days})^2$ days.  The average absolute shift is 0.58
days.  Two Cepheids had $\Delta P > 1 $ days, and these were the two
with the longest periods ($26.99$ and $36.95$ days).  Our mean
magnitudes tend to be fainter than those of M06, although the average
differences are comparable to their dispersion.  Two Cepheids had
differences in two or more filters greater than $0.20$ magnitudes.
Cepheid 39 is near the galaxy's center, while Cepheid 31 is one of our
faintest Cepheids.  Figure \ref{fig:lc_comp} displays the LBT
lightcurves of these Cepheids, overlaid with the M06 lightcurves
shifted to a common phase.  For comparison, the Cepheid with the
smallest difference in mean magnitudes (Cepheid 40) is also shown.
The \emph{HST} calibration points are clearly offset from the M06
lightcurves for Cepheid 39, and there is some suggestion of the same
effect in the $V$ band for Cepheid 31.  The mean magnitudes from M06
were calculated by numerically integrating the \citet{Stetson1996}
Cepheid lightcurve templates to find the phase-weighted mean
magnitude, which may result in small offsets from the mean magnitudes
determined in our fitting procedure.  If we exclude these 2 Cepheids,
the average differences between the mean magnitudes drop to $-0.02 \pm
0.08$ mag, $-0.05 \pm 0.04 $ mag, and $-0.08\pm 0.07$ mag in $B$, $V$,
and $I$.

\section{DISTANCE FITTING PROCEDURE AND PL RELATIONS}
We model the mean magnitude of Cepheid $i$ in filter $F$ as
\begin{align}
  \langle m\rangle_{iF}^{PL} &= L_F(P_{i}) + E_i(B-V) R_F \nonumber\\
  &+ \boldsymbol{\gamma}(Z_i-Z_{LMC}) + \Delta \mu_{LMC}
\label{equ:model}\end{align}
where $L_F(P_{i}) = a_F + b_F\log P_i $ is the LMC PL relation,
$E_i(B-V)$ is the unique reddening for each Cepheid, $R_F
=A_{iF}/E_i(B-V)$ is determined by the extinction law, $Z_i -Z_{LMC}$
is the metallicity of the Cepheid relative to the LMC, $\boldsymbol{\gamma}$
is the metallicity effect (the structure of $\boldsymbol{\gamma}$ is
discussed in \S \ref{sec:metallicity effect}), and $\Delta \mu_{LMC}$ is the
distance modulus between NGC 4258 and the LMC.  Our strategy is to
solve for the $E_i(B-V)$, $\boldsymbol{\gamma}$, and $\Delta \mu_{LMC}$ by
minimizing the function
\begin{align}
  \chi^2 = \sum_{i} \sum_{F} \left( \frac{\hat m_{Fi} -\langle
      m\rangle_{Fi}^{PL}}{\sigma_{Fi}} \right)^2.
\end{align}
We fit all the mean magnitude data simultaneously.  However, only
$\sim50$\% of our sample has complete 4-band photometry.  Missing
measurements for any filter were assigned a mean magnitude
corresponding to the Cepheid lying on the relevant PL relation at the
M06 distance modulus.  An uncertainty of $\sigma_{Fi} = 10^{6}$ mag
was assigned to this value so that it makes no contribution to the
likelihood, while simplifying the `bookkeeping' of the fit.  When we
account for the number of degrees of freedom in the model, we do not
include these dummy measurements, nor do we incorporate them in our
calculation of the covariances between residuals in different filters.
In addition to comparing the results of the LBT Cepheid sample to
those of M06, we combine both data sets, fitting all 122 Cepheids
simultaneously.

The standard error on one parameter is the point where $\Delta \chi^2
= 1$.  However, the PL relations have intrinsic scatter due to the
finite width of the instability strip, which our model must account
for.  While intrinsic scatter can bias parameter estimates
(\citealt{Weiner2006}, \citealt{Kelly2007}), the effect, by
definition, decreases with sample size.  Since we are primarily
concerned with estimating $\Delta\mu$ and $\boldsymbol{\gamma}$, our sample
size is reasonably large, and the intrinsic scatter is of minimal
interest, we treat this problem by simply rescaling the mean magnitude
uncertainties so that $\chi^2/dof = 1$.  The rescaling factor
propagates directly to the parameter uncertainties, and ensures that
the fits are consistent with the scatter.  To make sure that this
method does not mask any other systematic effects, we check these
estimates by bootstrap resampling the Cepheids over $10^4$ trials, and
reporting the median and the symmetric 68\% confidence intervals of
the distribution.  In practice, we found that the bootstrapping
uncertainty estimates are consistent with estimates based on
$\Delta\chi^2$.

We take PL relations from three sources.  First, the Optical
Gravitational Lensing Experiment (OGLE) has been monitoring the LMC
for over 20 years and has published several iterations of LMC PL
relations.  OGLE II published PL relations in BVI
(\citealt{OGLEII}).\footnote{ We employ the updated coefficients,
  which can be found at
  \url{ftp://sirius.astrouw.edu.pl/ogle/ogle2/varstars/lmc/cep/catalog/README.PL}}
More recently, \citet{Ngeow2009} matched OGLE III fundamental mode
Cepheids to 2MASS data, and published PL relations in $VIJHK$, as well
as several Spitzer bands.  Both samples were quite large (over 1300
Cepheids).  In addition, \citet{Persson2004} determined near-IR PL
relations ($JHK$) using 92 Cepheids.  Table \ref{tab:PLs} summarizes
these PL relations.  All PL relations have been extinction corrected,
and while the various models are nearly consistent, there is some
tension.  In particular, the $I$ band PL relations from OGLE II and
\citet{Ngeow2009} are discrepant at the 3$\sigma$ level, which
\citet{Ngeow2009} attributed to different treatments of extinction.  To
characterize the dependence of our results on the PL relations, we
tested various combinations of these models.

Several studies also suggest that there may be a ``break'' in the PL
relations at $\sim 10$ days, so that shorter-period Cepheids follow a
different PL relation than Cepheids with $P>10$ days
(e.g. \citealt{Sandage2004,Kanbur2004,Ngeow2009}).  Since our sample
is restricted to longer-period Cepheids, the functional form of the PL
relation does not affect our fitting procedure.  However, it is
possible that modified PL relations derived from long period Cepheids
are more appropriate for our sample.  We therefore also experimented
with the the PL relations of LMC Cepheids with $P>10$ days, determined
by \cite{Sandage2004} and \cite{Ngeow2009}.

A complication for our choice of near-IR PL relations is that the
\emph{HST} F160W filter/detector combination has an effective
wavelength of $\sim 1.5\mu$m, slightly offset from that of $H$ band at
$\sim1.6\mu$m.  This calls for a modification to the published near IR
PL relation.  Near the $J$ and $H$ bands, linear interpolation as a
function of wavelength should yield a reasonable estimation of
intermediate wavelength PL relations, as shown by
\citeauthor{Pejcha2012} (2012, see their equation 10 and figure 7).
Table \ref{tab:PLs} includes this modification to the near-IR PL
relations, and the errors reported there are determined by adding the
errors of the $J$ band and $H$ band PL coefficients in quadrature.
This is actually an over-estimate of the uncertainty, since the F160W
effective wavelength is $\sim\! 3$ times closer to the $H$ band than
$J$ band filter, which implies that $\sigma_{\text{F160W}}^2 =
(0.25\sigma_J)^2 + (0.75\sigma_H)^2$.  However, we have used the more
conservative over-estimate, particularly since it is still much
smaller than the intrinsic scatter of the PL relations.  In order to
further explore the interpolation uncertainty, we modified the F160W
PL relation during our bootstrap resampling routine, allowing the
interpolated slope and zeropoint to vary by random Gaussian deviates
scaled by the errors in Table 4.  This method samples a range of PL
relations intermediate between the J and H bands---in fact, inspection
of Table 4 indicates that the P04 PL relations are consistent within 1
to 2.3$\sigma$, so this procedure effectively includes the
uncertainties of using either the J or H band PL relations themselves.

We adopt a \citet{Cardelli1989} extinction law to set the ratio of
total to selective extinction in each band, $R_F = A_F/E(B-V)$.  For
our first estimate, we adopt the ``standard'' model of $R_V = 3.1$.
This sets the reddening vector $\boldsymbol{R}_F = (4.11, 3.10, 1.85, 0.64)$
for $B$, $V$, $I$, and F160W, respectively.  However, it is not necessary to
fix $R_V$ to a specific value because we have 4 band photometry, and
we explore how changing this parameter affects our distance modulus by
varying it on a grid from $R_V=$\ 2.5 to 6.5.  For each value of
$R_V$, we calculate the extinction vector $\boldsymbol{R}_F$ from the
\citet{Cardelli1989} extinction law and refit the data, using the
over-all minimum value of $\chi^2$ as an estimate of the best fit.

\subsection{Metallicity dependence}
For studying the effects of metallicity on Cepheid mean magnitudes, we
only need accurate differential metallicities -- the absolute
metallicities are unimportant.  In the context of the present study,
the relative mean metallicity of LMC Cepheids and our sample impacts
the determined distance modulus, while the variation of metallicities
within our sample determines the metallicity dependence of the Cepheid
mean magnitudes.  As given in Equation \ref{equ:model}, the generic
form of the metallicity effect is $\boldsymbol{\gamma}(Z_{i}-Z_{LMC})$
where $Z_i$ is the estimated metallicity of the NGC 4258 Cepheid $i$
and $Z_{LMC}$ is the reference metallicity of the LMC.  For the
Cepheids in NGC 4258, we estimate their metallicity based on a linear
fit to H II region abundances with radius, combined with the radial
position of the Cepheid.  The deprojected galactocentric radius $\rho
= \left(x^2 + y ^2\right)^{1/2}$ of a Cepheid is given by
\begin{align}\left(
  \begin{matrix}
    x\\y
  \end{matrix}\right) &= 
\left(\begin{matrix}
\cos\phi & \sin\phi\\
\frac{-\sin\phi}{\cos i}& \frac{\cos\phi }{\cos i}
\end{matrix}\right)
\left(\begin{array}{l}
\phantom{(} \delta - \delta_0 \phantom{)\cos\delta} \\ (\alpha-\alpha_0)\cos\delta
\end{array}\right)
\end{align}
where $\delta$ and $\alpha$ are the Declination and Right Ascension of
a given Cepheid, and $\delta_0 = 47^{\circ} 18' 14\farcs30$ and
$\alpha_0 = 12^h 18^m 57.50^s$ are the Declination and Right Ascension
of the center of NGC 4258.  We adopted a position angle $\phi =
150^{\circ} $, an inclination angle $i = 72^{\circ}$, and an isophotal
radius $\rho_0 = 7\farcm92$ (\citealt{vanAlbada1980}).

H II region metallicities for NGC 4258 are available from Z94 and
\citeauthor{Bresolin2011} (2011a, hereafter B11), but there are several
complications.  The first is the paucity of H II regions and the
resulting uncertainties in any estimate of the linear trend.  In our
quantitative results, we address this issue using the approach of G11.
Given a set of H II regions, we fit a linear trend directly to the
abundances and use that gradient for the fits in Equation 8.  We
include the uncertainties in the gradient by bootstrap resampling over
the H II regions (as well as the Cepheids) and refitting the linear
trend of the bootstrap-resampled data.  This method naturally includes
all the statistical uncertainties associated with the metallicity and
its slope.  We also allow the LMC metallicity to vary by a Gaussian
deviate of 0.08 when we do the bootstrap resampling (see G11).

The second problem is the systematic question of which metallicity
scale to use.  Metallicity gradients can only be compared using
samples with the same absolute calibrations.  Most H II region
metallicity estimates are based on ``strong line'' estimates, which
have significant uncertainties in their absolute calibrations.  In NGC
4258, there are only 4 H II regions with ``direct'' measurements using
detections of the faint [OIII]$\lambda$4363 auroral line.  The
original Z94 (strong line) oxygen abundance gradient is
\begin{align}
Z_i=12 + \log (\text{O/H})= 9.17-0.49\rho_i /\rho_0,
\end{align}
while B11 found a significantly shallower gradient (after converting
to our standard $\rho_0 = 7\farcm92$ isophotal radius) of
\begin{align}
Z_i=12 + \log (\text{O/H}) = 8.87 -0.20\rho_i/\rho_0
\end{align}
for the strong line method (from the photoionization models of
\citealt{McGaugh1991} and \citealt{Naray2004}) and
\begin{align}
Z_i=12 + \log (\text{O/H}) = 8.49 -0.18\rho_i/\rho_0 \label{equ:eTgrad}
\end{align}
based on strong lines calibrated to the few auroral line measurements
(the empirical system of \citealt{Pilyugin2005}).  If we transform all
the B11 strong line measurements to the Z94 system, following
\citet{Kewley2008}, we find
\begin{align}
Z_i=12 + \log (\text{O/H}) =9.06 - 0.28\rho_i/\rho_0.
\end{align}

When we change between these calibration scales, we must also be sure
that the reference metallicity for the LMC is on the same scale.  The
traditional value of $Z_{LMC}=8.5$ is essentially a strong line
estimate and should not be used with the auroral line calibrated
metallicities.  Essentially, the auroral line measurements provide
well-constrained estimates of the electron temperatures, so a
comparable measurement for the LMC is the estimate of $Z_{LMC}=8.25$
from detailed models of the 30 Doradus region by
\citet{Pellegrini2011}.

Combining all these issues, we will consider three different
metallicity models, combined with the bootstrap resampling methods
given above.  First, in the Z94-1 model, we simply combine the
original Z94 data with $Z_{LMC}=8.5$ (so as to better compare our
results with M06).  Second, in the Z94-2 model, we transform the B11
H II region data to the Z94 system and use the combined set of H II
regions, again with $Z_{LMC} = 8.5$.  Finally, in the B11-e model, we
transform the Z94 H II region data to the auroral B11 system and use
$Z_{LMC} = 8.25$.  After bootstrap re-sampling the H II regions, we
found gradients consistent with those presented in B11, with $Z_i =
($\Zzero$)-(0.30\pm0.05) \rho_i/\rho_o$\ in the Z94-2 system and $Z_i
= ($\Bzero$) -(0.19\pm0.04)\rho_i/\rho_o$\ in the B11-e system.

\section{RESULTS}
\begin{figure*}
      \includegraphics[width=1.0\textwidth]{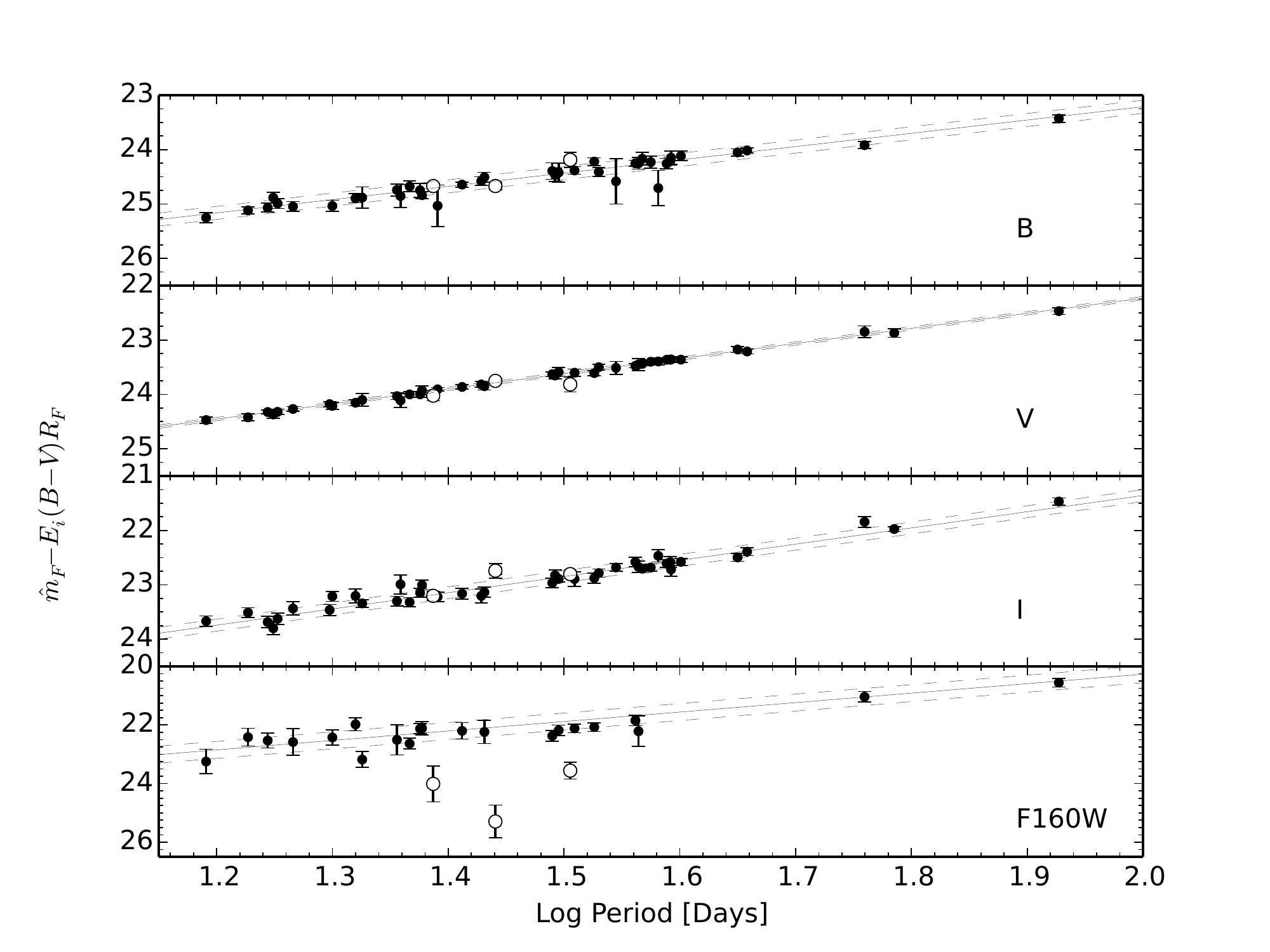} 
      \caption{Four-band fit to PL relations from the LMC (solid
        lines), with extinction-corrected mean magnitudes, for
        Cepheids in NGC 4258.  $\hat m_F$, $E_i(B-V)$, and $R_F$ are
        defined in Equation \ref{equ:model}.  All variations in the
        direction of the extinction vector have been modeled out,
        leading to the reduced scatter (dashed lines, see text for
        details).  The open circles mark the Cepheids that were
        removed due to large residuals from the F160W band PL
        relation, and are not included in the calculation of the
        scatter.}
    \label{fig:PLfit1}
\end{figure*}

We start with the ``standard'' model, in which we fit individual
Cepheid extinctions.  Then, in \S 5.1, we allow the extinction law to
vary, and in \S 5.2 we examine the role of metallicity.  Figure
\ref{fig:PLfit1} displays the extinction-corrected mean magnitudes and
the best-fit PL relations, adjusted to the appropriate distance
modulus.  We found $\Delta \mu_{LMC}=10.72\pm0.04$ mag, after
rescaling the uncertainties by a factor of 1.84 to make $\chi^2/dof =
1$.  After correcting for extinction and examining the residuals of
the Cepheid mean magnitudes from the fit, we found three Cepheids
significantly offset from the F160W PL relation.  Based on their
measurement uncertainties, Cepheids 5, 32, and 38 are $3.8\sigma$,
$7.7\sigma$, and $7.9\sigma$ outliers from the F160W PL relation,
respectively, and all are over 1.5 mag fainter than the F160W PL
relation.  The outliers are displayed as open circles in Figure
\ref{fig:PLfit1}.  If we clip these Cepheids from the sample and refit
the data, we find $\Delta \mu_{LMC}=$\standardmu\ mag, and we only
need to rescale the uncertainties by a factor of 1.38.  Using this
smaller rescaling factor, we fix the uncertainties on the remaining
Cepheids and exclude Cepheids 5, 32 and 38 from the rest of the study,
bringing our final LBT sample to 40 Cepheids.  Bootstrap resampling
these 40 Cepheids yields $\Delta \mu_{LMC}=$\bootstandardmu\ mag.

The best fit combines the OGLE II $BVI$ PL relations \citep{OGLEII}
with the \citet{Persson2004} F160W interpolated PL relation.  Table
\ref{tab:results1} summarizes the effects of different combinations of
PL relations on the distance modulus.  We found that the distance
modulus is relatively insensitive to the choice of PL relations.  The
only exception is using the \citet{Ngeow2009} $I$ band PL relation,
which drives $\Delta \mu_{LMC}$ down by $\sim 0.10$ magnitudes.  As
previously mentioned, there is moderate tension in the $I$ band PL
relation determinations from the OGLE II and \citet{Ngeow2009}
studies, with the differences attributed to varying treatments of
extinction.  Using the OGLE II $BVI$ PL relations improves
$\chi^2/dof$ by $0.01-0.11 $.  Our fit has 98 degrees of freedom,
which gives $\langle \sigma_{\chi^2/dof}\rangle = (2/dof)^{1/2} =
0.14$, so we cannot statistically distinguish between the PL
relations.  We also do not find a strong preference for the
long-period PL relations from \cite{Sandage2004} and \cite{Ngeow2009},
although we cannot rule out these PL relations, either. Comparing all
of these cases, we adopt PL relations that produce the global minimum
value of $\chi^2$, corresponding to the OGLE II optical and
\citet{Persson2004} interpolated F160W PL relations, which are are
shown with the 1$\sigma$ rms scatter of the data in Figure
\ref{fig:PLfit1}.  Since the F160W band is the \emph{HST} analog of
the standard near IR $H$ band, we refer to it as the ``$H$'' band for
the remainder of this section, although the effective wavelength of
this filter is slightly offset from that of the $H$ band.

The LMC PL relations are nominally corrected to zero extinction,
although there could be some residual contribution due to imperfect
corrections.  Only 2 NGC 4258 Cepheids had a reddening lower than the
foreground Galactic reddening of $E(B-V) =0.016$ mag, estimated from
the \citet{Schlegel1998} dust map.  Cepheid 9 had $E(B-V) =
0.01\pm0.01$ mag, and Cepheid 41 had $E(B-V) = -0.03\pm0.02$ mag.
While a negative reddening is unphysical, these values could be
explained by photometric errors rather than systematic effects (e.g.,
a blended blue star).

\begin{figure*}
  \includegraphics[width=1.00\textwidth]{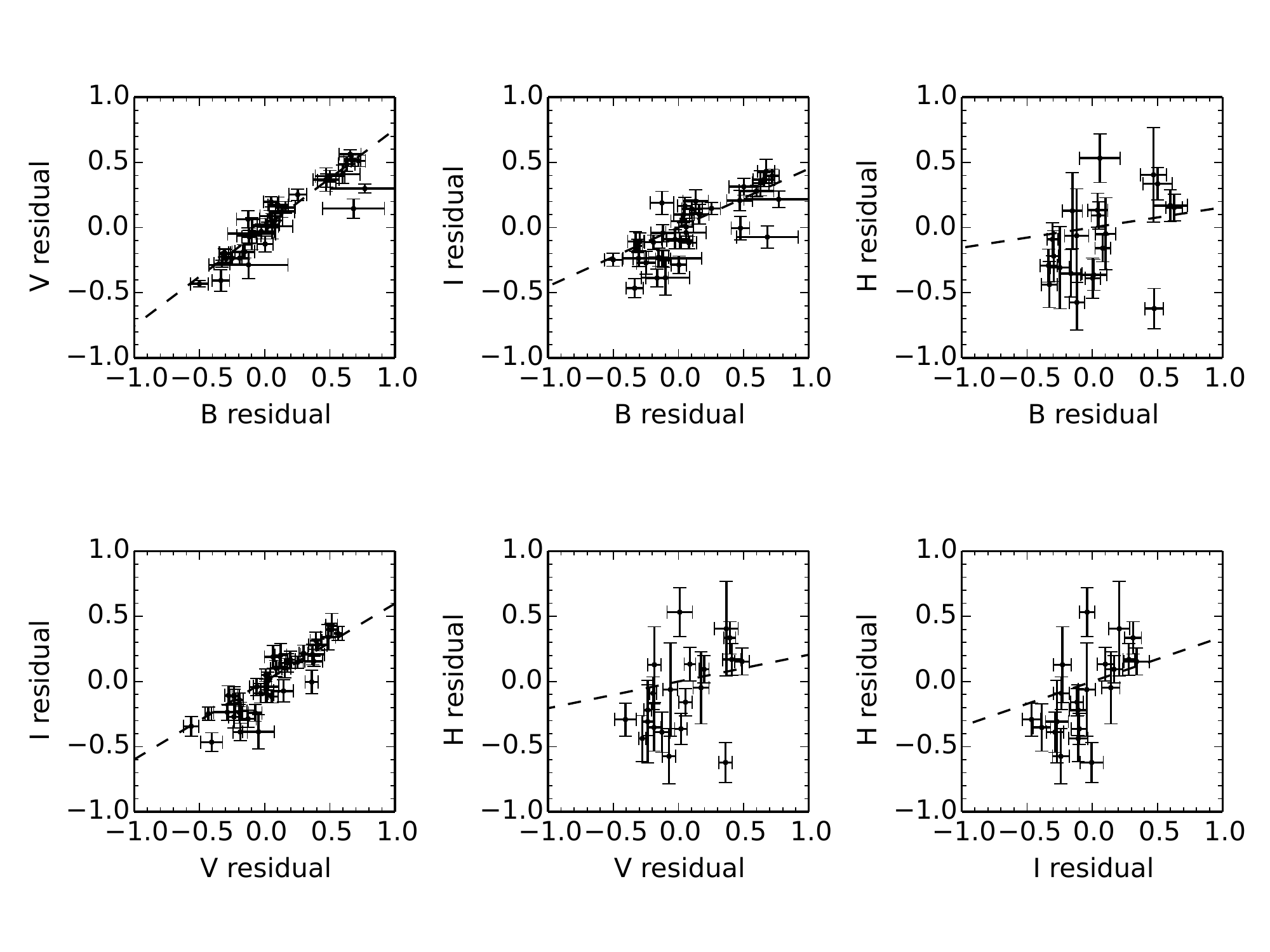}  
  \caption{Projected residuals from the PL relations after fitting a
    mean extinction and the distance modulus.  The residuals are
    dominated by the effects of differential extinction and thus are
    parallel to the correlations expected for a \citet{Cardelli1989}
    extinction law with $R_V=3.1$, shown by the dashed lines.}
  \label{fig:extinctionresidue}
\end{figure*}

\begin{figure*}
    \includegraphics[width=1\textwidth]{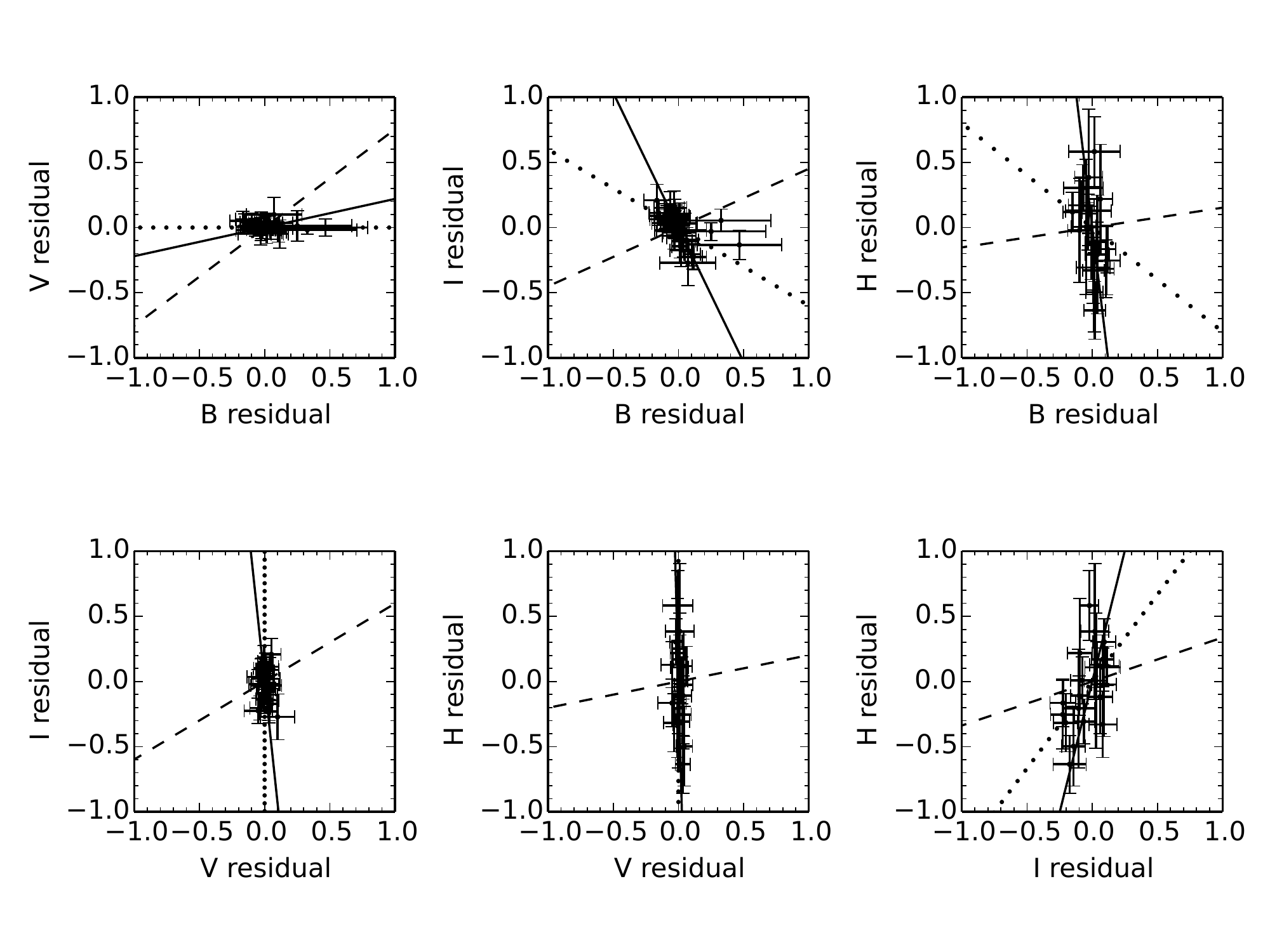}
    \caption{Projected residuals from the PL relations after fitting
      the individual extinctions and the mean distance.  The solid
      lines show the direction of the dominant component from our
      principle component analysis.  The dashed lines show the
      directions of the correlations due to differential extinction
      for $R_V=3.1$.  By definition, all correlations in this
      direction have been removed by fitting the individual
      extinctions. The dotted lines show the slope of the residuals
      expected for a change in the extinction law, $\partial
      \boldsymbol{R}_F/\partial R_V$.  Since the \citet{Cardelli1989}
      extinction law in normalized to the V band, the V component is
      zero by definition.}
    \label{fig:residue}
\end{figure*}

It is instructive to compare our fitting procedure to a simpler model
which employs a single mean extinction applied to all Cepheids (as
well as the common distance modulus). Figure
\ref{fig:extinctionresidue} shows the residuals of this fit, which are
highly correlated in the direction of the reddening vector, as
expected for differential extinction.  We can characterize the
residuals using the covariance matrix
\begin{align}
  c_{ij}= \text{covar}\left[ (\hat m_{i}-\langle m\rangle_i^{PL})(\hat
    m_{j}-\langle m\rangle_j^{PL})\right]
\end{align}
and the vector
\begin{align}
  c_{ii}^{1/2} = \left(\begin{array}{rrrr}B&V&I&H\\ \sigma_B &
      \sigma_V& \sigma_I& \sigma_H\end{array} \right)
\end{align}
where $i$ and $j$ run over the filters $B$, $V$, $I$, and $H$, and
$\sigma_{ii}$ is the rms of the residuals in the appropriate filter
(in magnitudes).  For this model, with only a mean extinction (rather
than individual extinction solutions), the formal covariances between
filters are
\begin{align}
  \frac{c_{ij}} {(c_{ii}c_{jj})^{1/2}} =\left(\begin{array}{rrrr}
    1.00  &0.88  &0.74  &0.32\\
    0.88  &1.00  &0.87  &0.37\\
    0.74  &0.87  &1.00  &0.63\\
    0.32  &0.37  &0.63  &1.00\\
  \end{array}\right).
\end{align}
with
\begin{align}
 c_{ii}^{1/2} = \left(\begin{array}{rrrr}B&V&I&H\\0.36&0.29&0.24&0.33\end{array}\right).
\end{align}
Here, the vector shows the amplitudes of the residuals for each filter
and the normalized matrix $c_{ij}/(c_{ii}c_{jj})^{1/2}$ shows the
strength of the correlations, where values of $0$, $1$, and $-1$ mean
no correlation, perfect correlation, and perfect anti-correlation,
respectively.  For comparison, Figure \ref{fig:residue} displays the
residuals of our fit using individual extinctions for each Cepheid.
These residuals have no component parallel to the reddening vector
because the fitting procedure will model out all variations in this
direction.  The formal covariance matrix now shows much weaker
correlations with
\begin{align}
  \frac{c_{ij}} {(c_{ii}c_{jj})^{1/2}} =\left( \standardcovar\right)\label{equ:covar1}
\end{align}
and
\begin{align}
  c_{ii}^{1/2} = \left(\standardrms\right).
\end{align}
Note, however, that the extinction corrections have done little to
reduce the variance of the $H$ band residuals.

The remaining correlations represent the components of all systematic
effects that cannot be modeled as extinction.  There is no simple,
intuitive means of interpreting these residuals since they may be due
to multiple systematic effects, each with a component degenerate with
extinction and distance.  G11 approached this problem by projecting
their residuals onto a vector $\boldsymbol{E}_2$ orthogonal to the
reddening vector $\boldsymbol{R}_F$ and distance vector
$\boldsymbol{\mu} = (1,1,1,1)$.  In this reduced error space, they
found a color dependence of the PL relations, which was also
correlated with galactocentric radius, and was therefore interpreted
as a metallicity effect.  However, with 4 band photometry, this method
would require 2 orthogonal vectors, which has no intuitive physical
interpretation.  Instead, we perform a principle component analysis
(PCA) on the residuals, and find eigenvalues of 0.092, 0.016, 0.005,
and 0.001.  This implies that there is a single, preferred direction
of the residuals, which lies in the direction
\begin{align}
  \boldsymbol{p}_1 =\left(\begin{matrix}B&V&I&H\\-0.12& -0.03&
      0.24& 0.96\end{matrix}\right).
\end{align}
The largest part of this component is a consequence of the $H$ band
residuals, which is problematic because of our incomplete photometric
coverage.  While the correlations between the $H$ band and the other
filters only make use of the 21 Cepheids for which we have 4-band
data, the aggregate covariance matrix includes all 40 Cepheids in
$BVI$.  Thus, the structure of $\boldsymbol{p}_1$ is extremely
sensitive to the residuals from the sub-sample of 21 Cepheids with 4
band photometry, and projections of the residuals onto
$\boldsymbol{p}_1$ are not strictly defined for the 19 Cepheids
without $H$ band measurements.  As an alternative to using incomplete
4-band photometry, we estimated $\boldsymbol{p}_1$ by refitting the
$BVI$ data alone and calculating the covariance matrix from nearly
complete 3-band photometry (only 2 Cepheids are missing $B$ band
measurements).  In this alternative fit, we find that $\mu = 10.72\pm
0.03$ mag, in agreement with our previous estimate ($\chi^2/dof =
0.83$), and the covariance matrix becomes
\begin{align}
\frac{c_{ij}} {(c_{ii}c_{jj})^{1/2}} &=
\left(\begin{array}{rrr}
    1.00    &-0.20 &-0.53  \\
    -0.20 & 1.00    &-0.21 \\
    -0.53 &-0.21 & 1.00    \\
  \end{array}\right)
\end{align}
and
\begin{align}
  c_{ii}^{1/2} &= \left(
\begin{matrix}B&V&I\\0.12& 0.03& 0.11&\end{matrix}
\right).
\end{align}
These covariances are in good agreement with those in equation
\ref{equ:covar1}, and the eigenvalues of this matrix are 0.021, 0.007,
and 0.001, again implying a single dominant direction of the
residuals.  However, the direction of this principal component is
\begin{align}
  \boldsymbol{p}_{BVI} =
  \left(\begin{matrix}B&V&I\\0.75&-0.01&-0.66\end{matrix}\right)
\end{align}
 substantially different from the
hyperplane defined by the $BVI$ components of $\boldsymbol{p}_1$.  Despite
this change, $\boldsymbol{p}_{BVI}$ does share some of the characteristics of
$\boldsymbol{p}_1$, particularly a small $V$ component and a noticeable
anti-correlation between $B$ and $I$.

\begin{figure*}
    \includegraphics[width=1\textwidth]{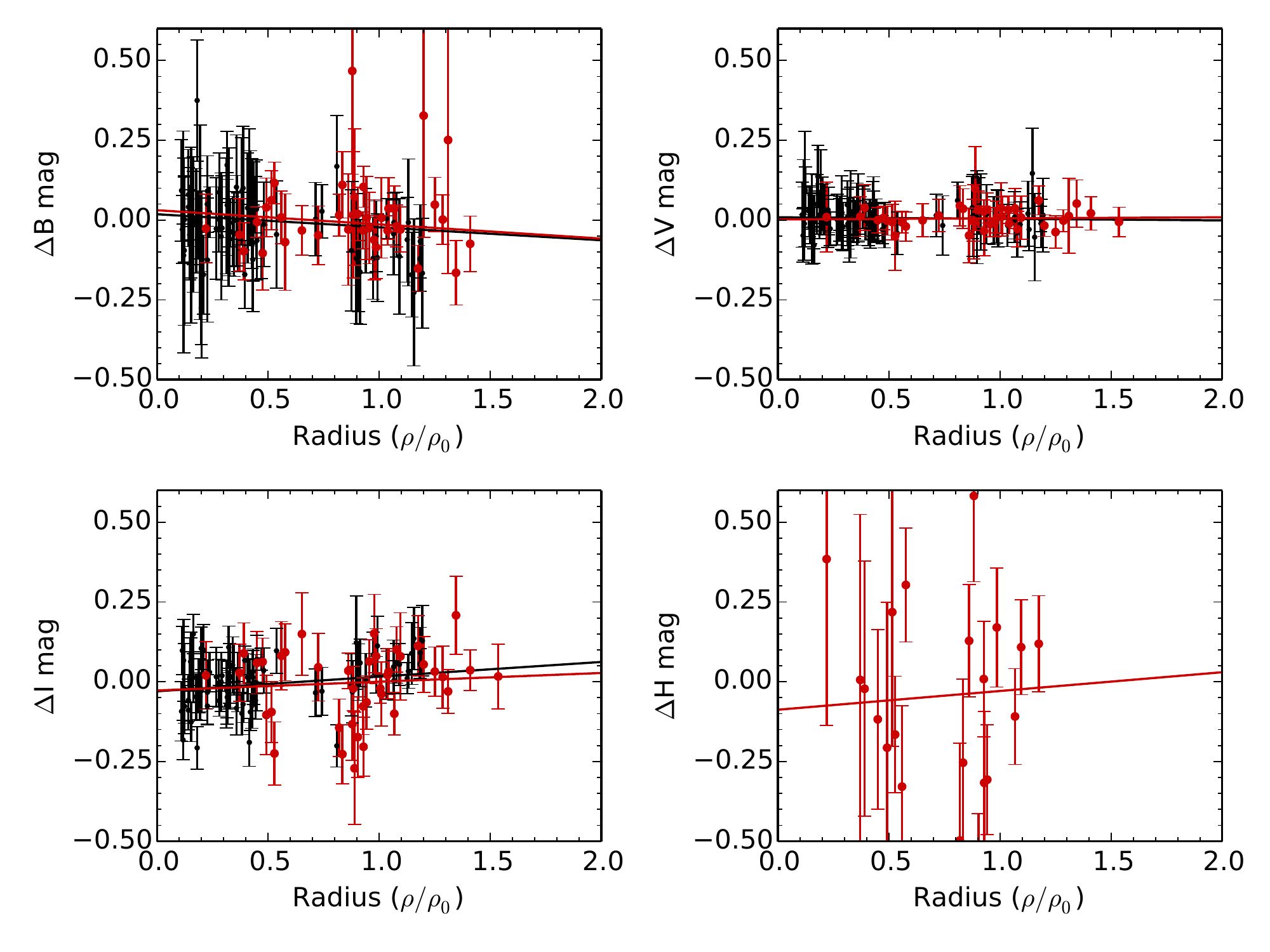}
    \caption{Residuals from PL relations as a function of
      galactocentric radius, in units of the isophotal radius,
      $\rho_0$.  The LBT Cepheids are in red, the M06 Cepheids are the
      smaller black points.  The red solid line is the result of a
      linear least-squares fit to the LBT Cepheids ($N=40$), the black
      line is a fit to all Cepheids ($N=122$).  Note the stronger
      trends in the B and I bands when using the full sample.  See
      Table \ref{tab:radial} for results of the linear regression.}
    \label{fig:radiusresidual}
\end{figure*}

The result of $\Delta \mu_{LMC}=$\standardmu\ mag using the LBT
Cepheids and individual extinctions is in excellent agreement with the
distance determined by M06 from the inner field Cepheids of $10.71\pm
0.04_{stat}\pm 0.05_{sys}$ mag, but is in tension with the outer field
distance of $10.87\pm0.05_{stat}\pm0.05_{sys}$ mag.  These distances
were derived by averaging the reddening-free distance modulus of each
Cepheid in the relevant field, and M06 attributed the difference
between the fields to their different locations, and hence
metallicities.  The LBT Cepheids provide a means of testing the
relation between inferred distance and galactocentric position, since
they are drawn from a wide range of azimuthal angles and radial
distances.  The LBT Cepheids run from 0.22-1.54$\rho_0$, a much larger
range of radii than the location of the outer field (centered at
1.02$\rho_0$).  First, we searched for trends in the PL residuals as a
function of galactocentric radius, and the results are shown in Figure
\ref{fig:radiusresidual}, while Table \ref{tab:radial} summarizes the
results of performing a linear least squares fits to the residuals as
a function of radius.  There is a slight negative slope in the $B$
band residuals, and a slight positive slope in the $I$ band residuals.
This suggests that the Cepheid spectral energy distributions (SEDs)
are shifting towards the blue with increasing galactocentric radius.
G11 found a similar result in M81, and they note that this could be
due to less line-blanketing in metal-poor Cepheids near the galaxy's
periphery.  As with the G11 result, \cite{Kochanek1997} found that
metal-poor Cepheids tend to be bluer than metal-rich Cepheids, which
matches theoretical expectations (\citealt{Chiosi1993,Marconi2005}).
In order to check for any impact on the distance modulus, we tried
binning the LBT Cepheids in radius, and fit all Cepheids with $0.22
\le \rho_i<0.60$, $0.60\le\rho_i<1.00$, and $1.00\le\rho_i<1.54$
separately.  We found $\Delta\mu_{LMC}=$\radinmu\ mag in the first bin
(10 Cepheids), $\Delta\mu_{LMC}=$\radmidmu\ mag in the second bin (15
Cepheids), and $\Delta\mu_{LMC}=$\radoutmu\ mag in the third bin (15
Cepheids).  The dependence of the distance modulus on galactocentric
radius is therefore unclear at this stage, and we return to this issue
in \S 5.2.

Next, we fit the combined sample of LBT and M06 Cepheids, using the
LBT results for the Cepheids in common.  We used the final sample of
89 Cepheids used by M06, giving us a total of 122 Cepheids (seven
Cepheids were matched between samples).  We first fit the M06 Cepheids
separately, so as to estimate the intrinsic scatter of this data about
the PL relations, and found a rescaling factor of 4.42 based on
$\chi^2/dof$ for this fit.  After rescaling the M06 uncertainties (the
LBT Cepheids have already been rescaled based on the initial fit), we
found $\Delta \mu_{LMC} =$\allstandardmu\ mag, and $\chi^2/dof
=$\allstandardrchi.  Bootstrapping the full Cepheid sample yields
$\Delta\mu_{LMC} =$\bootallstandardmu\ mag.  These values are in good
agreement with those determined from the LBT Cepheids alone, and the
covariance matrix of this fit has a slightly different structure,
\begin{align}\label{equ:M06covar}
  \frac{c_{ij}} {(c_{ii}c_{jj})^{1/2}} =\left(
    \allstandardcovar\right)
\end{align}
and
\begin{align}\label{equ:M06covar2}
 c_{ii}^{1/2} = \left(\allstandardrms\right).
\end{align}
Figure \ref{fig:radiusresidual} also shows the residuals of the M06
data from the PL relations as a function of galactocentric radius.
Using the combined data set, we find trends in the $B$ and $I$ bands
consistent with those found with the LBT Cepheids alone but at a
higher level of significance.  In addition, if we exclude $H$ band data
of the LBT Cepheids and recalculate the covariance matrix, we find
\begin{align}\label{equ:M06covarBVI}
  \frac{c_{ij}} {(c_{ii}c_{jj})^{1/2}} =\left(\begin{array}{rrr}
       1.00  &-0.02 &-0.65\\
       -0.02  &1.00  &-0.37\\
      -0.65  &-0.37  &1.00\\
    \end{array}\right)
\end{align}
and
\begin{align}\label{equ:M06covar2BVI}
 c_{ii}^{1/2} = \left(\begin{array}{rrr}B&V&I\\0.11&0.04&0.09 \end{array}\right).
\end{align}
The vector defining the principal component is $\boldsymbol{p}_{BVI} =
(0.80,0.04,-0.60)$, which is remarkably similar to the direction
derived from the $BVI$ measurements of the LBT Cepheids alone.  We
return to the interpretation $\boldsymbol{p}_1$ and $\boldsymbol{p}_{BVI}$ in \S
5.2.

We also experimented with fits to a broken PL relation, since the M06
sample contains 12 Cepheids with $P<10$ days.  Using the
\cite{Sandage2004} broken PL relations, we find that $\Delta\mu_{LMC}
= 10.71\pm0.01$ mag with $\chi^2/dof = 1.01$, while the the broken PL
relations from \cite{Ngeow2009} yield $\Delta\mu_{LMC} = 10.62\pm0.01$
mag with $\chi^2/dof = 1.19$.  These results are in good agreement
with our previous fits (again accounting for the \citealt{Ngeow2009}
$I$ band PL relation, see Table 5).  Since there are only 12 Cepheids
with $P<10$ days in the final M06 sample, it is likely that the 110
longer-period Cepheids overwhelm the fit.  To check this hypothesis, we
excluded the 12 short-period Cepheids, and refit the longer period
Cepheids using both the broken and linear PL relations.  We found that
the distance modulus does not change when using the broken PL
relations with the longer-period Cepheids alone, and that it decreases
by only 0.01 mag when using the linear PL relations.  These
consistencies indicate that short-period Cepheids and any break in the
PL relations have a minimal impact on our procedure.

\subsection{Model 2 -- varying the extinction law}
\begin{figure*}
  \includegraphics[width=1\textwidth]{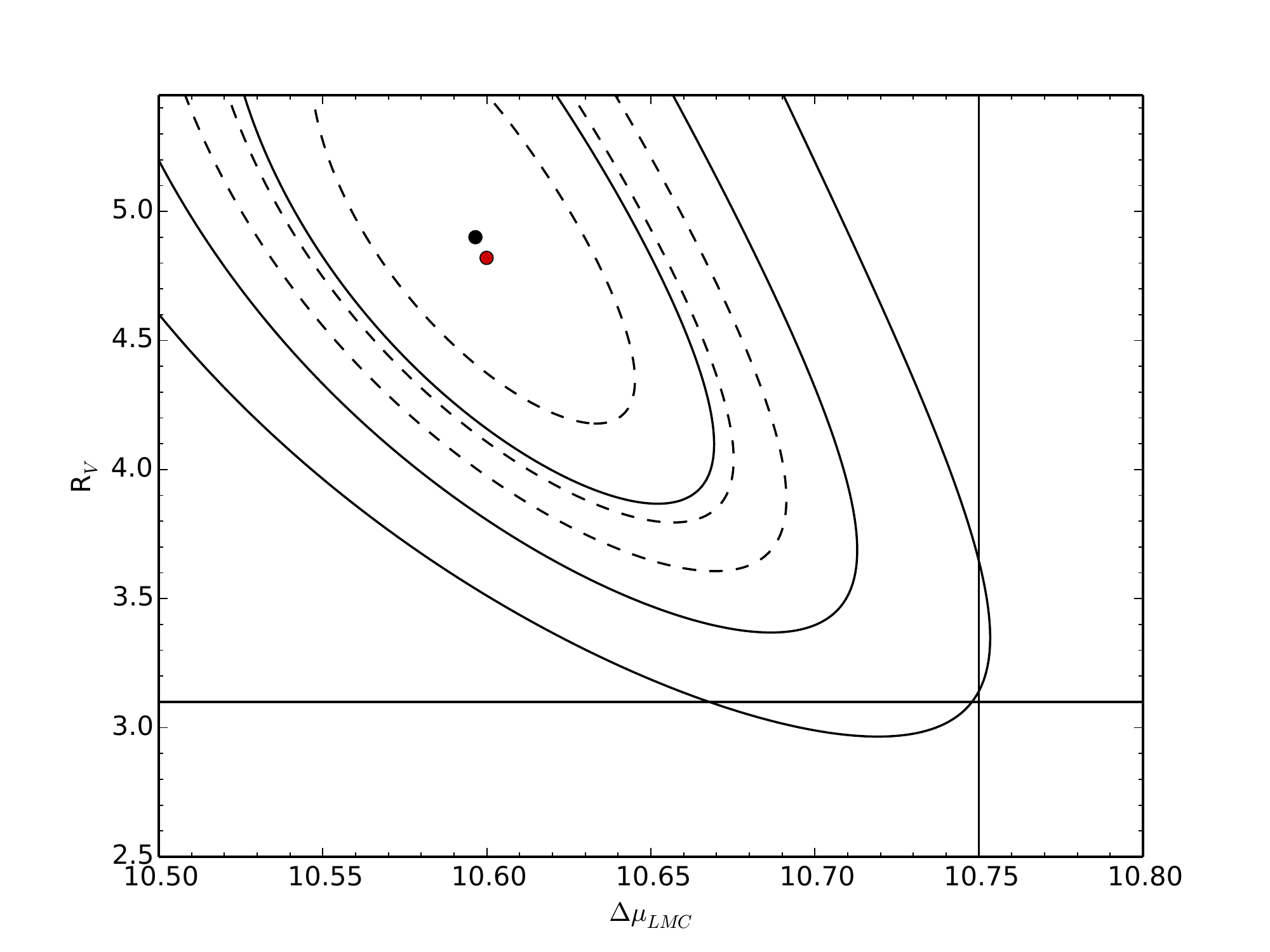}
  \caption{Contours of $\Delta\chi^2$ as a function of
    $\Delta\mu_{LMC}$ and $R_V$, using the LBT Cepheids only (N=40).
    The solid curves are the two-parameter $1$, $2$, and $3\sigma$
    confidence intervals.  The dashed curves are the same but for one
    parameter. The solid lines mark the standard value of $R_V=3.1$
    and the corresponding distance modulus.  The red point is the
    median of the bootstrap resampling distribution.}
  \label{fig:RV}
\end{figure*}

Next, we refit the LBT Cepheids alone while allowing the extinction
law to vary, and Figure \ref{fig:RV} displays the best fit value of
$R_V$ and $\Delta\mu_{LMC}$, along with contours of the $\chi^2$
surface.  We find that the best fit parameters are $R_V =$ \RVbest,
with $\Delta\mu_{LMC}=$\extlawmu\ mag.  This value of $R_V$ improved
the fit by $\Delta\chi^2 =9.60$.  Bootstrap resampling the Cepheids
yields $R_V =$ \bootRV\ and $\Delta\mu_{LMC}$=\bootextlawmu\ mag.  As
we would expect from the modest reduction in $\chi^2$, we do not see a
significant decrease in the strength of correlations between residuals
in different bands:
\begin{align}
  \frac{c_{ij}} {(c_{ii}c_{jj})^{1/2}} &=\left( \extlawcovar\right)
\end{align}
and
\begin{align}
  c_{ii}^{1/2} &= \left(\extlawrms\right).
\end{align}
Varying $R_V$ corresponds to fitting residuals in the direction
\begin{align}
  \frac{\partial \boldsymbol{R}_F}{\partial R_V} =
  \left(\begin{matrix}B&V&I&H\\-0.05& 0.00& 0.03&
      0.04\end{matrix}\right)\left(\frac{R_V}{3.1}\right)^{-2},
\end{align}
and, as we see in Figure \ref{fig:residue}, this vector is not aligned
with the residuals.  Quantitatively, the dot product of the direction
associated with $\boldsymbol{p}_1$ and the derivative of the
extinction law with respect to $R_V$ is only 8\%, and so we would not
expect changes in the extinction law to absorb very much of the
principal component.  If we include the M06 Cepheids, we find
$\Delta\mu_{LMC}=$ \allextlawmuM\ mag with $R_V=$ \RVbestM.  For the
\citet{Cardelli1989} extinction law and $R_V=$\RVbest, the reddening
vector is (5.91, 4.90, 3.26, 1.16) in $B$, $V$, $I$ and F160W,
respectively.  If $R_V$ were instead assumed to be 3.1, the
extinctions in the $I$ and F160W bands would be underestimated by 56\%
and 55\%, respectively.  While the magnitude of extinction in the
near-IR is still several times smaller than in the optical, the grayer
extinction law changes the distance modulus by 0.11 mag, which
corresponds to a 5\% change in the distance.

It would be dangerous to interpret our large value of $R_V = 4.9$ as
an indication of the physical properties of the dust in NGC
4258. $R_V>4.5$ is a condition realized for some sight-lines (both
within the Milky Way and towards extra-galactic sources), but such
extinction laws are usually associated with molecular clouds.  For
example, \cite{DeMarchi2014a} found $R_V\sim 5.6$ for a sight-line
offset 6 arcminutes from 30 Dor in the LMC, and $R_V\sim 4.5$ within
the nebula/HII region itself (although with a different functional
form than the \cite{Cardelli1989} parameterization, see
\citealt{DeMarchi2014b}).  On the other hand, \cite{Gordon2003} found
a mean value of $R_V = 3.41$ over the entire LMC, and
\cite{Pejcha2012} found a mean value of $R_V = 3.127$ for an aggregate
sample of Galactic, LMC, and Small Magallenic Cloud Cepheids.  Several
of these studies also found some discrepancy between the Milky Way
extinction law and those inferred for the LMC
\citep{DeMarchi2014a,DeMarchi2014b,Gordon2003}, which highlights the
systematic issues associated with assuming a universal extinction law.

With this discussion in mind, we find it unlikely that a physical
reason explains why the mean extinction law across the entire disk of
NGC 4258 would be parameterized by $R_V=4.9$.  The large value of
$R_V$ likely indicates that the Cepheid data prefer some kind of color
correction.  This could be due to a number of factors, for example,
systematic photometric errors in a single pass-band, or mistakes in
the adopted PL relations.  In fact, multiple systematic effects,
including true variations in the extinction law, may be operating
simultaneously.  Given the extremely gray extinction law implied by
$R_V=$ \RVbestshort, it seems probable that at least one other
systematic effect is at work in our sample's colors.  We pursue the
question of peculiar Cepheid colors further in \S5.2.

\subsection{Model 3 -- metallicity effects}\label{sec:metallicity effect}
In general, metallicity may affect both the mean magnitudes and the
colors of the Cepheids.  This means that we must allow the distance
modulus as well as the mean magnitudes in each filter to vary based on
the Cepheid metallicities, which can be accomplished by setting
\begin{align}
  \boldsymbol{\gamma} = (\gamma_1\boldsymbol{\mu} + \gamma_2\boldsymbol{c} )(Z_i-Z_{LMC})
\end{align}
where $\boldsymbol{\mu} = (1,1,1,1)$ corresponds to metallicity
effects that change the distance but not the color, and
$\boldsymbol{c}$ is a vector quantifying the magnitude of the
metallicity effect in each pass band.  Most Cepheid studies have only
examined the effects of $\gamma_1$ and simply assume that $\gamma_2
\equiv 0$.  So as to compare our results with these studies, we first
set $\gamma_2\equiv 0$ and solve for $\gamma_1$.  In subsequent
models, we solve for both parameters.  However, $\boldsymbol{c}$\ is
not known \emph{a priori}, and some component of this vector is
probably degenerate with extinction.  We know that our residuals are
dominated by a single principal component, either $\boldsymbol{p}_1$
or $\boldsymbol{p}_{BVI}$, and we will therefore experiment with
models where $\boldsymbol{c} = \boldsymbol{p}_1$ or
$\boldsymbol{p}_{BVI}$.

For the LBT Cepheids alone, Table \ref{tab:metaldistance1} shows the
results of incorporating $\gamma_1$ with $\gamma_2 \equiv 0$ and
testing models with $R_V $ equal to both 3.1 and 4.9.  For simplicity,
we only discuss the results for $R_V=3.1$, although the results for
$R_V=4.9$ can be found in Table \ref{tab:metaldistance1}.  The
inferred values of $\Delta\mu_{LMC}$ are \Zstandardonemu\ mag,
\Zstandardtwomu\ mag, and \Bstandardmu\ mag in the Z94-1, Z94-2, and
B11-e models, respectively.  The changing distance modulus is a
consequence of systematic uncertainties associated with absolute
metallicity measurements. The Z94-2 gradient
($\Delta\mu_{LMC}=$\Zstandardtwomu\ mag) is shallower than that of
Z94-1, which places NGC 4258 at a higher mean metallicity, and our
model forces the distance modulus to absorb this offset.  The
shallower gradient also drives correlations between $\gamma_1$ and
$\Delta\mu_{LMC}$, which results in a flatter slope around the minimum
value of $\chi^2$ and a larger uncertainty on
$\Delta\mu_{LMC}$. However, in the B11-e model, the difference between
the metallicity zeropoints is smaller than in Z94-2, which balances
the effects of the shallower gradient.

\begin{figure*}                                                      
    \includegraphics[width=1\textwidth]{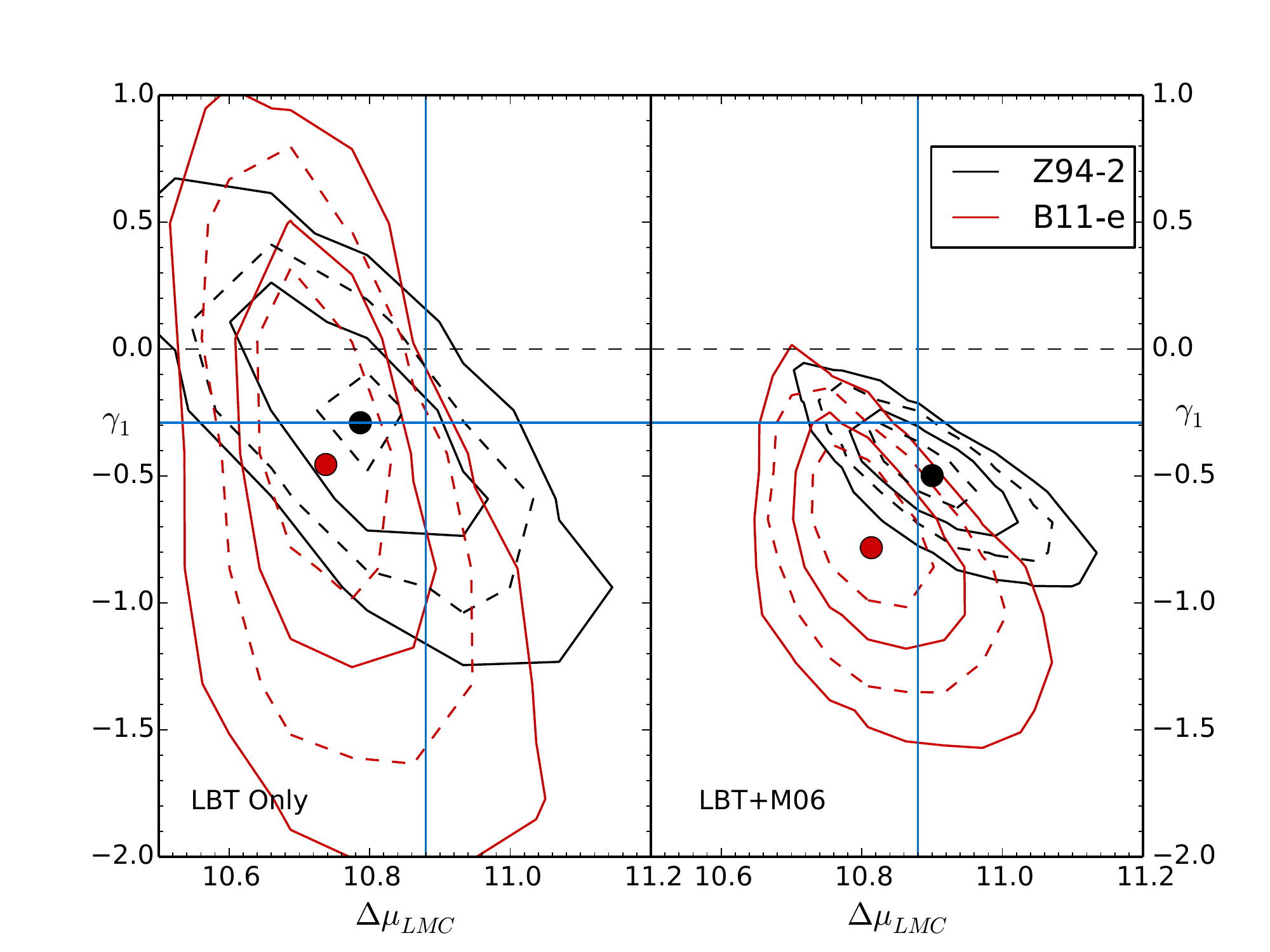}       
    \caption{Distributions of $\Delta\mu$ and $\gamma_1$ obtained from
      bootstrap resampling.  The left panel uses only the LBT
      Cepheids, while the right panel shows the combined LBT+M06
      sample (N=122).  The black lines are for the Z94-2 metallicity
      system, while the red contours are for the B11-e system.  The
      solid contours are the 68\% and 95\% limits for two parameters,
      while the dashed contours mark the limits for a single
      parameter.  The large circles mark the medians of the
      distributions, and the blue lines show the measurement from
      M06.}
    \label{fig:gamma1}
\end{figure*}

We emphasize that these issues are a consequence of the assumed
metallicity scale, while the effect of the relative Cepheid
metallicities is quantified by $\gamma_1$.  We find that $\gamma_1$=
\Zstandardonegammaa\ mag/dex, \Zstandardtwogammaa\ mag/dex, and
\Bstandardgammaa\ mag/dex in the Z94-1, Z94-2, and B11-e models,
respectively. A shallower metallicity gradient implies a larger value
of $\gamma_1$, as seen in the Z94-2 and B11-e models, since larger
values of $\gamma_1$ can be used to compensate for a smaller range of
Cepheid metallicities.  The parameter uncertainties based on rescaled
$\Delta\chi^2$ statistics continue to agree well with the bootstrap
models, as can be seen in Table \ref{tab:metaldistance1}.  Figure
\ref{fig:gamma1} shows the distribution of $\Delta\mu_{LMC}$ and
$\gamma_1$ obtained from our bootstrap estimates.  However, adding
this metallicity term does not significantly reduce or change the
covariance matrix of the residuals, since $\gamma_1$ is a correction
applied in the direction $(1, 1, 1, 1)$, which cannot absorb any
significant part of $\boldsymbol{p}_1$ or $\boldsymbol{p}_{BVI}$.

The sign of $\gamma_1$ is defined so that metal-rich Cepheids are
intrinsically more luminous than their metal-poor counterparts
(Equation \ref{equ:model}), as is typically found in other Cepheid
studies (e.g. M06 and \citealt{Shappee2011}).  In order to compare our
results with those of M06, we fit the LBT Cepheids alone, using the
same metallicity gradient and extinction law.  We find a smaller
metallicity dependence of \Zstandardonegammaa\ mag/dex, compared to
their reported value of \macrimetal\ mag/dex.  If we instead fit the
combined LBT+M06 Cepheid sample, we find that $\gamma_1
=$\allZstandardonegammaa\ mag/dex with $\Delta\mu_{LMC}
=$\allZstandardonemu\ mag, and $\chi^2/dof =$\allZstandardonerchi, in
good agreement with the original findings of M06.  We find
$\Delta\mu_{LMC}=$\allZstandardtwomu\ mag and $\gamma_1
=$\allZstandardtwogammaa\ mag/dex for the Z94-2 model
($\Delta\mu=$\bootallZstandardtwomu\ mag, $\gamma_1
=$\bootallZstandardtwogammaa\ mag/dex with bootstrap resampling), and
in the B11-e model, we find $\Delta\mu_{LMC}=$\allBstandardmu\ mag and
$\gamma_1 =$\allBstandardgammaa\ mag/dex
($\Delta\mu_{LMC}=$\bootallBstandardmu\ mag, $\gamma_1
=$\bootallBstandardgammaa\ mag/dex with bootstrap resampling).  The
differences in the results between the rescaled $\chi^2/dof$ method
and bootstrap resampling method may be due to outliers in the M06 data
set.  If we employ a 5-$\sigma$ iterative clipping routine, we reject
19 M06 Cepheids, and find for the combined data set that
$\Delta\mu_{LMC} = 10.89\pm0.06$\ mag and $\gamma_1 = -0.42\pm0.14$\
mag/dex in the Z94-2 model, with bootstrap resampling yielding
$\Delta\mu_{LMC} = 10.90\pm0.08$\ mag and $\gamma_1 = -0.49\pm0.17$\
mag/dex.  In the B11-e model, we find $\Delta\mu_{LMC} = 10.81\pm
0.03$\ mag and $\gamma_1 = -0.65\pm0.22$\ mag/dex with the trimmed
sample, while bootstrapping yields $\Delta\mu_{LMC} = 10.81\pm0.08$\
mag and $\gamma_1 = -0.77\pm0.29$\ mag/dex.  These values and their
uncertainties are in much better agreement, although using the full
set of 122 Cepheids still yields consistent results.  We therefore
retain the full set of M06 Cepheids for all of our models.

Figure \ref{fig:gamma1} shows the bootstrap resampling distributions
of $\Delta\mu_{LMC}$ and $\gamma_1$ for the combined sample.
Generally speaking, the combined fit pulls the distance modulus
$\sim0.15$ mag higher and shrinks the uncertainties on $\gamma_1$ such
that $\gamma_1=0$ is statistically ruled out.  As can be seen in
Figure \ref{fig:radiusresidual}, this is due to trends in the
residuals from the PL relations with galactocentric radius that become
more robust with the full sample.

\begin{figure*}
    \includegraphics[width=1\textwidth]{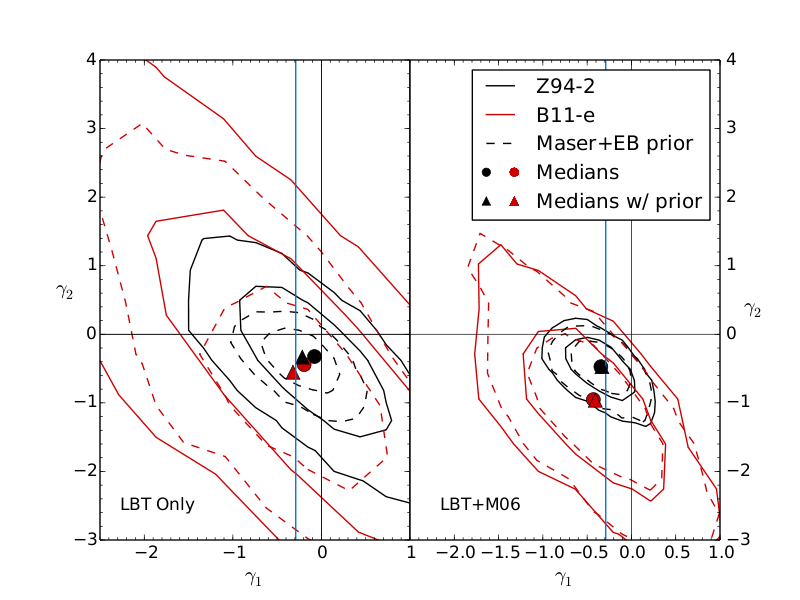}
    \caption{Distributions of $\gamma_1$ and $\gamma_2$ obtained from
      bootstrap resampling.  The left panel uses only the LBT Cepheids
      and the right panel shows the combined LBT+M06 sample (N=122).
      The black lines are for the Z94-2 metallicity system, while the
      red contours are for the B11-e system.  The vertical blue lines
      mark the value of $\gamma_1$ found by M06.  The solid contours
      are the 68\% and 95\% limits for two parameters, while the
      dashed contours are the limits obtained by imposing a prior on
      $\Delta\mu_{LMC}$, based on the \citet{Pietrzy2013} eclipsing
      binary distance for the LMC and the maser distance of NGC 4258
      from \citet{Humphreys2013}.}
    \label{fig:gamma2}
\end{figure*}

Table \ref{tab:metaldistance2} shows the results of fitting for
$\gamma_2$ with $\boldsymbol{c} = \boldsymbol{p}_1$, and Figure
\ref{fig:gamma2} shows the bootstrapping distributions of $\gamma_1$
and $\gamma_2$.  The LBT Cepheids alone do little to constrain either
$\gamma_1$ or $\gamma_2$.  However, we find that the combined LBT+M06
fit rules out $\gamma_1 = \gamma_2 = 0$ at $\gtrsim 99\%$ confidence,
due to the shape of the error ellipse.  In the Z94-1, Z94-2, and B11-e
models, respectively, we find that $\gamma_1 = $
\allZstandardonegtwogammaa\ mag/dex, \allZstandardtwogtwogammaa\
mag/dex, and \allBstandardgtwogammaa\ mag/dex, while $\gamma_2 =
$\allZstandardonegtwogammab\ mag/dex, \allZstandardtwogtwogammab\
mag/dex, and \allBstandardgtwogammab\ mag/dex.  This is a
$1.8$-$2\sigma$ detection of Cepheid color shifts with metallicity,
and incorporating $\gamma_2$ obviates the need for a direct correction
of the distance modulus.  Because of the way in which
$\boldsymbol{p}_1$ is defined, this correction implies that an
increasing Cepheid metallicity results in a fainter $BV$ magnitude and
brighter $I$/F160W magnitude.  However, we also note that the
magnitude of the correction increases towards the near IR, contrary to
theoretical predictions.  In addition, the correction to the F160W
band mean magnitudes is four times larger than in any other band,
which is a result of our poor photometric coverage at this wavelength
and the problematic definition of $\boldsymbol{p}_1$.  Interestingly,
changing the extinction law to $R_V=$ \RVbestshort\ (and thereby
changing the Cepheid colors) makes $\gamma_2$ consistent with 0, while
$\gamma_1$ becomes significant at the $2.3$-$3.1\sigma$ level.  This
implies that Cepheid mean magnitudes must depend on metallicity, even
if this effect is degenerate with a systematic color correction
required by the data.

Figure \ref{fig:gamma2} also shows the results of imposing a prior on
$\Delta\mu_{LMC}$, based on the maser distance to NGC 4258
\citep[$\mu_{N4258} = 29.40 \pm 0.06$\ mag]{Humphreys2013} and the
eclipsing binary distance to the LMC \citep[$\mu_{LMC} = 18.49\pm
0.05$\ mag]{Pietrzy2013}.  Our prior takes the form of a Gaussian
probability distribution with mean $10.91$ mag and width $0.08$ mag,
derived from adding the uncertainties of the independent distance
estimates in quadrature.  The results from the previous models are
only discrepant with this value by about 2$\sigma$, and so the prior
does little to constrain the data.  The best fit parameters, estimated
from bootstrap resampling, can be found in Table \ref{tab:priors}.  We
find that there is a slight effect on the LBT Cepheid sample, which
increases the distance modulus by $\sim 0.05$ mag and slightly narrows
the error ellipse for $\gamma_1$ and $\gamma_2$.  The prior also
shifts the values of $\gamma_1$ and $\gamma_2$ in a way that is
consistent with the values inferred from the combined LBT+M06 sample,
although the model still does not result in a significant detection of
either parameter.  The prior has no effect on the results of the
combined sample, since they are already consistent with the value of
$10.91\pm 0.08$\ mag.

\begin{figure*}
    \includegraphics[width=1\textwidth]{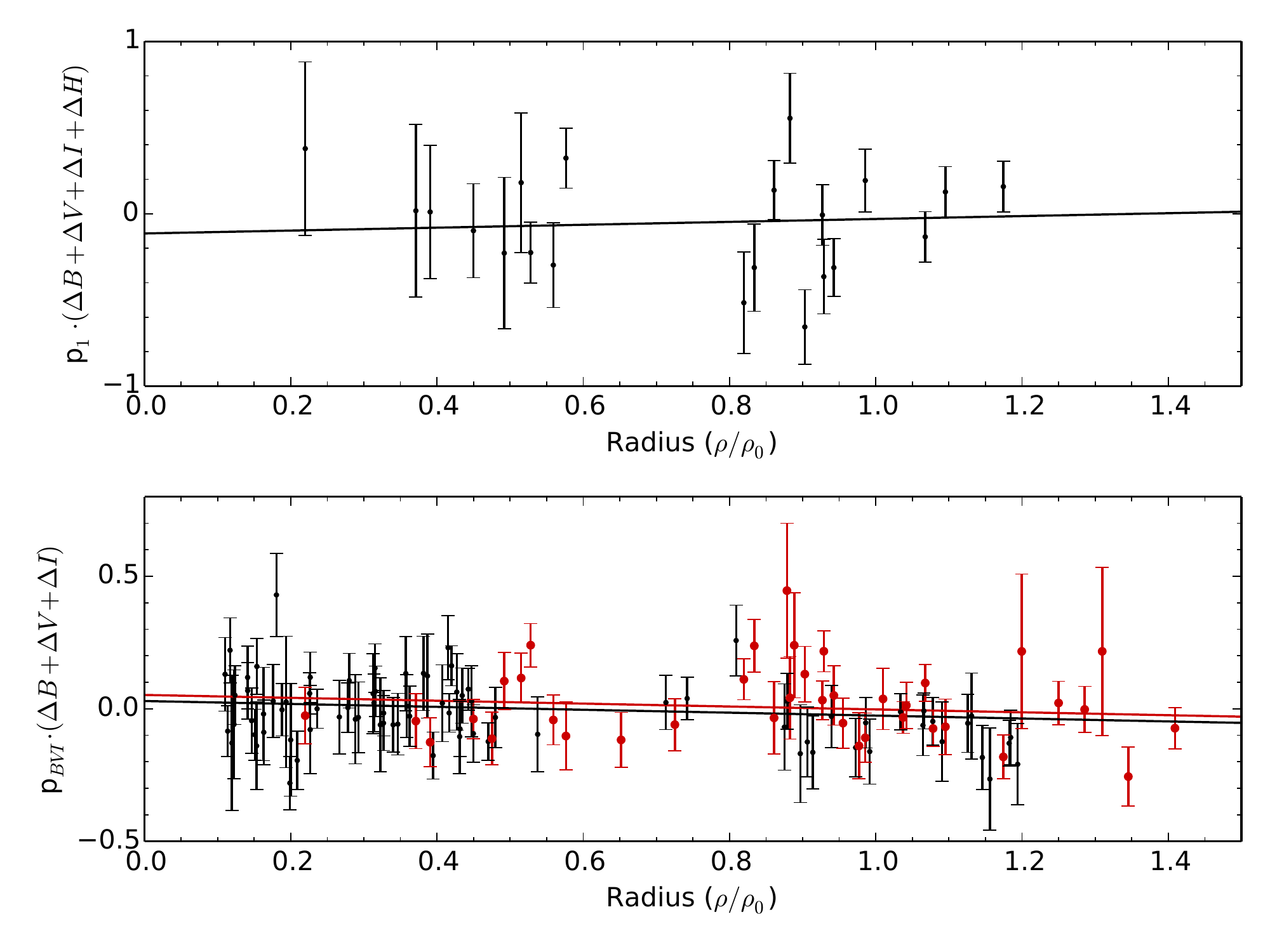}
    \caption{Residuals from the PL relations projected onto their
      principal components, as a function of galactocentric radius.
      The top panel includes all four filters, for which we only have
      21 Cepheids.  The bottom panel uses BVI only, and includes the
      LBT Cepheids in red and the M06 Cepheids as the small black
      points.  The solid lines are the results of linear least-squares
      fits.}
    \label{fig:pcradius}
\end{figure*}

After inspecting the covariance matrix for this fit, we found that
this model does little to reduce the covariances between different
bands.  Essentially, this is because the residuals from model 1,
projected onto the principle component $\boldsymbol{p}_1$, do not
correlate very strongly with galactocentric radius, as shown in Figure
\ref{fig:pcradius}.  As noted in \S5, $\boldsymbol{p}_1$ is only
defined for 21 Cepheids, and most of the information in the data come
from the $BVI$ filters.  In Figure \ref{fig:pcradius} we also show the
residuals projected onto $\boldsymbol{p}_{BVI}$ as a function of
galactocentric radius for both the LBT and M06 Cepheids.  After
performing a linear least-squares fit, we again see a more significant
slope, changing from $-0.05\pm0.05$ using the LBT Cepheids alone to
$-0.05\pm 0.03$\ mag/$\rho$\ when using all 122 Cepheids, consistent
with the trends in Figure \ref{fig:radiusresidual}.

\begin{figure*}
    \includegraphics[width=1\textwidth]{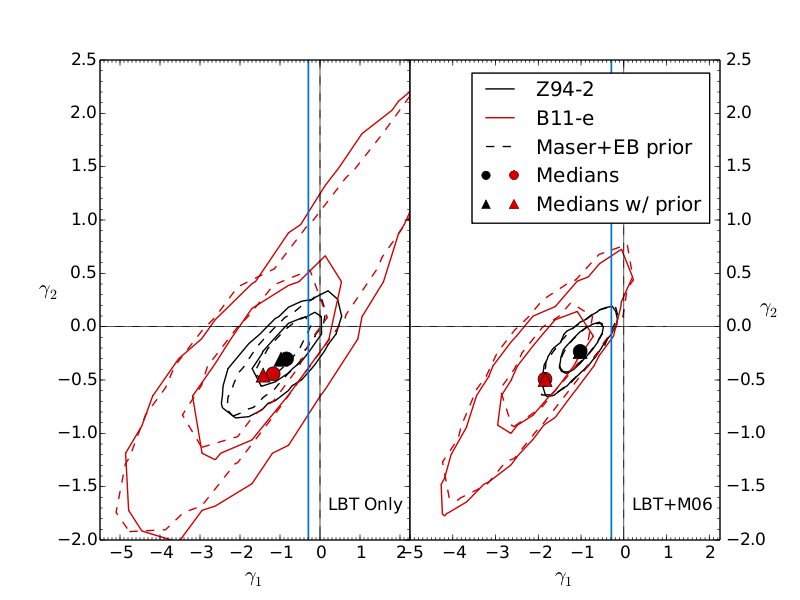}
    \caption{Distribution of $\gamma_1$ and $\gamma_2$ obtained from
      bootstrap resampling and setting $\boldsymbol{c}=\boldsymbol{p}_{BVI}$, using
      only the BVI data.  The left panel uses the 40 LBT Cepheids and
      the right panel shows the combined LBT+M06 sample (N=122).  The
      black contours represent the Z94-2 metallicity system, while the
      red contours are the B11-e system.  The vertical blue lines mark
      the value of $\gamma_1$ found by M06.  The solid contours are
      the 68\% and 95\% limits for two parameters, while the dashed
      contours are the limits obtained by imposing a prior on
      $\Delta\mu_{LMC}$, based on the \citet{Pietrzy2013} eclipsing
      binary distance for the LMC and the maser distance of NGC 4258
      from \citet{Humphreys2013}.}
    \label{fig:redo_g1_g2}
\end{figure*}

Table \ref{tab:redo_gamma2} shows fits for $\gamma_1$ and $\gamma_2$
with $\boldsymbol{c} = \boldsymbol{p}_{BVI}$, and Figure
\ref{fig:redo_g1_g2} shows the bootstrap resampling distribution.  At
this stage, we only discuss the combined LBT+M06 sample, and we find
that $\gamma_1 =$\allZstandardonegtwoBVIgammaa\ mag/dex,
\allZstandardtwogtwoBVIgammaa\ mag/dex, and
\allBstandardgtwoBVIgammaa\ mag/dex, with $\gamma_2
=$\allZstandardonegtwoBVIgammab\ mag/dex,
\allZstandardtwogtwoBVIgammab\ mag/dex, and \allBstandardgtwogammab\
mag/dex, in the Z94-1, Z94-2, and B11-e models, respectively
($\Delta\mu_{LMC} =$\allZstandardonegtwoBVImu\ mag,
\allZstandardtwogtwoBVImu\ mag, and \allBstandardgtwoBVImu\ mag for
these fits). While fits with 3-band photometry call for some
adjustment to the distance modulus with metallicity, they are unable
to tightly constrain any color effects.

In Figure \ref{fig:redo_g1_g2} we also show the results of imposing a
prior on $\Delta\mu_{LMC}$, and we find shifts in $\gamma_1$ and
$\gamma_2$ using the LBT data alone that move these parameters closer
to the the values obtained by fitting the larger sample.  Furthermore,
the LBT Cepheid sample in the Z94-2 metallicity model produces a
metallicity effect of $\gamma_1=$ \bootZstandardtwogtwoBVIpriorgammaa,
detected at the 1.9$\sigma$ level ($\gamma_2$ =
\bootZstandardtwogtwoBVIpriorgammab), but all other results are still
consistent with $\gamma_1 = \gamma_2 = 0$.  The prior again has no
effect on the combined Cepheid sample, since the results from these
models are already consistent within the uncertainties.

\section{DISCUSSION AND CONCLUSIONS}
\subsection{Metallicity dependence of the PL relations}
Our sample of 40 LBT Cepheids does not provide substantial evidence
for a metallicity-dependent adjustment to the distance modulus of NGC
4258.  The basic cause is the lack of any strong correlations of the
residuals from the PL relations with galactocentric radius, used as a
proxy for metallicity.  This is exacerbated by the high inclination of
NGC 4258's disk ($\sim72^{\circ}$) and the shallow metallicity
gradient (\citealt{Bono2008}, \citealt{Bresolin2011}).  However, if we
combine our sample with the Cepheids found by M06, we find a
statistically significant metallicity dependence of the mean
magnitudes, at a value consistent with their estimate.

While the detection of the metallicity dependence using the combined
data set is a robust feature of all the adopted metallicity scales,
systematic uncertainties in the metallicity scales themselves limit
the physical interpretation of this effect.  We find mean magnitude
corrections ranging from $\gamma_1=$\gammamin\ mag/dex to \gammamax\
mag/dex, depending on the metallicity system used to estimate the
Cepheid compositions.  Regardless of assumptions about how well the
oxygen abundance gradient tracks the physical metallicity of the
Cepheids, the broad range of parameter estimates for $\gamma_1$
illustrates how strongly metallicity measurement uncertainties affect
estimates of the metallicity dependence of the PL relations, and
perhaps accounts for the wide range of values found in the literature.
Because the metallicity system used by Z94 is very prevalent, we
report a final metallicity effect of $\gamma_1
=$\bootallZstandardtwogammaa\ mag/dex, inferred from our bootstrap
resampling of the H II regions and Cepheids in NGC 4258 using the
Z94-2 model.  This value takes appropriate measure of the
uncertainties in both the metallicity gradient of NGC 4258 and the
Cepheid mean magnitudes, is readily comparable with other studies, and
is easily translatable into other metallicity systems.  However, we
note that there may be reasons to believe that other metallicity
scales (e.g., the empirical electron temperature scale of B11) may be
a more physical estimate of this effect.

There is a strong indication that both the LBT sample and the combined
LBT+M06 sample prefer some adjustment to the Cepheid colors.  This is
evidenced by a decrease in $\chi^2$ when using a grayer extinction law
($R_V = $\RVbestshort), as well as the appearance of a single
principle component in the residuals of our initial fitting procedure
(fits for individual extinctions only).  However, we are unable to
measure this component so as to uniquely determine its cause.  Because
of correlations between galactocentric radius and residuals from the B
and I band PL relations (see Figure \ref{fig:radiusresidual}), part of
this effect can be attributed to the varying metallicity of the
Cepheids.  However, our limited near-IR photometric coverage means
that there is considerable uncertainty as to the magnitude of this
effect in the F160W band, and we do not detect the metallicity-color
correction using 3 band BVI photometry alone.  On the other hand, a
grayer extinction law removes the need for metallicity-dependent color
corrections (although corrections to the distance modulus are still
found), but the very large value of $R_V=$ \RVbestshort\ suggests that
this adjustment may be due to some other systematic effect beyond
variations in the extinction law.  For example, the adopted PL
relations directly determine the expected Cepheid colors, so any
errors in their determination (for example, due to interpolation or
de-extinction procedures) could mimic variations in the extinction
law.  In order to disentangle these systematic effects, it appears
that a larger sample with complete 4 band photometry is needed.

While this study has been predominately concerned with the effects of
the extinction law and metallicity on Cepheid colors, other systematic
effects exist that are expected to contribute to the problem.  These
include the difficulties of obtaining precise photometry of Cepheids
in crowded fields, and, more importantly, the unknown systematic
effects of blending due to stars physically associated with the
Cepheids.  To combat these issues, future studies will require a more
thorough characterization of the Cepheid SEDs, with high quality data
in many pass bands.

\subsection{Calibration of the Cepheid PL relations}
\begin{figure*}
  \begin{center}1
    \includegraphics[width=1\textwidth]{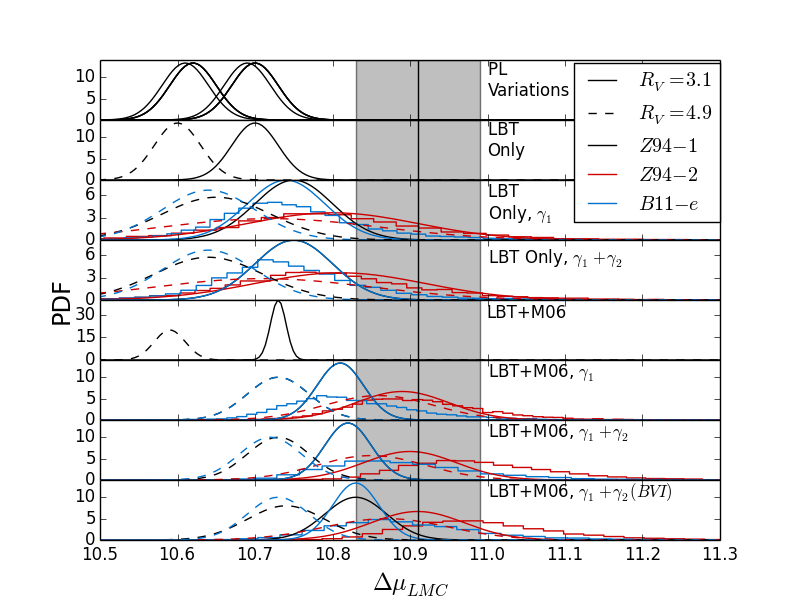}
    \caption{Probability distribution functions for $\Delta\mu_{LMC}$
      from all models.  The vertical black line marks the value of
      $\Delta\mu_{LMC} = 10.91\pm0.08$, based on the
      \citet{Pietrzy2013} eclipsing binary distance for the LMC and
      maser distance of NGC 4258 (\citet{Humphreys2013}).  The shaded
      gray region marks the 1$\sigma$ uncertainty associated with this
      value.}
    \label{fig:mu_agg}
  \end{center}
\end{figure*}

Calibrating the Cepheid PL relation is equivalent to determining an
absolute distance to the LMC.  This can be accomplished by means of
the \citet{Humphreys2013} geometric maser distance to NGC 4258.
Taking $\mu_{N4258}=29.40\pm0.06$ mag ($7.6\pm0.23$ Mpc), we calculate
$\mu_{LMC}$ for each fit in Tables \ref{tab:metaldistance1},
\ref{tab:metaldistance2}, and \ref{tab:redo_gamma2}.  Figure
\ref{fig:mu_agg} shows an alternative means of visualizing the data,
by displaying the probability density functions (PDFs) for all
estimates of $\Delta\mu_{LMC}$.  The PDFs are taken to be univariate
Gaussians, except for the bootstrapping estimates for which we show
the (normalized) posterior distributions.  The vertical black line and
shaded gray region mark $\Delta\mu_{LMC} = 10.91\pm 0.08$ mag,
determined from $\mu_{N4258} = 29.40 \pm 0.06$
(\citealt{Humphreys2013}) and $\mu_{LMC} = 18.49\pm 0.05$
(\citealt{Pietrzy2013}), with the uncertainties added in quadrature.

Although there are small differences in $\chi^2$ for each fit, we have
no strong evidence in favor of any of the particular models that we
tried.  However, we also note that including a metallicity effect
tends to shift the Cepheid distance towards the value inferred by
independent determinations, in some cases to within 0.01--0.02 mag
(0.2--0.3$\sigma$).  A conservative way to combine all of the results
is to simply combine all the models with equal weight.  Thus, if
$P_i(\Delta\mu_{LMC})$ is the probability distribution for model $i$,
we define the joint PDF as $(\sum P_i )/N$.  This gives a
particularly simple form for the mean and variance of the joint
probability distribution
\begin{align}
  \langle\Delta \bar\mu_{LMC} \rangle = \sum_i^N\frac{ \langle
    \Delta\mu_{LMC}\rangle_i}{N}
\end{align}
and
\begin{align}
  \langle \left(\Delta\bar\mu_{LMC} - \langle \Delta\bar\mu_{LMC}\rangle
  \right)^2\rangle = \\ \sum_i^N\frac{ \left(\langle \Delta\mu_{LMC}\rangle_i - \langle
      \Delta\bar\mu_{LMC}\rangle \right)^2 + \sigma_i^2}{N}
\end{align}
where $\langle\Delta \bar\mu_{LMC} \rangle$ is the average of the means of
the individual PDFs, and its variance is the quadrature sum of the rms
scatter and the arithmetic mean of the intrinsic widths.  We can
interpret
\begin{align}
\sigma_{stat} = \sqrt{\sum_i^N\frac{\sigma_i^2}{N}}
\end{align}
as an estimate of our statistical error, and
\begin{align}
  \sigma_{sys} = \sqrt{\frac{1}{N}\sum_i^N \left(\langle \Delta\mu_{LMC}\rangle_i -
      \langle \Delta\bar\mu_{LMC}\rangle \right)^2}
\end{align}
as an estimate of our systematic uncertainties.

We include all PDFs shown in Figure \ref{fig:mu_agg}, but separate the
PDFs derived from the LBT Cepheids only and the combined LBT+M06
sample.  If we use only the LBT Cepheid PDFs, we find $\langle
\Delta\bar \mu_{LMC}\rangle =$\muLBT, while for the combined LBT+M06
PDFs we find $\langle \Delta\bar \mu_{LMC}\rangle =$\muM.  These
values translate into LMC distances of \muLMCLBT\ and \muLMCM.  We
choose to adopt the value from the combined LBT+M06 sample, which
corresponds to an LMC distance of \distLMCM\ kpc (6\%
uncertainty). The smaller value of $\mu_{LMC}$ is driven by the
stronger metallicity dependence found for the combined data.  While it
is trivial to derive the absolute PL relations from this distance, we
provide a calibration in Table \ref{tab:absPLs} for completeness.  The
uncertainty is dominated by the error on $\mu_{LMC}$, yielding
calibrations accurate to 13\% in luminosity.

\subsection{Summary}
We have identified 81 Cepheids in the maser-host Galaxy NGC 4258 using
data collected over 5 years from the LBT.  Using image subtraction and
empirical lightcurve templates, we were able to accurately phase the
Cepheids, and we efficiently calibrated the Cepheid mean magnitudes
using \emph{HST}.  Our final sample consists of 40 Cepheids, limited
by the available \emph{HST} data, with photometry in (up to) four
different pass bands.  Our sample was fit to PL relations determined
from LMC Cepheids, using several models that explored uncertainty in
the PL relations, the effects of extinction, the form of the
extinction law, and metallicity on the determined distance modulus.
Our key results are as follows:
\begin{enumerate}
\item While the LBT data set does not support a statistically
  significant metallicity dependence, combining the LBT Cepheids with
  those from M06 yields a robust detection.  The possible values of
  the observed effect are largely compatible with previously
  determined values from the literature, but uncertainties in the
  underlying metallicity scale make interpretation of the absolute
  effect difficult.  We report a final value of $\gamma_1
  =$\bootallZstandardtwogammaa\ mag/dex, which uses the prevalent
  metallicity system of \citet{Zaritsky1994} and takes appropriate
  account of uncertainties in both the Cepheid mean magnitudes and the
  metallicity gradient of NGC 4258.
\item There is evidence for color corrections to the PL relations,
  which are consistent with either a grayer extinction law in NGC 4258
  compared to the Milky way ($R_V$ =\RVbest), or a
  metallicity-dependent correction to the Cepheid mean magnitudes.
  While both effects are of physical interest, we lack sufficient 4
  band photometric data to disentangle these possibilities from other
  systematic effects.
\item Despite the degeneracy of the color corrections with a
  metallicity term, the data rule out the possibility that there is no
  metallicity effect at $\gtrsim99\%$ confidence, as seen in the
  right-hand panel of Figure \ref{fig:gamma2}.  Furthermore,
  incorporating a metallicity adjustment to the PL relations helps to
  reconcile our Cepheid distance with independent distances to the LMC
  and NGC 4258.
\item We report a final distance modulus between NGC 4258 and the LMC
  of \muM\ mag.  Coupled with the maser distance from
  \citet{Humphreys2013}, this implies that the LMC has a distance
  modulus of $\mu_{LMC}=$ \muLMCM\ mag (\distLMCM\ kpc).
\end{enumerate}

The LBT is an international collaboration among institutions in the
United States, Italy and Germany. LBT Corporation partners are: The
Ohio State University, and The Research Corporation, on behalf of The
University of Notre Dame, University of Minnesota and University of
Virginia; The University of Arizona on behalf of the Arizona
university system; Istituto Nazionale di Astrofisica, Italy; LBT
Beteiligungsgesellschaft, Germany, representing the Max-Planck
Society, the Astrophysical Institute Potsdam, and Heidelberg
University.

MMF thanks Kevin Croxall for helpful conversations about metallicity
measurements and calibration issues.  LMM \& AGR acknowledge support by
NASA through \emph{HST} program GO-11570 from the Space Telescope Science
Institute, which is operated by AURA, Inc., under NASA contract NAS
5-26555.

\bibliography{N4258.bbl}

\begin{thebibliography}{}

\bibitem[\protect\citeauthoryear{{Abazajian}, {Adelman-McCarthy},
  {Ag{\"u}eros}, {Allam}, {Allende Prieto}, {An}, {Anderson}, {Anderson},
  {Annis}, {Bahcall} \& et al.}{{Abazajian} et~al.}{2009}]{SDSSDR72009}
{Abazajian} K.~N.,  {Adelman-McCarthy} J.~K.,  {Ag{\"u}eros} M.~A.,  {Allam}
  S.~S.,  {Allende Prieto} C.,  {An} D.,  {Anderson} K.~S.~J.,  {Anderson}
  S.~F.,  {Annis} J.,  {Bahcall} N.~A.,    et al. 2009, \apjs, 182, 543

\bibitem[\protect\citeauthoryear{{Alard} \& {Lupton}}{{Alard} \&
  {Lupton}}{1998}]{Alard1998}
{Alard} C.,  {Lupton} R.~H.,  1998, \apj, 503, 325

\bibitem[\protect\citeauthoryear{{Anderson et~al.}}{2008}]{BAO2014}
{Anderson} L.,  {Aubourg} {\'E}.,  {Bailey} S.,  {Beutler} F.,  {Bhardwaj} V.,
  {Blanton} M.,  {Bolton} A.~S.,  {Brinkmann} J.,  {Brownstein} J.~R.,
  {Burden} A.,  {Chuang} C.-H.,  {Cuesta} A.~J.,  {Dawson} K.~S.,  {Eisenstein}
  D.~J.,  {Escoffier} S.,  {Gunn} J.~E.,  {Guo} H.,  {Ho} S.,  {Honscheid} K.,
  {Howlett} C.,  {Kirkby} D.,  {Lupton} R.~H.,  {Manera} M.,  {Maraston} C.,
  {McBride} C.~K.,  {Mena} O.,  {Montesano} F.,  {Nichol} R.~C.,  {Nuza} S.~E.,
   {Olmstead} M.~D.,  {Padmanabhan} N.,  {Palanque-Delabrouille} N.,  {Parejko}
  J.,  {Percival} W.~J.,  {Petitjean} P.,  {Prada} F.,  {Price-Whelan} A.~M.,
  {Reid} B.,  {Roe} N.~A.,  {Ross} A.~J.,  {Ross} N.~P.,  {Sabiu} C.~G.,
  {Saito} S.,  {Samushia} L.,  {S{\'a}nchez} A.~G.,  {Schlegel} D.~J.,
  {Schneider} D.~P.,  {Scoccola} C.~G.,  {Seo} H.-J.,  {Skibba} R.~A.,
  {Strauss} M.~A.,  {Swanson} M.~E.~C.,  {Thomas} D.,  {Tinker} J.~L.,
  {Tojeiro} R.,  {Maga{\~n}a} M.~V.,  {Verde} L.,  {Wake} D.~A.,  {Weaver}
  B.~A.,  {Weinberg} D.~H.,  {White} M.,  {Xu} X.,  {Y{\`e}che} C.,  {Zehavi}
  I.,    {Zhao} G.-B.,  2014, \mnras, 441, 24

\bibitem[\protect\citeauthoryear{{Bertin} \& {Arnouts}}{{Bertin} \&
  {Arnouts}}{1996}]{Bertin1996}
{Bertin} E.,  {Arnouts} S.,  1996, \aaps, 117, 393

\bibitem[\protect\citeauthoryear{{Bird}, {Stanek} \& {Prieto}}{{Bird}
  et~al.}{2009}]{Bird2009}
{Bird} J.~C.,  {Stanek} K.~Z.,    {Prieto} J.~L.,  2009, \apj, 695, 874

\bibitem[\protect\citeauthoryear{{Bonanos}, {Castro}, {Macri} \&
  {Kudritzki}}{{Bonanos} et~al.}{2011}]{Bonanos2011}
{Bonanos} A.~Z.,  {Castro} N.,  {Macri} L.~M.,    {Kudritzki} R.-P.,  2011,
  \apjl, 729, L9

\bibitem[\protect\citeauthoryear{{Bono}, {Caputo}, {Fiorentino}, {Marconi} \&
  {Musella}}{{Bono} et~al.}{2008}]{Bono2008}
{Bono} G.,  {Caputo} F.,  {Fiorentino} G.,  {Marconi} M.,    {Musella} I.,
  2008, \apj, 684, 102

\bibitem[\protect\citeauthoryear{{Bono}, {Caputo}, {Marconi} \&
  {Musella}}{{Bono} et~al.}{2010}]{Bono2010}
{Bono} G.,  {Caputo} F.,  {Marconi} M.,    {Musella} I.,  2010, \apj, 715, 277

\bibitem[\protect\citeauthoryear{{Bresolin}}{{Bresolin}}{2011}]{Bresolin2011}
{Bresolin} F.,  2011, \apj, 729, 56

\bibitem[\protect\citeauthoryear{{Cardelli}, {Clayton} \& {Mathis}}{{Cardelli}
  et~al.}{1989}]{Cardelli1989}
{Cardelli} J.~A.,  {Clayton} G.~C.,    {Mathis} J.~S.,  1989, \apj, 345, 245

\bibitem[\protect\citeauthoryear{{Chavez}, {Macri} \& {Pellerin}}{{Chavez}
  et~al.}{2012}]{Chavez2012}
{Chavez} J.~M.,  {Macri} L.~M.,    {Pellerin} A.,  2012, \aj, 144, 113

\bibitem[\protect\citeauthoryear{{Chiosi}, {Wood} \& {Capitanio}}{{Chiosi}
  et~al.}{1993}]{Chiosi1993}
{Chiosi} C.,  {Wood} P.~R.,    {Capitanio} N.,  1993, \apjs, 86, 541

\bibitem[\protect\citeauthoryear{{De Marchi} \& {Panagia}}{{De Marchi} \&
  {Panagia}}{2014}]{DeMarchi2014b}
{De Marchi} G.,  {Panagia} N.,  2014, \mnras, 445, 93

\bibitem[\protect\citeauthoryear{{De Marchi}, {Panagia} \& {Girardi}}{{De
  Marchi} et~al.}{2014}]{DeMarchi2014a}
{De Marchi} G.,  {Panagia} N.,    {Girardi} L.,  2014, \mnras, 438, 513

\bibitem[\protect\citeauthoryear{{Efstathiou}}{{Efstathiou}}{2014}]{Efstathiou%
2014}
{Efstathiou} G.,  2014, \mnras, 440, 1138

\bibitem[\protect\citeauthoryear{{Flaherty}, {Pipher}, {Megeath}, {Winston},
  {Gutermuth}, {Muzerolle}, {Allen} \& {Fazio}}{{Flaherty}
  et~al.}{2007}]{Flaherty2007}
{Flaherty} K.~M.,  {Pipher} J.~L.,  {Megeath} S.~T.,  {Winston} E.~M.,
  {Gutermuth} R.~A.,  {Muzerolle} J.,  {Allen} L.~E.,    {Fazio} G.~G.,  2007,
  \apj, 663, 1069

\bibitem[\protect\citeauthoryear{{Freedman} \& {Madore}}{{Freedman} \&
  {Madore}}{2010}]{Freedman2010}
{Freedman} W.~L.,  {Madore} B.~F.,  2010, \araa, 48, 673

\bibitem[\protect\citeauthoryear{{Freedman} \& {Madore}}{{Freedman} \&
  {Madore}}{2011}]{Freedman2011a}
{Freedman} W.~L.,  {Madore} B.~F.,  2011, \apj, 734, 46

\bibitem[\protect\citeauthoryear{{Freedman}, {Madore}, {Scowcroft}, {Burns},
  {Monson}, {Persson}, {Seibert} \& {Rigby}}{{Freedman}
  et~al.}{2012}]{Freedman2012}
{Freedman} W.~L.,  {Madore} B.~F.,  {Scowcroft} V.,  {Burns} C.,  {Monson} A.,
  {Persson} S.~E.,  {Seibert} M.,    {Rigby} J.,  2012, \apj, 758, 24

\bibitem[\protect\citeauthoryear{{Fukugita}, {Ichikawa}, {Gunn}, {Doi},
  {Shimasaku} \& {Schneider}}{{Fukugita} et~al.}{1996}]{Fukugita1996}
{Fukugita} M.,  {Ichikawa} T.,  {Gunn} J.~E.,  {Doi} M.,  {Shimasaku} K.,
  {Schneider} D.~P.,  1996, \aj, 111, 1748

\bibitem[\protect\citeauthoryear{{Gerke}, {Kochanek}, {Prieto}, {Stanek} \&
  {Macri}}{{Gerke} et~al.}{2011}]{Gerke2011}
{Gerke} J.~R.,  {Kochanek} C.~S.,  {Prieto} J.~L.,  {Stanek} K.~Z.,    {Macri}
  L.~M.,  2011, \apj, 743, 176

\bibitem[\protect\citeauthoryear{{Gerke}, {Kochanek} \& {Stanek}}{{Gerke}
  et~al.}{2014}]{Gerke2014}
{Gerke} J.~R.,  {Kochanek} C.~S.,    {Stanek} K.~Z.,  2014, ArXiv e-prints

\bibitem[\protect\citeauthoryear{{Giallongo et~al.}}{2008}]{Giallongo2008}
{Giallongo} E.,  {Ragazzoni} R.,  {Grazian} A.,  {Baruffolo} A.,  {Beccari} G.,
   {de Santis} C.,  {Diolaiti} E.,  {di Paola} A.,  {Farinato} J.,  {Fontana}
  A.,  {Gallozzi} S.,  {Gasparo} F.,  {Gentile} G.,  {Green} R.,  {Hill} J.,
  {Kuhn} O.,  {Pasian} F.,  {Pedichini} F.,  {Radovich} M.,  {Salinari} P.,
  {Smareglia} R.,  {Speziali} R.,  {Testa} V.,  {Thompson} D.,  {Vernet} E.,
  {Wagner} R.~M.,  2008, \aap, 482, 349

\bibitem[\protect\citeauthoryear{{Gordon}, {Clayton}, {Misselt}, {Landolt} \&
  {Wolff}}{{Gordon} et~al.}{2003}]{Gordon2003}
{Gordon} K.~D.,  {Clayton} G.~C.,  {Misselt} K.~A.,  {Landolt} A.~U.,
  {Wolff} M.~J.,  2003, \apj, 594, 279

\bibitem[\protect\citeauthoryear{{Gould}}{{Gould}}{1994}]{Gould1994}
{Gould} A.,  1994, \apj, 426, 542

\bibitem[\protect\citeauthoryear{{Harris} \& {Zaritsky}}{{Harris} \&
  {Zaritsky}}{1999}]{Harris1999}
{Harris} J.,  {Zaritsky} D.,  1999, \aj, 117, 2831

\bibitem[\protect\citeauthoryear{{Hill}, {Green}, {Ashby}, {Brynnel},
  {Cushing}, {Little}, {Slagle} \& {Wagner}}{{Hill} et~al.}{2010}]{Hill2010}
{Hill} J.~M.,  {Green} R.~F.,  {Ashby} D.~S.,  {Brynnel} J.~G.,  {Cushing}
  N.~J.,  {Little} J.,  {Slagle} J.~H.,    {Wagner} R.~M.,  2010, in Society of
  Photo-Optical Instrumentation Engineers (SPIE) Conference Series Vol.~7733 of
  Society of Photo-Optical Instrumentation Engineers (SPIE) Conference Series,
  {The Large Binocular Telescope}

\bibitem[\protect\citeauthoryear{{Humphreys}, {Reid}, {Moran}, {Greenhill} \&
  {Argon}}{{Humphreys} et~al.}{2013}]{Humphreys2013}
{Humphreys} E.~M.~L.,  {Reid} M.~J.,  {Moran} J.~M.,  {Greenhill} L.~J.,
  {Argon} A.~L.,  2013, \apj, 775, 13

\bibitem[\protect\citeauthoryear{{Kanbur} \& {Ngeow}}{{Kanbur} \&
  {Ngeow}}{2004}]{Kanbur2004}
{Kanbur} S.~M.,  {Ngeow} C.-C.,  2004, \mnras, 350, 962

\bibitem[\protect\citeauthoryear{{Kelly}}{{Kelly}}{2007}]{Kelly2007}
{Kelly} B.~C.,  2007, \apj, 665, 1489

\bibitem[\protect\citeauthoryear{{Kennicutt et~al.}}{1998}]{Kennicutt1998}
{Kennicutt} Jr. R.~C.,  {Stetson} P.~B.,  {Saha} A.,  {Kelson} D.,  {Rawson}
  D.~M.,  {Sakai} S.,  {Madore} B.~F.,  {Mould} J.~R.,  {Freedman} W.~L.,
  {Bresolin} F.,  {Ferrarese} L.,  {Ford} H.,  {Gibson} B.~K.,  {Graham} J.~A.,
   {Han} M.,  {Harding} P.,  {Hoessel} J.~G.,  {Huchra} J.~P.,  {Hughes}
  S.~M.~G.,  {Illingworth} G.~D.,  {Macri} L.~M.,  {Phelps} R.~L.,
  {Silbermann} N.~A.,  {Turner} A.~M.,    {Wood} P.~R.,  1998, \apj, 498, 181

\bibitem[\protect\citeauthoryear{{Kewley} \& {Ellison}}{{Kewley} \&
  {Ellison}}{2008}]{Kewley2008}
{Kewley} L.~J.,  {Ellison} S.~L.,  2008, \apj, 681, 1183

\bibitem[\protect\citeauthoryear{{Kochanek}}{{Kochanek}}{1997}]{Kochanek1997}
{Kochanek} C.~S.,  1997, \apj, 491, 13

\bibitem[\protect\citeauthoryear{{Kuzio de Naray}, {McGaugh} \& {de
  Blok}}{{Kuzio de Naray} et~al.}{2004}]{Naray2004}
{Kuzio de Naray} R.,  {McGaugh} S.~S.,    {de Blok} W.~J.~G.,  2004, \mnras,
  355, 887

\bibitem[\protect\citeauthoryear{{Macri}, {Calzetti}, {Freedman}, {Gibson},
  {Graham}, {Huchra}, {Hughes}, {Madore}, {Mould}, {Persson} \&
  {Stetson}}{{Macri} et~al.}{2001}]{Macri2001}
{Macri} L.~M.,  {Calzetti} D.,  {Freedman} W.~L.,  {Gibson} B.~K.,  {Graham}
  J.~A.,  {Huchra} J.~P.,  {Hughes} S.~M.~G.,  {Madore} B.~F.,  {Mould} J.~R.,
  {Persson} S.~E.,    {Stetson} P.~B.,  2001, \apj, 549, 721

\bibitem[\protect\citeauthoryear{{Macri}, {Stanek}, {Bersier}, {Greenhill} \&
  {Reid}}{{Macri} et~al.}{2006}]{Macri2006}
{Macri} L.~M.,  {Stanek} K.~Z.,  {Bersier} D.,  {Greenhill} L.~J.,    {Reid}
  M.~J.,  2006, \apj, 652, 1133

\bibitem[\protect\citeauthoryear{{Mager}, {Madore} \& {Freedman}}{{Mager}
  et~al.}{2013}]{Mager2013}
{Mager} V.~A.,  {Madore} B.~F.,    {Freedman} W.~L.,  2013, \apj, 777, 79

\bibitem[\protect\citeauthoryear{{Marconi}, {Musella} \&
  {Fiorentino}}{{Marconi} et~al.}{2005}]{Marconi2005}
{Marconi} M.,  {Musella} I.,    {Fiorentino} G.,  2005, \apj, 632, 590

\bibitem[\protect\citeauthoryear{{Markwardt}}{{Markwardt}}{2009}]{Markwardt200%
9}
{Markwardt} C.~B.,  2009, in {Bohlender} D.~A.,  {Durand} D.,   {Dowler} P.,
  eds, Astronomical Data Analysis Software and Systems XVIII Vol.~411 of
  Astronomical Society of the Pacific Conference Series, {Non-linear
  Least-squares Fitting in IDL with MPFIT}.
p.~251

\bibitem[\protect\citeauthoryear{{McGaugh}}{{McGaugh}}{1991}]{McGaugh1991}
{McGaugh} S.~S.,  1991, \apj, 380, 140

\bibitem[\protect\citeauthoryear{{Mochejska}, {Macri}, {Sasselov} \&
  {Stanek}}{{Mochejska} et~al.}{2000}]{Mochejska2000}
{Mochejska} B.~J.,  {Macri} L.~M.,  {Sasselov} D.~D.,    {Stanek} K.~Z.,  2000,
  \aj, 120, 810

\bibitem[\protect\citeauthoryear{{Ngeow}, {Kanbur}, {Neilson}, {Nanthakumar} \&
  {Buonaccorsi}}{{Ngeow} et~al.}{2009}]{Ngeow2009}
{Ngeow} C.-C.,  {Kanbur} S.~M.,  {Neilson} H.~R.,  {Nanthakumar} A.,
  {Buonaccorsi} J.,  2009, \apj, 693, 691

\bibitem[\protect\citeauthoryear{{Nishiyama}, {Tamura}, {Hatano}, {Kato},
  {Tanab{\'e}}, {Sugitani} \& {Nagata}}{{Nishiyama}
  et~al.}{2009}]{Nishiyama2009}
{Nishiyama} S.,  {Tamura} M.,  {Hatano} H.,  {Kato} D.,  {Tanab{\'e}} T.,
  {Sugitani} K.,    {Nagata} T.,  2009, \apj, 696, 1407

\bibitem[\protect\citeauthoryear{{Pejcha} \& {Kochanek}}{{Pejcha} \&
  {Kochanek}}{2012}]{Pejcha2012}
{Pejcha} O.,  {Kochanek} C.~S.,  2012, \apj, 748, 107

\bibitem[\protect\citeauthoryear{{Pellegrini}, {Baldwin} \&
  {Ferland}}{{Pellegrini} et~al.}{2011}]{Pellegrini2011}
{Pellegrini} E.~W.,  {Baldwin} J.~A.,    {Ferland} G.~J.,  2011, \apj, 738, 34

\bibitem[\protect\citeauthoryear{{Persson}, {Madore}, {Krzemi{\'n}ski},
  {Freedman}, {Roth} \& {Murphy}}{{Persson} et~al.}{2004}]{Persson2004}
{Persson} S.~E.,  {Madore} B.~F.,  {Krzemi{\'n}ski} W.,  {Freedman} W.~L.,
  {Roth} M.,    {Murphy} D.~C.,  2004, \aj, 128, 2239

\bibitem[\protect\citeauthoryear{{Pietrzy{\'n}ski et~al.}}{2013}]{Pietrzy2013}
{Pietrzy{\'n}ski} G.,  {Graczyk} D.,  {Gieren} W.,  {Thompson} I.~B.,
  {Pilecki} B.,  {Udalski} A.,  {Soszy{\'n}ski} I.,  {Koz{\l}owski} S.,
  {Konorski} P.,  {Suchomska} K.,  {Bono} G.,  {Moroni} P.~G.~P.,  {Villanova}
  S.,  {Nardetto} N.,  {Bresolin} F.,  {Kudritzki} R.~P.,  {Storm} J.,
  {Gallenne} A.,  {Smolec} R.,  {Minniti} D.,  {Kubiak} M.,  {Szyma{\'n}ski}
  M.~K.,  {Poleski} R.,  {Wyrzykowski} {\L}.,  {Ulaczyk} K.,  {Pietrukowicz}
  P.,  {G{\'o}rski} M.,    {Karczmarek} P.,  2013, \nat, 495, 76

\bibitem[\protect\citeauthoryear{{Pilyugin} \& {Thuan}}{{Pilyugin} \&
  {Thuan}}{2005}]{Pilyugin2005}
{Pilyugin} L.~S.,  {Thuan} T.~X.,  2005, \apj, 631, 231

\bibitem[\protect\citeauthoryear{{Planck Collaboration}, {Ade}, {Aghanim},
  {Armitage-Caplan}, {Arnaud}, {Ashdown}, {Atrio-Barandela}, {Aumont},
  {Baccigalupi}, {Banday} \& et al.}{{Planck Collaboration}
  et~al.}{2014}]{Planck2013}
{Planck Collaboration} {Ade} P.~A.~R.,  {Aghanim} N.,  {Armitage-Caplan} C.,
  {Arnaud} M.,  {Ashdown} M.,  {Atrio-Barandela} F.,  {Aumont} J.,
  {Baccigalupi} C.,  {Banday} A.~J.,    et al. 2014, \aap, 571, A16

\bibitem[\protect\citeauthoryear{{Riess}, {Macri}, {Casertano}, {Lampeitl},
  {Ferguson}, {Filippenko}, {Jha}, {Li} \& {Chornock}}{{Riess}
  et~al.}{2011}]{Riess2011}
{Riess} A.~G.,  {Macri} L.,  {Casertano} S.,  {Lampeitl} H.,  {Ferguson} H.~C.,
   {Filippenko} A.~V.,  {Jha} S.~W.,  {Li} W.,    {Chornock} R.,  2011, \apj,
  730, 119

\bibitem[\protect\citeauthoryear{{Riess}, {Macri}, {Casertano}, {Sosey},
  {Lampeitl}, {Ferguson}, {Filippenko}, {Jha}, {Li}, {Chornock} \&
  {Sarkar}}{{Riess} et~al.}{2009}]{Riess2009}
{Riess} A.~G.,  {Macri} L.,  {Casertano} S.,  {Sosey} M.,  {Lampeitl} H.,
  {Ferguson} H.~C.,  {Filippenko} A.~V.,  {Jha} S.~W.,  {Li} W.,  {Chornock}
  R.,    {Sarkar} D.,  2009, \apj, 699, 539

\bibitem[\protect\citeauthoryear{{Romaniello}, {Primas}, {Mottini},
  {Pedicelli}, {Lemasle}, {Bono}, {Fran{\c c}ois}, {Groenewegen} \&
  {Laney}}{{Romaniello} et~al.}{2008}]{Romaniello2008}
{Romaniello} M.,  {Primas} F.,  {Mottini} M.,  {Pedicelli} S.,  {Lemasle} B.,
  {Bono} G.,  {Fran{\c c}ois} P.,  {Groenewegen} M.~A.~T.,    {Laney} C.~D.,
  2008, \aap, 488, 731

\bibitem[\protect\citeauthoryear{{Sandage}, {Tammann} \& {Reindl}}{{Sandage}
  et~al.}{2004}]{Sandage2004}
{Sandage} A.,  {Tammann} G.~A.,    {Reindl} B.,  2004, \aap, 424, 43

\bibitem[\protect\citeauthoryear{{Schlegel}, {Finkbeiner} \&
  {Davis}}{{Schlegel} et~al.}{1998}]{Schlegel1998}
{Schlegel} D.~J.,  {Finkbeiner} D.~P.,    {Davis} M.,  1998, \apj, 500, 525

\bibitem[\protect\citeauthoryear{{Shappee} \& {Stanek}}{{Shappee} \&
  {Stanek}}{2011}]{Shappee2011}
{Shappee} B.~J.,  {Stanek} K.~Z.,  2011, \apj, 733, 124

\bibitem[\protect\citeauthoryear{{Sirianni}, {Jee}, {Ben{\'{\i}}tez},
  {Blakeslee}, {Martel}, {Meurer}, {Clampin}, {De Marchi}, {Ford}, {Gilliland},
  {Hartig}, {Illingworth}, {Mack} \& {McCann}}{{Sirianni}
  et~al.}{2005}]{Sirianni2005}
{Sirianni} M.,  {Jee} M.~J.,  {Ben{\'{\i}}tez} N.,  {Blakeslee} J.~P.,
  {Martel} A.~R.,  {Meurer} G.,  {Clampin} M.,  {De Marchi} G.,  {Ford} H.~C.,
  {Gilliland} R.,  {Hartig} G.~F.,  {Illingworth} G.~D.,  {Mack} J.,
  {McCann} W.~J.,  2005, \pasp, 117, 1049

\bibitem[\protect\citeauthoryear{{Stanek} \& {Udalski}}{{Stanek} \&
  {Udalski}}{1999}]{Stanek1999}
{Stanek} K.~Z.,  {Udalski} A.,  1999, ArXiv Astrophysics e-prints

\bibitem[\protect\citeauthoryear{{Stetson}}{{Stetson}}{1996}]{Stetson1996}
{Stetson} P.~B.,  1996, \pasp, 108, 851

\bibitem[\protect\citeauthoryear{{Subramanian} \& {Subramaniam}}{{Subramanian}
  \& {Subramaniam}}{2013}]{Subramanian2013}
{Subramanian} S.,  {Subramaniam} A.,  2013, \aap, 552, A144

\bibitem[\protect\citeauthoryear{{Udalski}, {Soszynski}, {Szymanski}, {Kubiak},
  {Pietrzynski}, {Wozniak} \& {Zebrun}}{{Udalski} et~al.}{1999}]{OGLEII}
{Udalski} A.,  {Soszynski} I.,  {Szymanski} M.,  {Kubiak} M.,  {Pietrzynski}
  G.,  {Wozniak} P.,    {Zebrun} K.,  1999, \actaa, 49, 223

\bibitem[\protect\citeauthoryear{{van Albada}}{{van
  Albada}}{1980}]{vanAlbada1980}
{van Albada} G.~D.,  1980, \aap, 90, 123

\bibitem[\protect\citeauthoryear{{Weiner}, {Willmer}, {Faber}, {Harker},
  {Kassin}, {Phillips}, {Melbourne}, {Metevier}, {Vogt} \& {Koo}}{{Weiner}
  et~al.}{2006}]{Weiner2006}
{Weiner} B.~J.,  {Willmer} C.~N.~A.,  {Faber} S.~M.,  {Harker} J.,  {Kassin}
  S.~A.,  {Phillips} A.~C.,  {Melbourne} J.,  {Metevier} A.~J.,  {Vogt} N.~P.,
    {Koo} D.~C.,  2006, \apj, 653, 1049

\bibitem[\protect\citeauthoryear{{Zaritsky}, {Kennicutt} Jr. \&
  {Huchra}}{{Zaritsky} et~al.}{1994}]{Zaritsky1994}
{Zaritsky} D.,  {Kennicutt} Jr. R.~C.,    {Huchra} J.~P.,  1994, \apj, 420, 87

\end{thebibliography}
\clearpage
\begin{table*}
\begin{minipage}{140mm}
\caption{NGC 4258 Cepheids}
\label{tab:cephdata}
\begin{tabular}{lrrrrrrrrrrr}
ID & Period & RA & Dec & $\hat B$ & $\sigma_B$ & $\hat V$ & $\sigma_V$ & $\hat I$ & $\sigma_I$ & $\hat H$ & $\sigma_H$\\
\hline
1  & 23.75 & 184.84408 & +47.22925 & 25.31 & 0.10 & 24.42 & 0.05 & 23.40 & 0.07 & 22.22 & 0.12 \\
2  & 22.85 & 184.79454 & +47.31188 & 25.24 & 0.19 & 24.40 & 0.12 & 23.17 & 0.13 & ...   & ... \\
3  & 84.59 & 184.68515 & +47.28090 & 24.03 & 0.07 & 22.92 & 0.05 & 21.74 & 0.05 & 20.65 & 0.10 \\
4  & 32.29 & 184.85758 & +47.22968 & 25.60 & 0.07 & 24.52 & 0.06 & 23.44 & 0.10 & 22.30 & 0.10 \\
5  & 24.38 & 184.73314 & +47.27493 & 25.14 & 0.05 & 24.38 & 0.07 & 23.42 & 0.06 & 24.08 & 0.43 \\
6  & 38.14 & 184.76166 & +47.34665 & 25.48 & 0.24 & 23.98 & 0.08 & 22.82 & 0.09 & ...   & ... \\
7  & 31.29 & 184.80028 & +47.20735 & 25.61 & 0.14 & 24.48 & 0.07 & 23.43 & 0.05 & 22.36 & 0.12 \\
8  & 35.07 & 184.61344 & +47.35845 & 24.76 & 0.30 & 23.65 & 0.11 & 22.76 & 0.07 & ...   & ... \\
9  & 38.78 & 184.72059 & +47.39205 & 24.28 & 0.08 & 23.38 & 0.03 & 22.62 & 0.05 & ...   & ... \\
10 & 39.12 & 184.82310 & +47.28103 & 24.90 & 0.10 & 23.92 & 0.05 & 23.06 & 0.09 & ...   & ... \\
11 & 16.87 & 184.77157 & +47.22368 & 25.54 & 0.06 & 24.74 & 0.05 & 23.70 & 0.07 & 22.48 & 0.21 \\
12 & 57.48 & 184.76648 & +47.24888 & 24.03 & 0.07 & 22.93 & 0.09 & 21.89 & 0.08 & 21.05 & 0.13 \\
13 & 23.26 & 184.71926 & +47.25198 & 25.36 & 0.08 & 24.52 & 0.05 & 23.63 & 0.07 & 22.74 & 0.13 \\
14 & 25.82 & 184.69883 & +47.35581 & 24.91 & 0.05 & 24.07 & 0.03 & 23.29 & 0.07 & 22.24 & 0.20 \\
15 & 39.90 & 184.73410 & +47.40800 & 25.00 & 0.08 & 24.03 & 0.05 & 22.98 & 0.05 & ...   & ... \\
16 & 19.95 & 184.72128 & +47.26102 & 25.32 & 0.09 & 24.43 & 0.06 & 23.34 & 0.07 & 22.47 & 0.18 \\
17 & 33.89 & 184.87187 & +47.22579 & 24.73 & 0.08 & 23.74 & 0.05 & 22.93 & 0.06 & ... & ... \\
18 & 45.51 & 184.84633 & +47.24654 & 24.30 & 0.05 & 23.42 & 0.04 & 22.52 & 0.06 & ... & ... \\
19 & 33.58 & 184.71568 & +47.24173 & 24.98 & 0.06 & 24.18 & 0.04 & 23.22 & 0.07 & 22.19 & 0.10 \\
20 & 30.89 & 184.77245 & +47.24221 & 25.52 & 0.12 & 24.48 & 0.05 & 23.47 & 0.07 & 22.55 & 0.12 \\
21 & 18.45 & 184.68757 & +47.33440 & 25.32 & 0.08 & 24.47 & 0.04 & 23.56 & 0.09 & 22.62 & 0.32 \\
22 & 31.06 & 184.70008 & +47.40291 & 24.99 & 0.11 & 24.05 & 0.06 & 23.06 & 0.08 & ... & ... \\
23 & 39.06 & 184.73588 & +47.39786 & 24.83 & 0.07 & 23.83 & 0.03 & 22.86 & 0.07 & ... & ... \\
24 & 23.84 & 184.72887 & +47.37800 & 25.30 & 0.07 & 24.27 & 0.06 & 23.21 & 0.07 & 22.19 & 0.15 \\
25 & 26.99 & 184.69672 & +47.33284 & 25.27 & 0.08 & 24.42 & 0.06 & 23.48 & 0.07 & 22.35 & 0.28 \\
26 & 15.52 & 184.69183 & +47.32215 & 25.60 & 0.08 & 24.73 & 0.05 & 23.82 & 0.07 & 23.30 & 0.29 \\
27 & 21.17 & 184.76529 & +47.22284 & 25.48 & 0.16 & 24.55 & 0.10 & 23.61 & 0.06 & 23.27 & 0.19 \\
28 & 17.89 & 184.75834 & +47.33870 & 25.76 & 0.08 & 24.90 & 0.05 & 23.97 & 0.08 & ... & ... \\
29 & 26.83 & 184.74570 & +47.25310 & 25.84 & 0.07 & 24.77 & 0.05 & 23.77 & 0.09 & ... & ... \\
30 & 17.74 & 184.74875 & +47.39152 & 25.48 & 0.09 & 24.82 & 0.07 & 24.07 & 0.09 & ... & ... \\
31 & 36.95 & 184.71281 & +47.35475 & 25.49 & 0.09 & 24.43 & 0.04 & 23.30 & 0.06 & ... & ... \\
32 & 32.01 & 184.76900 & +47.28149 & 25.66 & 0.12 & 24.93 & 0.11 & 23.47 & 0.07 & 23.79 & 0.20 \\
33 & 60.98 & 184.69380 & +47.27621 & ...   & ...  & 23.64 & 0.08 & 22.44 & 0.05 & ... & ... \\
34 & 20.89 & 184.74066 & +47.23927 & 25.91 & 0.08 & 24.92 & 0.05 & 23.66 & 0.09 & 22.14 & 0.15 \\
35 & 24.59 & 184.85940 & +47.24534 & 26.03 & 0.27 & 24.66 & 0.04 & 23.67 & 0.07 & ... & ... \\
36 & 17.54 & 184.77167 & +47.24397 & 25.29 & 0.07 & 24.49 & 0.03 & 23.79 & 0.08 & 22.56 & 0.18 \\
37 & 36.44 & 184.75716 & +47.22511 & 24.54 & 0.05 & 23.69 & 0.03 & 22.71 & 0.07 & 21.89 & 0.13 \\
38 & 27.59 & 184.74555 & +47.26070 & 25.35 & 0.07 & 24.27 & 0.04 & 23.05 & 0.10 & 25.40 & 0.39 \\
39 & 36.67 & 184.71982 & +47.31071 & 25.31 & 0.10 & 24.25 & 0.09 & 23.15 & 0.08 & 22.38 & 0.36 \\
40 & 22.68 & 184.69844 & +47.33315 & 25.23 & 0.10 & 24.40 & 0.06 & 23.52 & 0.07 & 22.58 & 0.36 \\
41 & 19.84 & 184.70879 & +47.43572 & ...   & ...  & 24.06 & 0.06 & 23.39 & 0.08 & ... & ... \\
42 & 37.58 & 184.63776 & +47.40795 & 24.84 & 0.09 & 23.86 & 0.04 & 22.96 & 0.05 & ... & ... \\
43 & 44.65 & 184.61830 & +47.40159 & 25.35 & 0.07 & 24.15 & 0.05 & 23.08 & 0.06 & ... & ... \\
\hline
\end{tabular}

Periods, coordinates, and mean apparent magnitudes
($\hat B$, $\hat V$, $\hat I$, and $F160W$) of Cepheids identified
with the LBT and calibrated with HST.
\end{minipage}
\end{table*}

\clearpage
\begin{table*}
\begin{center}
\begin{minipage}{50mm}
\caption{NGC 4258 Cepheids (Unmatched) }
\label{tab:cephdata_nocal}\begin{tabular}{lrrr}
  ID & Period & RA & Dec\\
  & (days)   & (degrees) & (degrees) \\
\hline
44 & 32.26 & 184.62030 & 47.33033 \\
45 & 42.80 & 184.73132 & 47.26414 \\
46 & 17.43 & 184.77511 & 47.21588 \\
47 & 47.49 & 184.65047 & 47.29608 \\
48 & 22.09 & 184.78754 & 47.17365 \\
49 & 53.31 & 184.65292 & 47.28472 \\
50 & 25.10 & 184.85163 & 47.14524 \\
51 & 33.19 & 184.85609 & 47.16104 \\
52 & 16.53 & 184.76931 & 47.32193 \\
53 & 15.04 & 184.73743 & 47.26392 \\
54 & 17.86 & 184.76862 & 47.24926 \\
55 & 33.22 & 184.71355 & 47.32232 \\
56 & 16.99 & 184.83127 & 47.16879 \\
57 & 18.84 & 184.66743 & 47.28373 \\
58 & 17.72 & 184.73623 & 47.27552 \\
59 & 89.52 & 184.70233 & 47.26043 \\
60 & 13.40 & 184.77087 & 47.30534 \\
61 & 16.80 & 184.68944 & 47.40506 \\
62 & 26.30 & 184.71478 & 47.23186 \\
63 & 61.73 & 184.80089 & 47.30431 \\
64 & 46.93 & 184.71934 & 47.34864 \\
65 & 28.04 & 184.72227 & 47.29616 \\
66 & 44.27 & 184.74955 & 47.19773 \\
67 & 14.48 & 184.85145 & 47.19703 \\
68 & 14.15 & 184.66035 & 47.29288 \\
69 & 48.19 & 184.73100 & 47.23930 \\
70 & 28.26 & 184.64476 & 47.33938 \\
71 & 18.75 & 184.70437 & 47.46564 \\
72 & 16.82 & 184.60664 & 47.40835 \\
73 & 14.30 & 184.51651 & 47.48418 \\
74 & 19.28 & 184.61142 & 47.40765 \\
75 & 34.50 & 184.60635 & 47.41172 \\
76 & 21.70 & 184.66006 & 47.46426 \\
77 & 12.75 & 184.47592 & 47.40205 \\
78 & 43.02 & 184.64133 & 47.40476 \\
79 & 56.71 & 184.60351 & 47.39328 \\
80 & 29.82 & 184.44645 & 47.40800 \\
81 & 17.40 & 184.62580 & 47.39856 \\
\hline
\end{tabular}

Periods and coordinates for Cepheids with no identifiable match in HST
images.
\end{minipage}
\end{center}
\end{table*}
\begin{table*}
\begin{minipage}{110mm}
\caption{Matched Cepheids}
\label{tab:matches}
\begin{tabular}{lr rrrr}
  Cepheid &   LBT Period &  P$_{M06}$-P$_{LBT}$        &  $B_{M06}-B_{LBT}$ 
  &   $V_{M06}-V_{LBT}$        &  $I_{M06}-I_{LBT}$       \\
  (LBT ID) &  (Days)     &  (Days)     &  (mag)     &   (mag)     &  (mag)    \\
\hline
  26  &15.52&$-0.23$ &$-0.15  $&$-0.07   $&\phantom{$-$}$ 0.02 $    \\
  21  &18.45&$-0.06$ &\phantom{$-$}$ 0.06  $&\phantom{$-$}$ 0.04   $&$-0.02 $    \\
  40  &22.68&\phantom{$-$}$ 0.11$ &\phantom{$-$}$ 0.00  $&$-0.09   $&$-0.10 $    \\
  14  &25.82&$-0.31$ &$-0.05  $&$-0.06   $&$-0.18 $    \\
  25  &26.99&\phantom{$-$}$ 1.25$ &\phantom{$-$}$ 0.04  $&$-0.07   $&$-0.09 $    \\
  39  &36.67&\phantom{$-$}$ 0.03$ &$-0.58  $&$-0.43   $&$-0.24 $    \\
  31  &36.95&$-2.06$ &$-0.24  $&$-0.22   $&$-0.15 $    \\
  \hline
  average  &&0.58  &$-0.13$&$-0.13$&$-0.11$\\
$\sigma$&&0.72&0.21&0.14&0.08\\
\hline
\end{tabular}

Comparison of periods and mean magnitudes for Cepheids found in common
between this study and M06.
\end{minipage}
\end{table*}

\begin{table*}
\begin{minipage}{70mm}
\caption{PL Relations}
\label{tab:PLs}
\begin{tabular}{lrrrrr}
  Study&  Band&  $a_F$&   $\sigma_a$&  $b_F$&  $\sigma_b$\\
\hline
  OII&$B$&$17.368$&0.031& $-2.439$&0.046\\
  &$V$&$17.066$&0.021& $-2.779$&0.031\\
  &$I$&$16.594$&0.014& $-2.979$&0.021\\
  N09&$V$&$17.115$&0.015& $-2.769$&0.023\\
  &$I$&$16.629$&0.010& $-2.961$&0.015\\
  &$J$&$16.293$&0.009& $-3.115$&0.014\\
  &F160W &$16.122$&0.012& $-3.182$&0.019\\
  &$H$&$16.063$&0.008& $-3.206$&0.013\\
  P04 &$J$&$16.336$&0.064& $-3.153$&0.051\\
  &F160W&$16.145$&0.083& $-3.213$&0.066\\
  &$H$&$16.079$&0.053& $-3.234$&0.042\\
  \hline
  ($P>10$ days)\\
  \hline
  S04 &$B$ & 17.136&0.177 &$-2.151$&0.134 \\
      &$V$ & 16.906&0.135 &$-2.567$&0.102\\
      &$I$ & 16.456&0.111 &$-2.822$&0.084\\
  N09 &$V$ & 17.122&0.195 &$-2.746$&0.165\\
      &$I$ & 16.440&0.132 &$-2.775$&0.111\\
      &$J$ & 16.075&0.139 &$-2.909$&0.120\\
      &F160W &15.895 &0.179&$-2.968$ &0.154\\
      &$H$ & 15.832&0.113 &$-2.989$&0.096 \\
\hline
\end{tabular}

The OGLE II (\citealt{OGLEII}, OII), \citeauthor{Ngeow2009} (2009,
N09), and \citeauthor{Persson2004} (2004, P04) extinction corrected PL
relations for the LMC, of the form $L_F(P) = a_F +b_F\log P$.  PL
relations for F160W were derived by linear interpolation of the J and
H band coefficients, as a function of effective wavelength. Also
included are the \citet[S04]{Sandage2004} and N09 PL relations derived
for long period Cepheids alone ($P>10$ days).
\end{minipage}
\end{table*}

\begin{table*}
\begin{minipage}{90mm}
\caption{Extinction-only Distance Moduli}
\label{tab:results1}
\begin{tabular}{lllrr}
  &PL relations&  &$\Delta\mu_{LMC}$&  $\chi^2/dof$\\
  \hline
  OII  $BVI$ &        &P04 F160W&   $10.70\pm0.03$&1.00 \\
  OII  $BVI$ &N09 F160W   &     &   $10.70\pm0.03$&1.01 \\
  OII  $BI $ &N09 $V$   &P04 F160W& $10.70\pm0.03$&1.09 \\
  OII  $BI $ &N09 $V$,F160W  &     &$10.69\pm0.03$&1.10 \\
  OII  $BV $ &N09 $I$   &P04 F160W& $10.62\pm0.03$&1.06 \\
  OII  $BV $ &N09 $I$,F160W  &     &$10.62\pm0.03$&1.05 \\
  OII  $B  $ &N09 $VI$  &P04 F160W& $10.62\pm0.03$&1.02 \\
  OII  $B  $ &N09 $VI$,F160W &     &$10.61\pm0.03$&1.01 \\
  \hline
  ($P>10$ days)   \\
  \hline
  S04$BVI$ & N09 F160W      & &$10.67\pm0.03$ &1.04\\
  S04$B$   & N09 $VI$,F160W & &$10.60\pm0.03$ &1.30\\
  \hline
\end{tabular}

OII is OGLE II, \citealt{OGLEII}, N09 is \citealt{Ngeow2009}, P04 is
\citealt{Persson2004}, and S04 is \citealt{Sandage2004}.
\end{minipage}
\end{table*}

\begin{table*}
\begin{minipage}{90mm}
\caption{Radial Fits}
\label{tab:radial}
\begin{tabular}{lrrrr}
  Band &  $a$&  $\sigma_a$&  $b$&  $\sigma_b$\\
\hline
  LBT Cepheids Only ($N=40$)\\
  $B$&$0.03$&$\pm0.04 $  &$-0.04$&$\pm0.05 $ \\ 
  $V$&$ 0.00$&$\pm0.03 $ &$0.00$&$\pm0.03 $ \\ 
  $I$&$-0.03$&$\pm0.05 $ &$0.03$&$\pm0.05 $ \\ 
  $H$&$-0.09$&$\pm0.18 $ &$0.06$&$\pm0.20 $ \\ 
  \hline
  LBT+M06 Cepheids ($N=122$)\\
  $B$&$0.02$  &$\pm0.02 $ &$-0.04$&$\pm0.03 $ \\ 
  $V$&$0.01$  &$\pm0.01 $ &$-0.00$&$\pm0.02 $ \\ 
  $I$&$-0.03$ &$\pm0.01 $ &$0.05$ &$\pm0.02 $ \\ 
  $H$&$-0.11$ &$\pm0.20 $ &$0.06$ &$\pm0.22 $  \\
\hline
\end{tabular}

Results of linear fits ($a+b(\rho_i/\rho_0$)) to the residuals from
Model 1 as a function of galactocentric radius (see Figure
\ref{fig:radiusresidual}).
\end{minipage}
\end{table*}

\begin{table*}  
\begin{minipage}{150mm}
\caption{Metallicity Fits ($\gamma_1$; $\gamma_2 \equiv 0$)}
\label{tab:metaldistance1}
\begin{tabular}{lcccccc}
Model & Gradient Slope & $R_V$ & $\Delta\mu_{LMC}$ & $\gamma_1$ &  $\chi^2/dof$ & $\mu_{LMC}$ \\
\hline
LBT Cepheids Only ($N=40$)\\
Standard       & $-0.49\pm0.08$ & 3.10 & $10.70\pm0.03$ &    & 1.00 & $18.70\pm0.07$ \\
Extinction Law & $-0.49\pm0.08$ & 4.90 & $10.60\pm0.03$ &    & 0.90 & $18.81\pm0.07$ \\
Z94-1 & $-0.49\pm0.08$ & 3.10 & $10.75\pm0.05$ & $-0.18\pm0.20$ & 1.00 & $18.66\pm0.08$ \\
Z94-2 & $-0.28\pm0.04$ & 3.10 & $10.80\pm0.11$ & $-0.32\pm0.35$ & 1.00 & $18.60\pm0.13$ \\
B11-e & $-0.18\pm0.03$ & 3.10 & $10.74\pm0.05$ & $-0.50\pm0.54$ & 1.00 & $18.66\pm0.08$ \\
Z94-2 (boot) & \Zslope & 3.10 & \bootZstandardtwomu & \bootZstandardtwogammaa &  & $18.61\pm  0.13$  \\
B11-e (boot) & \Bslope & 3.10 & \bootBstandardmu    & \bootBstandardgammaa    &  & $18.66 \pm 0.10$ \\
Z94-1 & $-0.49\pm0.08$ & 4.90 & $10.65\pm0.07$ & $-0.22\pm0.25$ & 0.91 & $18.76\pm0.09$ \\
Z94-2 & $-0.28\pm0.04$ & 4.90 & $10.72\pm0.14$ & $-0.38\pm0.43$ & 0.91 & $18.69\pm0.15$ \\
B11-e & $-0.18\pm0.03$ & 4.90 & $10.64\pm0.06$ & $-0.59\pm0.67$ & 0.91 & $18.76\pm0.09$ \\
Z94-2 (boot) & \Zslope & 4.90 & \bootZextlawtwomu & \bootZextlawtwogammaa &  & $18.71\pm  0.15$ \\
B11-e (boot) & \Bslope & 4.90 & \bootBextlawmu    & \bootBextlawgammaa    &  & $18.77\pm  0.11$ \\
\hline
LBT+M06 Cepheids ($N=122$)\\
Standard       & & 3.10 & $10.73\pm0.01$ &  &  1.00 & $18.68\pm0.07$ \\
Extinction Law & & 4.90 & $10.59\pm0.02$ &  &  0.91 & $18.82\pm0.07$ \\
Z94-1 & $-0.49\pm0.08$ & 3.10 & $10.81\pm0.03$ & $-0.24\pm0.08$ & 0.97 & $18.59\pm0.07$ \\
Z94-2 & $-0.28\pm0.04$ & 3.10 & $10.89\pm0.06$ & $-0.42\pm0.14$ & 0.97 & $18.52\pm0.09$ \\
B11-e & $-0.18\pm0.03$ & 3.10 & $10.81\pm0.03$ & $-0.65\pm0.22$ & 0.97 & $18.59\pm0.07$ \\
Z94-2 (boot) & \Zslope & 3.10 & \bootallZstandardtwomu & \bootallZstandardtwogammaa &  & $18.43 \pm 0.11$ \\
B11-e (boot) & \Bslope & 3.10 & \bootallBstandardmu    & \bootallBstandardgammaa    &  & $18.53 \pm 0.11$ \\
Z94-1 & $-0.49\pm0.08$ & 4.90 & $10.73\pm0.04$ & $-0.40\pm0.10$ & 0.85 & $18.67\pm0.08$ \\
Z94-2 & $-0.28\pm0.04$ & 4.90 & $10.86\pm0.07$ & $-0.71\pm0.18$ & 0.85 & $18.55\pm0.10$ \\
B11-e & $-0.18\pm0.03$ & 4.90 & $10.73\pm0.04$ & $-1.10\pm0.28$ & 0.85 & $18.68\pm0.08$ \\
Z94-2 (boot) & \Zslope & 4.90 & \bootallZextlawtwomu & \bootallZextlawtwogammaa &  & $18.46\pm  0.14$ \\
B11-e (boot) & \Bslope & 4.90 & \bootallBextlawmu    & \bootallBextlawgammaa    &  & $18.61\pm  0.15$ \\
\hline
\end{tabular}
\end{minipage}
\end{table*}

\begin{table*}
\begin{minipage}{170mm}
\caption{Metallicity Fits [$\gamma_1$ \& $\gamma_2$; $\boldsymbol{c} = \boldsymbol{p}_1$]}
\label{tab:metaldistance2}
\begin{tabular}{lccccccc}
Model & Gradient Slope & $R_V$ & $\Delta\mu_{LMC}$ & $\gamma_1$ & $\gamma_2$ & $\chi^2/dof$ & $\mu_{LMC}$ \\
\hline
LBT Cepheids Only ($N=40$)\\
Z94-1 & $-0.49\pm0.08$ & 3.10 & $10.75\pm0.05$ & $0.00\pm0.29$ & $-0.29\pm0.33$ & 1.0 & $18.65\pm0.08$ \\
Z94-2 & $-0.28\pm0.04$ & 3.10 & $10.81\pm0.11$ & $-0.14\pm0.39$ & $-0.28\pm0.27$ & 1.0 & $18.60\pm0.13$ \\
B11-e & $-0.18\pm0.03$ & 3.10 & $10.75\pm0.05$ & $0.02\pm0.81$ & $-0.80\pm0.94$ & 1.0 & $18.66\pm0.08$ \\
Z94-2 (boot) & \Zslope & 3.10 & \bootZstandardtwogtwomu  & \bootZstandardtwogtwogammaa &\bootZstandardtwogtwogammab &  & $18.61 \pm0.13  $ \\
B11-e (boot) & \Bslope & 3.10 & \bootBstandardgtwomu    & \bootBstandardgtwogammaa    &\bootBstandardgtwogammab    &  & $18.66 \pm0.10  $ \\
Z94-1 & $-0.49\pm0.08$ & 4.90 & $10.64\pm0.07$ & $-0.56\pm0.44$ & $0.41\pm0.44$ & 0.91 & $18.76\pm0.09$ \\
Z94-2 & $-0.28\pm0.04$ & 4.90 & $10.71\pm0.14$ & $-0.67\pm0.53$ & $0.34\pm0.37$ & 0.91 & $18.69\pm0.15$ \\
B11-e & $-0.18\pm0.03$ & 4.90 & $10.64\pm0.06$ & $-1.56\pm1.24$ & $1.16\pm1.26$ & 0.91 & $18.76\pm0.09$ \\
Z94-2 (boot) & \Zslope & 3.10 & \bootZextlawtwogtwomu & \bootZextlawtwogtwogammaa & \bootZextlawtwogtwogammab & & $18.71 \pm0.15  $ \\
B11-e (boot) & \Bslope & 3.10 & \bootBextlawgtwomu    & \bootBextlawgtwogammaa    & \bootBextlawgtwogammab    & & $18.77 \pm0.11  $ \\
\hline
LBT+M06 Cepheids ($N=122$)\\
Z94-1 & $-0.49\pm0.08$ & 3.10 & $10.82\pm0.03$ & $-0.00\pm0.15$ & $-0.44\pm0.24$ & 0.96 & $18.58\pm0.07$ \\
Z94-2 & $-0.28\pm0.04$ & 3.10 & $10.90\pm0.06$ & $-0.20\pm0.18$ & $-0.44\pm0.22$ & 0.96 & $18.50\pm0.09$ \\
B11-e & $-0.18\pm0.03$ & 3.10 & $10.82\pm0.03$ & $0.02\pm0.43$ & $-1.23\pm0.68$ & 0.96 & $18.59\pm0.07$ \\
Z94-2 (boot) & \Zslope & 3.10 & \bootallZstandardtwogtwomu & \bootallZstandardtwogtwogammaa &\bootallZstandardtwogtwogammab &  & $18.42 \pm0.11  $ \\
B11-e (boot) & \Bslope & 3.10 & \bootallBstandardgtwomu    & \bootallBstandardgtwogammaa    &\bootallBstandardgtwogammab    &  & $18.52 \pm0.11  $   \\
Z94-1 & $-0.49\pm0.08$ & 4.90 & $10.73\pm0.04$ & $-0.61\pm0.25$ & $0.30\pm0.33$ & 0.85 & $18.68\pm0.08$ \\
Z94-2 & $-0.28\pm0.04$ & 4.90 & $10.85\pm0.07$ & $-0.84\pm0.26$ & $0.22\pm0.30$ & 0.86 & $18.56\pm0.10$ \\
B11-e & $-0.18\pm0.03$ & 4.90 & $10.72\pm0.04$ & $-1.70\pm0.71$ & $0.87\pm0.95$ & 0.85 & $18.68\pm0.08$ \\
Z94-2 (boot) & \Zslope & 3.10 & \bootallZextlawtwogtwomu & \bootallZextlawtwogtwogammaa & \bootallZextlawtwogtwogammab & & $18.47 \pm0.14  $ \\
B11-e (boot) & \Bslope & 3.10 & \bootallBextlawgtwomu    & \bootallBextlawgtwogammaa    & \bootallBextlawgtwogammab    & & $18.61 \pm0.15  $ \\
\hline
\end{tabular}
\end{minipage}
\end{table*}

\begin{table*}
\begin{minipage}{170mm}
\caption{Metallicity Fits [$\gamma_1$ \& $\gamma_2$; $\boldsymbol{c} = \boldsymbol{p}_{BVI}$]}
\label{tab:redo_gamma2}
\begin{tabular}{lrrrrrrr}
Model & Gradient Slope & $R_V$ & $\Delta\mu_{LMC}$ & $\gamma_1$ & $\gamma_2$ & $\chi^2/dof$ & $\mu_{LMC}$ \\
\hline
LBT Cepheids Only ($N=40$)\\
Z94-1 & $-0.49\pm0.08$ & 3.10 & $10.76\pm0.06$ & $-0.65\pm0.86$ & $-0.26\pm0.47$ & 0.84 & $18.64\pm0.09$ \\
Z94-2 & $-0.28\pm0.04$ & 3.10 & $10.82\pm0.12$ & $-0.84\pm0.82$ & $-0.29\pm0.40$ & 0.83 & $18.58\pm0.14$ \\
B11-e & $-0.18\pm0.03$ & 3.10 & $10.76\pm0.06$ & $-1.77\pm2.46$ & $-0.72\pm1.33$ & 0.84 & $18.65\pm0.09$ \\
Z94-2 (boot) & \Zslope & 3.10 & \bootZstandardtwogtwoBVImu  & \bootZstandardtwogtwoBVIgammaa &\bootZstandardtwogtwoBVIgammab &  & $18.60 \pm0.13  $  \\
B11-e (boot) & \Bslope & 3.10 & \bootBstandardgtwoBVImu    & \bootBstandardgtwoBVIgammaa    &\bootBstandardgtwoBVIgammab    &  & $18.65 \pm0.10  $ \\
ZZ94-1 & $-0.49\pm0.08$ & 4.90 & $10.64\pm0.07$ & $0.02\pm0.70$ & $0.12\pm0.28$ & 0.73 & $18.77\pm0.10$ \\
Z94-2 & $-0.28\pm0.04$ & 4.90 & $10.72\pm0.16$ & $-0.20\pm0.74$ & $0.10\pm0.23$ & 0.73 & $18.69\pm0.17$ \\
B11-e & $-0.18\pm0.03$ & 4.90 & $10.63\pm0.07$ & $0.09\pm1.97$ & $0.34\pm0.79$ & 0.73 & $18.77\pm0.10$ \\
Z94-2 (boot) & \Zslope & 3.10 & \bootZextlawtwogtwoBVImu & \bootZextlawtwogtwoBVIgammaa & \bootZextlawtwogtwoBVIgammab & & $18.71 \pm0.15  $ \\
B11-e (boot) & \Bslope & 3.10 & \bootBextlawgtwoBVImu    & \bootBextlawgtwoBVIgammaa    & \bootBextlawgtwoBVIgammab    & & $18.78 \pm0.11  $  \\
\hline
LBT+M06 Cepheids ($N=122$)\\
Z94-1 & $-0.49\pm0.08$ & 3.10 & $10.83\pm0.04$ & $-0.61\pm0.33$ & $-0.20\pm0.18$ & 0.9 & $18.57\pm0.07$ \\
Z94-2 & $-0.28\pm0.04$ & 3.10 & $10.91\pm0.06$ & $-0.84\pm0.35$ & $-0.22\pm0.19$ & 0.9 & $18.49\pm0.09$ \\
B11-e & $-0.18\pm0.03$ & 3.10 & $10.83\pm0.03$ & $-1.69\pm0.91$ & $-0.56\pm0.52$ & 0.9 & $18.58\pm0.07$ \\
Z94-2 (boot) & \Zslope & 3.10 & \bootallZstandardtwogtwoBVImu & \bootallZstandardtwogtwoBVIgammaa &\bootallZstandardtwogtwoBVIgammab &  &$18.42 \pm0.11  $  \\
B11-e (boot) & \Bslope & 3.10 & \bootallBstandardgtwoBVImu    & \bootallBstandardgtwoBVIgammaa    &\bootallBstandardgtwoBVIgammab    &  &$18.52 \pm0.11  $  \\
Z94-1 & $-0.49\pm0.08$ & 4.90 & $10.74\pm0.05$ & $-0.14\pm0.26$ & $0.13\pm0.11$ & 0.79 & $18.66\pm0.08$ \\
Z94-2 & $-0.28\pm0.04$ & 4.90 & $10.87\pm0.08$ & $-0.45\pm0.30$ & $0.13\pm0.11$ & 0.79 & $18.53\pm0.10$ \\
B11-e & $-0.18\pm0.03$ & 4.90 & $10.73\pm0.04$ & $-0.36\pm0.73$ & $0.36\pm0.31$ & 0.79 & $18.67\pm0.08$ \\
Z94-2 (boot) & \Zslope & 3.10 & \bootallZextlawtwogtwoBVImu & \bootallZextlawtwogtwoBVIgammaa & \bootallZextlawtwogtwoBVIgammab & &  $18.45 \pm0.15  $ \\
B11-e (boot) & \Bslope & 3.10 & \bootallBextlawgtwoBVImu    & \bootallBextlawgtwoBVIgammaa    & \bootallBextlawgtwoBVIgammab    & & $18.61 \pm0.15  $ \\
\hline
\end{tabular}
\end{minipage}
\end{table*}

\begin{table*}
\begin{minipage}{160mm}
\caption{Bootstrapping Fits with Prior}
\label{tab:priors}
\begin{tabular}{lrlrrrr}
Model &Gradient Slope & $\boldsymbol{c}$ & $\Delta\mu_{LMC}$ &$\gamma_1$ & $\gamma_2$ & $\mu_{LMC}$\\
\hline
LBT Cepheids Only ($N=40$)\\
Z94-2 (boot) & \Zslope & $\boldsymbol{p}_{1}$  & \bootZstandardtwogtwopriormu  & \bootZstandardtwogtwopriorgammaa &\bootZstandardtwogtwopriorgammab            & $18.56 \pm0.09  $  \\
B11-e (boot) & \Bslope & $\boldsymbol{p}_{1}$  & \bootBstandardgtwopriormu    & \bootBstandardgtwopriorgammaa    &\bootBstandardgtwopriorgammab                & $18.63 \pm0.09  $ \\
Z94-2 (boot) & \Zslope & $\boldsymbol{p}_{BVI}$ & \bootZstandardtwogtwoBVIpriormu  & \bootZstandardtwogtwoBVIpriorgammaa &\bootZstandardtwogtwoBVIpriorgammab   & $18.55 \pm0.08  $  \\
B11-e (boot) & \Bslope & $\boldsymbol{p}_{BVI}$ & \bootBstandardgtwoBVIpriormu    & \bootBstandardgtwoBVIpriorgammaa    &\bootBstandardgtwoBVIpriorgammab       & $18.62 \pm0.09  $ \\
\hline
LBT+M06 Cepheids ($N=122$)\\
Z94-2 (boot) & \Zslope & $\boldsymbol{p}_{1}$ & \bootallZstandardtwogtwopriormu & \bootallZstandardtwogtwopriorgammaa &\bootallZstandardtwogtwopriorgammab             &$18.43 \pm0.11  $  \\
B11-e (boot) & \Bslope & $\boldsymbol{p}_{1}$ & \bootallBstandardgtwopriormu    & \bootallBstandardgtwopriorgammaa    &\bootallBstandardgtwopriorgammab                &$18.52 \pm0.11  $  \\
Z94-2 (boot) & \Zslope & $\boldsymbol{p}_{BVI}$ & \bootallZstandardtwogtwoBVIpriormu & \bootallZstandardtwogtwoBVIpriorgammaa &\bootallZstandardtwogtwoBVIpriorgammab   &$18.42 \pm0.11  $  \\
B11-e (boot) & \Bslope & $\boldsymbol{p}_{BVI}$ & \bootallBstandardgtwoBVIpriormu    & \bootallBstandardgtwoBVIpriorgammaa    &\bootallBstandardgtwoBVIpriorgammab      &$18.52 \pm0.11  $  \\
\hline
\end{tabular}

The prior imposed on $\Delta\mu_{LMC}$ is a gaussian with mean $10.91$
and width $0.08$, based on the \citet{Pietrzy2013} eclipsing binary
distance for the LMC and the maser distance of NGC 4258 from
\citet{Humphreys2013}.  See the text for definitions of $\boldsymbol{c}$,
$\boldsymbol{p}_1$, and $\boldsymbol{p}_{BVI}$.  $R_V=3.1$ for all of these models.
\end{minipage}
\end{table*}

\begin{table*}
\begin{minipage}{70mm}
  \caption{Absolute PL relations}
  \label{tab:absPLs}
  \begin{tabular}{lrrrrr}
    Study&  Band&  $a$&   $\sigma_a$&  $b$&  $\sigma_b$\\
    \hline
    OII &$B$&$-1.20$&0.14& $-2.439$&0.046\\
        &$V$&$-1.50$&0.14& $-2.779$&0.031\\
        &$I$&$-1.98$&0.14& $-2.979$&0.021\\
    P04 &$J$&$-2.23$&0.14& $-3.153$&0.013\\
    &F160W  &$-2.42$&0.14& $-3.213$&0.013\\
    &$H$    &$-2.49$&0.14& $-3.234$&0.013\\
    \hline
  \end{tabular}

  The calibrated optical OGLE II (OII, \citealt{OGLEII}),
  and near-IR \citeauthor{Persson2004} (2004, P04) PL relations, using
  $\mu_{LMC} = $\muLMCM.  The PL relations are given in the form
  $L_F(P) = a_F +b_F\log P$.
\end{minipage}
\end{table*}

\end{document}